\documentclass[aps,prd,superscriptaddress, twocolumn,floatfix]{revtex4-1}
\usepackage[utf8]{inputenc} 
\usepackage[T1]{fontenc}    
\usepackage{hyperref}       
\usepackage{url}            
\usepackage{booktabs}       
\usepackage{amsfonts}       
\usepackage{nicefrac}       
\usepackage{microtype}      
\usepackage{multirow}
\usepackage{tabu}
\usepackage[table,xcdraw]{xcolor}
\usepackage{caption}
\usepackage{subcaption}
\usepackage{cleveref}
\usepackage{float}
\usepackage{stfloats}
\usepackage{amsmath}
\usepackage{graphicx}
\usepackage{xfrac}
\usepackage{booktabs}
\usepackage{afterpage, placeins}

\begin{document}

\title{\textsc{CaloGAN}: Simulating 3D High Energy Particle Showers in Multi-Layer Electromagnetic Calorimeters with Generative Adversarial Networks}

\author{Michela Paganini}
\email[]{michela.paganini@yale.edu}
\affiliation{Department of Physics, Yale University, New Haven, CT 06520, USA}
\affiliation{Lawrence Berkeley National Laboratory, Berkeley, CA, 94720, USA} 

\author{Luke de Oliveira}
\email[]{lukedeoliveira@lbl.gov}
\affiliation{Lawrence Berkeley National Laboratory, Berkeley, CA, 94720, USA} 

\author{Benjamin Nachman}
\email[]{bnachman@cern.ch}
\affiliation{Lawrence Berkeley National Laboratory, Berkeley, CA, 94720, USA} 

\date{\today}

\begin{abstract}
The precise modeling of subatomic particle interactions and propagation through matter is paramount for the advancement of nuclear and particle physics searches and precision measurements. The most computationally expensive step in the simulation pipeline of a typical experiment at the Large Hadron Collider (LHC) is the detailed modeling of the full complexity of physics processes that govern the motion and evolution of particle showers inside calorimeters. 
We introduce \textsc{CaloGAN}, a new fast simulation technique based on generative adversarial networks (GANs). We apply these neural networks to the modeling of electromagnetic showers in a longitudinally segmented calorimeter, and achieve speedup factors comparable to or better than existing full simulation techniques on CPU ($100\times$-$1000\times$) and even faster on GPU (up to $\sim10^5\times$). There are still challenges for achieving precision across the entire phase space, but our solution can reproduce a variety of geometric shower shape properties of photons, positrons and charged pions.
This represents a significant stepping stone toward a full neural network-based detector simulation that could save significant computing time and enable many analyses now and in the future. 
\end{abstract}

\maketitle

\section{Introduction}
\label{sec:intro}

The physics programs of all experiments based at the LHC rely heavily on detailed simulation for all aspects of event reconstruction and data analysis. Simulated particle collisions, decays, and material interactions are used to interpret the results of ongoing experiments and estimate the performance of new ones, including detector upgrades. 

State-of-the-art simulations are able to precisely model detector geometries and physical processes spanning distance scales as small as $10^{-20}$ m for the initial parton-parton scattering, all the way to the material interactions at meter length scales. These processes, which include nuclear and atomic interactions, such as ionization, as well as strong, weak, and electromagnetic processes, will alter the state of incoming particles as they propagate through and interact with layers of material in the various detector components. Detection techniques such as calorimetry exploit these physical interactions to detect the presence and measure the energy of particles such as photons, electrons and hadrons via their interactions with hundreds of thousands of detector components. Upon interaction with a calorimeter, a cascade (\textit{shower}) of secondary particles is produced and their energy is collected and transformed into electric signals.

Physics-based (\textit{full simulation}) modeling of particle showers in calorimeters (with \textsc{Geant4}~\cite{Geant} as the state of the art) is the most computationally demanding part of the whole simulation process, and can take minutes per event on modern, distributed high performance platforms~\cite{Aad:2010ah,1742-6596-396-6-062016}. The production of physics results is often limited by the absence of adequate Monte Carlo (MC) simulation, and the increase in luminosity at the LHC will only exacerbate the problem. For example, the ATLAS and CMS experiments at the high-luminosity phase of the LHC (HL-LHC) will each see about 3 billion top quark pair events~\cite{Czakon:2011xx,Botje:2011sn,Martin:2009bu,Gao:2013xoa,Ball:2012cx,Khachatryan:2016kzg,Aaboud:2016syx}; for a MC statistical uncertainty that is significantly below the data uncertainty, hundreds of billion simulated events would be required. This is not possible using full detector simulation techniques with existing computing resources. Currently, full MC simulation occupies 50-70$\%$ of the experiments' worldwide computing resources, equivalent to billions of CPU hours per year~\cite{Flynn:2002240,dashboard,Bozzi:1984010}. 

The relevance of the calorimeter simulation step has sparked the development of approximate, fast simulation solutions to mitigate its computational complexity. Fast simulation techniques rely on parametrized showers~\cite{Grindhammer:1993kw,ATLAS:1300517,Grindhammer:1989zg} for fluctuations, and look-up tables for low energy interactions~\cite{frozen}. For many applications, these techniques are sufficient. However, analyses that utilize the detailed structure of showers for particle identification as well as energy and direction calibration may not be able to rely on these simplified approaches~\cite{ATL-SOFT-PUB-2014-001}.

We introduce a Deep Learning model to enable high-fidelity fast simulation of particle showers in electromagnetic calorimeters. Previous work~\cite{deOliveira:2017pjk} assessed the viability of GAN-based simulation of \textit{jet-images}~\cite{Cogan:2014oua} -- sparse, structured, 2D representations of jet fragmentation analogous to a single-layer, idealized calorimeter -- and focused on providing architectural guidelines for this regime. Neural network-based generation, including GANs, Variational Auto-Encoders~\cite{VAE}, and Adversarial Auto-Encoders~\cite{AAE}, have also been tested in other areas of science, such as Cosmology~\cite{CosmoGAN1, CosmoGAN2}, Condensed Matter Physics~\cite{condmatter}, and Oncology~\cite{oncology}. The longitudinally segmented calorimeter simulation addressed in this work offers unique challenges due to the sparsity of hit cells, the non-uniform granularity among the detector layers, and their sequential structure. In addition to enabling physics analyses at the LHC, the \textsc{CaloGAN} may form a base for solving similar computationally intensive modeling problems in other domains of science, medicine, and technology.

The paper is organized as follows.  Section~\ref{sec:dataset} introduces the dataset of calorimeter showers and Sec.~\ref{sec:GANreview} briefly reviews the generic GAN setup. The \textsc{CaloGAN} is described in Sec.~\ref{sec:calogan} and first results of its performance are documented in Sec.~\ref{sec:performance}. The paper ends with conclusions and future outlook in Sec.~\ref{sec:conclusions}.

\section{Dataset}
\label{sec:dataset}
A detector simulation begins with a list of particles with lifetimes greater than $\mathcal{O}(\text{mm}/c)$.  For each particle, we are given its type (e.g. electron, pion, etc.), its energy, and its direction. The particle type determines when and how the particle interacts with the material along its trajectory. Material interactions with the detector factorize~\footnote{Energy losses factorize, but detector readout does not. Due to threshold and digitization effects, the energy readout from two energy deposits in different detector elements need not be the same as the recorded energy from the two deposits in the same element. In detector simulations, these non-linear effects are treated after accounting for the material interactions and are therefore beyond the scope of the \textsc{CaloGAN}. It may be interesting in future work to consider an end-to-end generator that includes these effects, but it may not save a lot of time since simulation is much more costly than reconstruction.}: the energy deposited in a calorimeter by various particles is the sum of the energy from each shower treated independently.

There are two flavors of calorimeters: electromagnetic and hadronic. Electromagnetic calorimeters are designed to stop electrons and photons, which have shallower and narrower showers compared with protons, neutrons, and charged pions. Hadronic calorimeters are thicker and deeper in order to capture penetrating radiation that forms irregular showers from nuclear interactions. In this first application of GANs to a longitudinally segmented calorimeter, we choose to focus only on electromagnetic showers. In addition to already providing the capability to simulate electrons and photons, the electromagnetic shower contains all of the new challenges described in Sec.~\ref{sec:intro}.

Transverse segmentation is critical for particle identification and energy calibration in an electromagnetic calorimeter. For example, the radiation pattern can be used to distinguish prompt photons from $\pi^0\rightarrow \gamma\gamma$, where the distance between the two photons is $\mathcal{O}(\text{cm})$ for a 10 GeV $\pi^0$ at one meter from the interaction point. Pion rejection and an excellent resolution for photons in the Higgs boson $H\rightarrow \gamma\gamma$ discovery channel were driving factors for the design of the ATLAS Liquid Argon (LAr) electromagnetic calorimeter~\cite{larcalo}, which will serve as an inspiration for the calorimeter used in this study. In particular, the calorimeter used in this study is a cube with size $480$ mm$^3$ with no material in front of it.  There are three instrumented layers in the radial ($z$) direction~\footnote{This is the direction that prompt neutral particles at $\eta=0$ would enter the calorimeter without any prior material interactions.} with thicknesses 90 mm, 347 mm, and 43 mm. The active material is LAr and the absorber material is lead. Only the total energy per layer, that includes both the active and inactive contributions is used in what follows.

In contrast to the complex accordion geometry in the actual ATLAS calorimeter, our simplified setup (built on the \textsc{Geant4} B4 example) uses flat alternating layers of lead and LAr that are 2 mm and 4 mm thick, respectively. Each of the three layers has a different segmentation, which is also not square in the first and third layers. In particular, the cells in the first layer are 160 mm $\times$ 5 mm, the cells in the second layer are 40 mm$^2$, and the cells in the third layer are $40\times 80$ mm$^2$. The short direction in the first layer ($\eta$) corresponds to what would be the $pp$ beam direction in a full experiment. In contrast, the short direction in the third layer ($\phi$) is perpendicular to $\eta$. Table~\ref{table:dimensions} summarizes the calorimeter geometry.

\afterpage{
\begin{table}
\centering
\begin{tabular}{@{}*{4}{p{.12\textwidth}@{}}} 
 \toprule
  Layer  & $z$ length & $\eta$ length & $\phi$ length \\ 
   & [mm] & [mm] & [mm] \\
 \midrule
 0 & 90 & 5 & 160 \\ 
 1 & 347 & 40 & 40 \\
 2 & 43 & 80 & 40 \\
 \bottomrule
\end{tabular}
\caption{Dimension of a calorimeter cell. The $z$ direction is the direction of particle propagation (radial direction in a full experiment), the $\eta$ direction would be along the $pp$ beam axis in a full experiment, and $\phi$ is perpendicular to $z$ and $\eta$.}
\label{table:dimensions}
\end{table}
}

\afterpage{
\begin{figure}
    \centering
    \begin{subfigure}[]{0.48\textwidth}
        \centering
        \includegraphics[width=\textwidth]{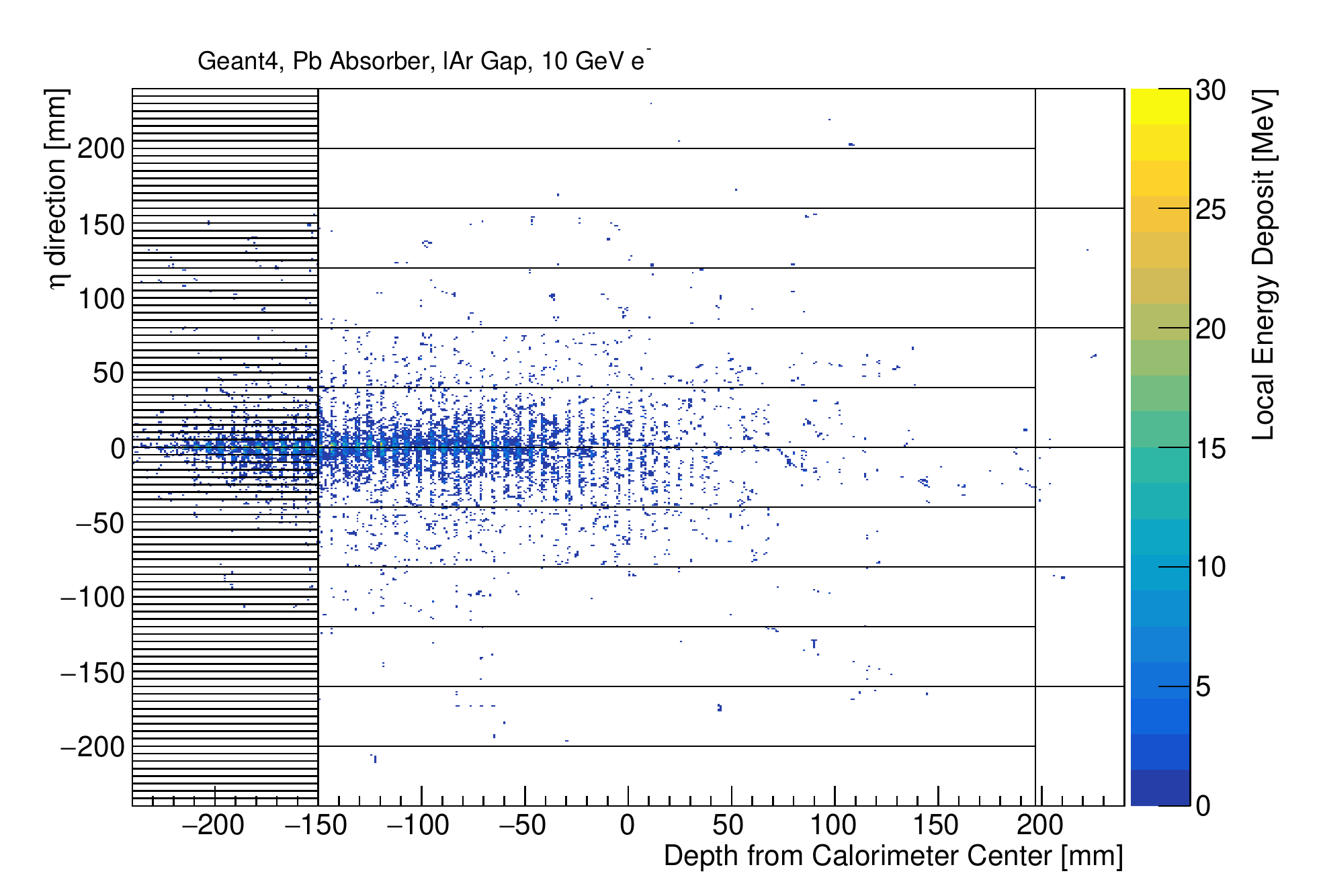}
        \caption{Each dot represents one energy deposit from \textsc{Geant4} and the color of the dot encodes the energy. The absorber-gap structure is clearly visible, where most of the energy is lost in the absorber.}
        \label{fig:calo_image2}
    \end{subfigure}%
    
    \begin{subfigure}[]{0.48\textwidth}
        \centering
        \includegraphics[width=\textwidth]{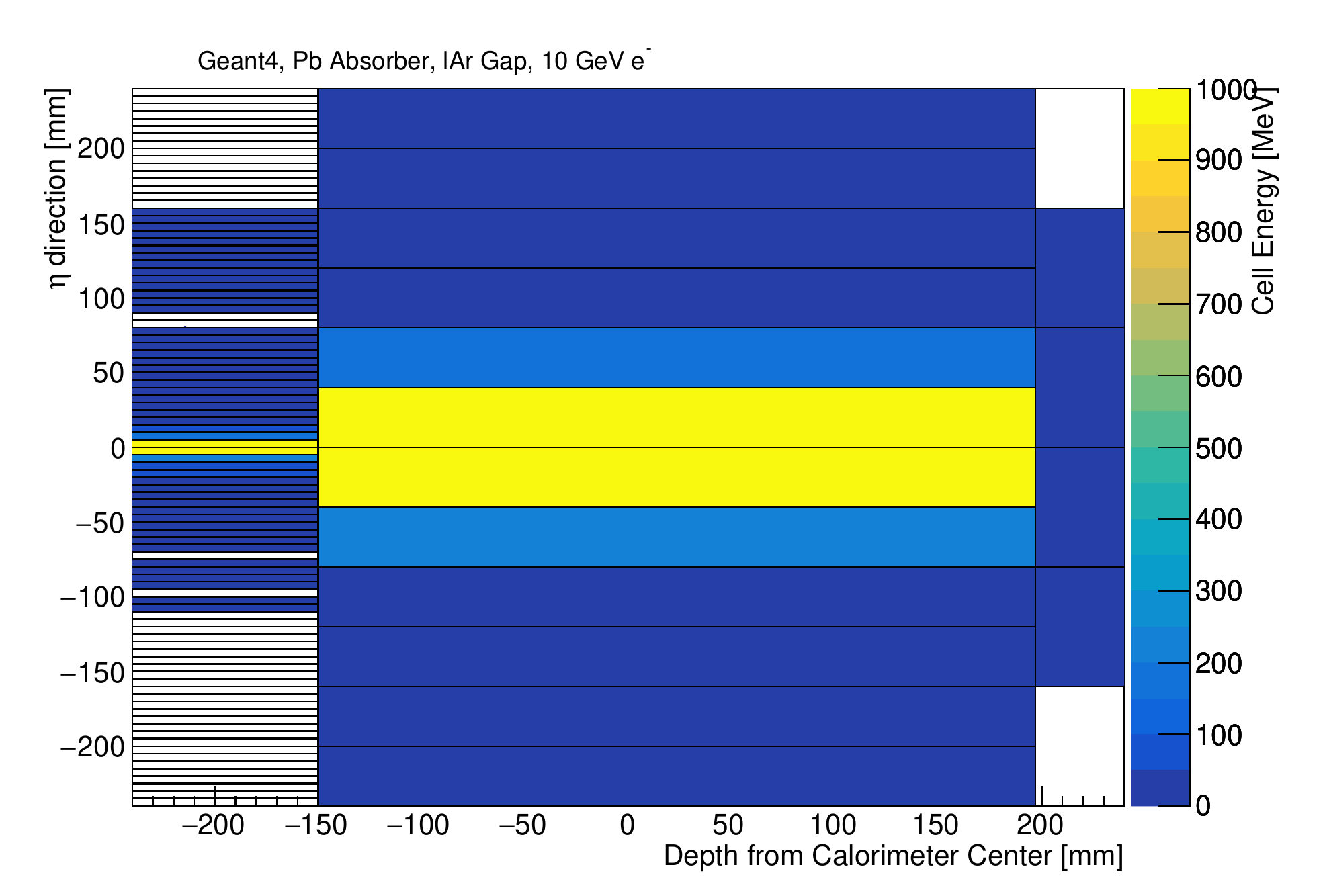}
        \caption{Discretized version of (a), in which energy depositions are assigned to individual, discrete detector cells.}
        \label{fig:calo_image1}
    \end{subfigure}
    \caption{The electromagnetic shower from one 10 GeV electron event. The boundaries of the cells are shown, projecting out the $\phi$ segmentation.}
    \label{fig:calo_image}
\end{figure}
}

\afterpage{
\begin{figure}
    \centering
    \includegraphics[width=0.48\textwidth, trim={12, 12, 12, 12}]{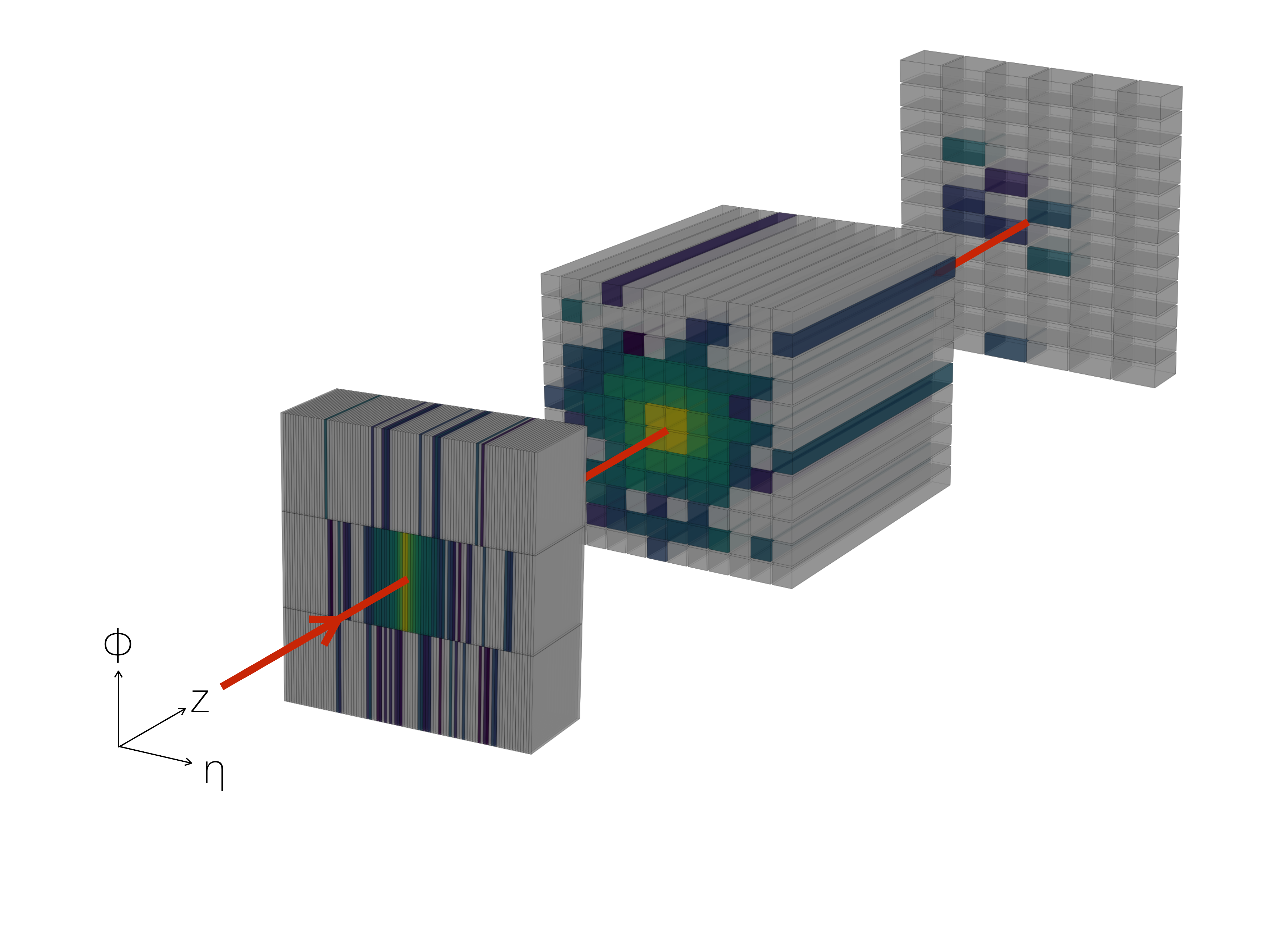}
    \caption{Three-dimensional representation of a 10 GeV $e^+$ incident perpendicular to the center of the detector. Not-to-scale separation among the longitudinal layers is added for visualization purposes.}
    \label{fig:3d}
\end{figure}

\begin{figure}
    \centering
    \includegraphics[width=0.15\textwidth]{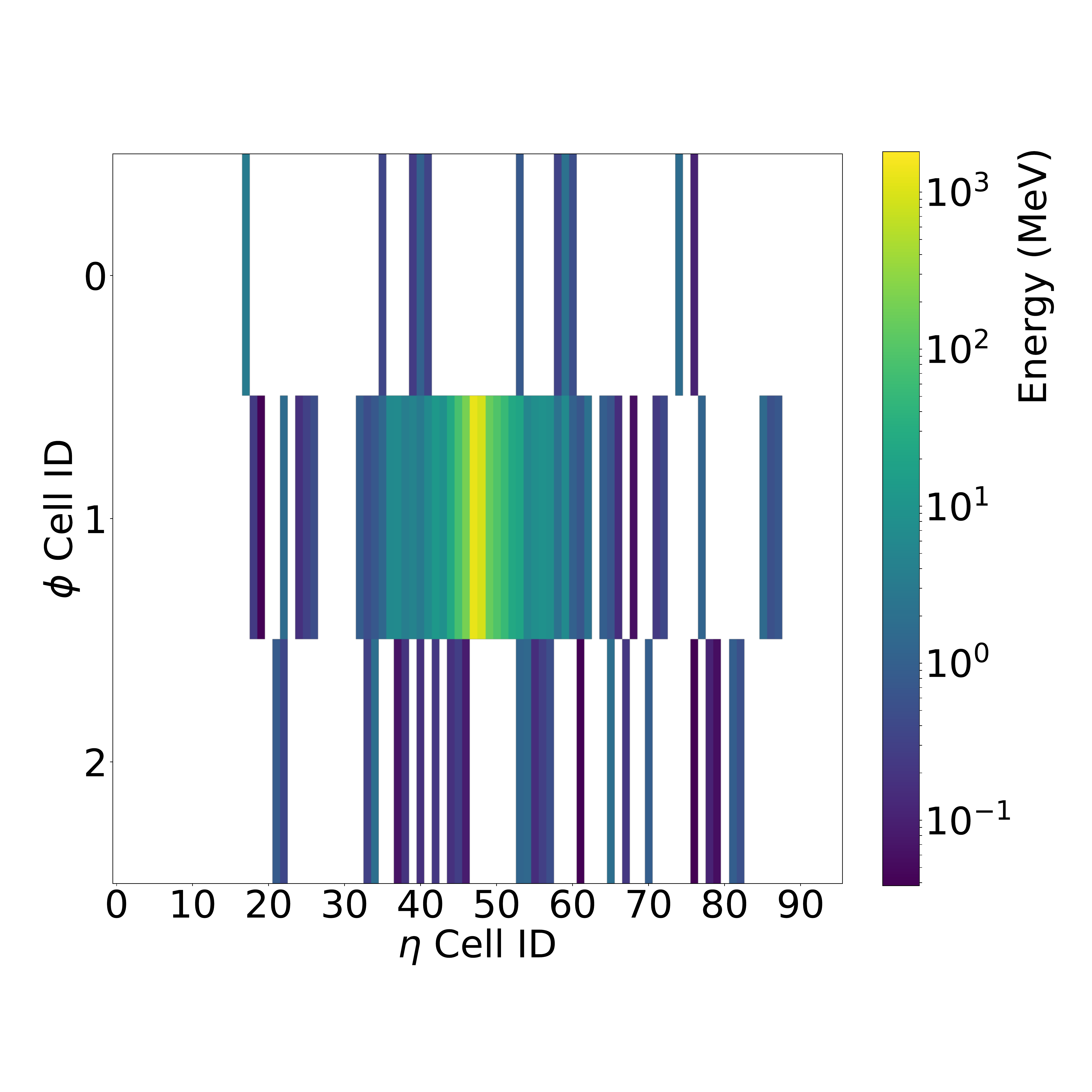}
    \includegraphics[width=0.15\textwidth]{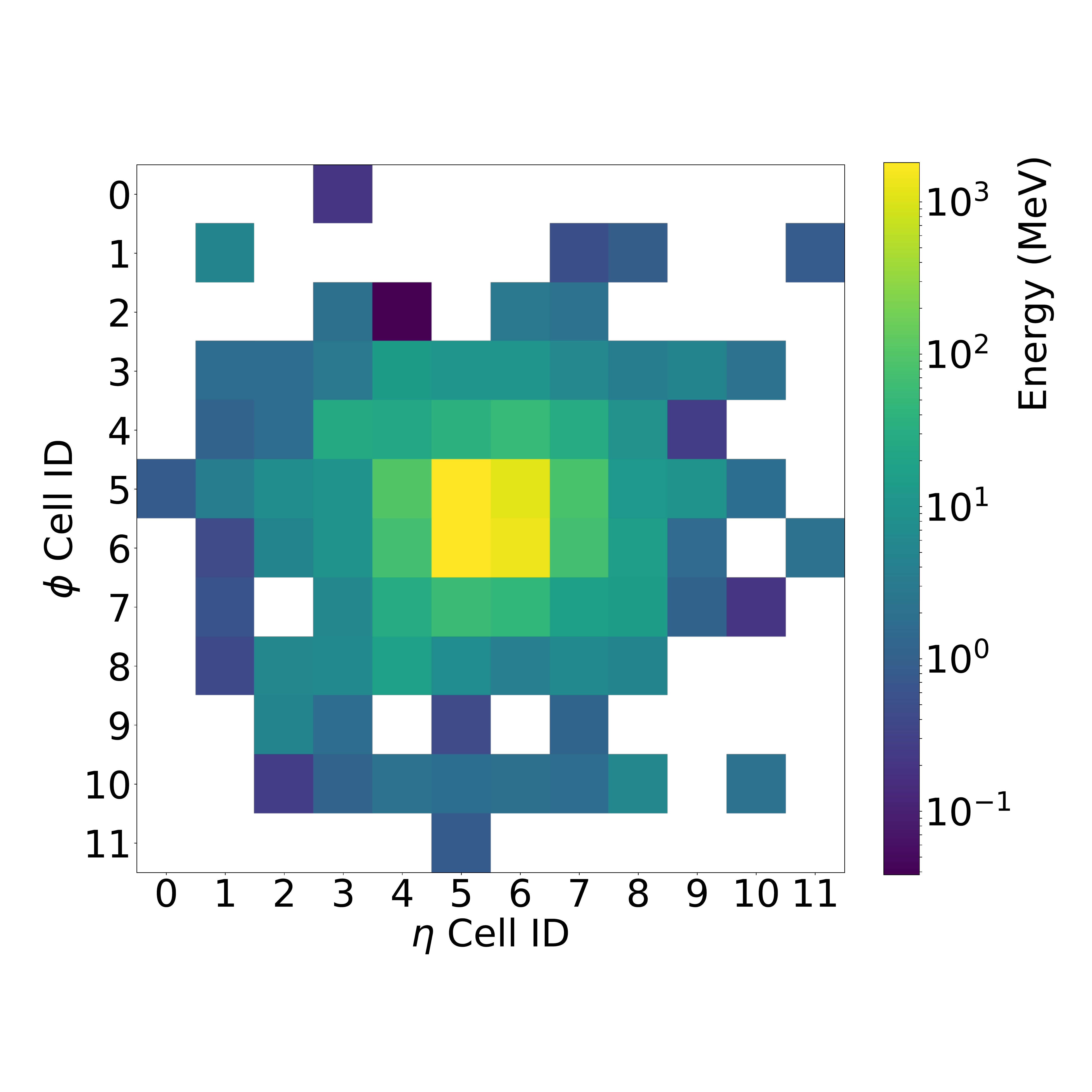}
    \includegraphics[width=0.15\textwidth]{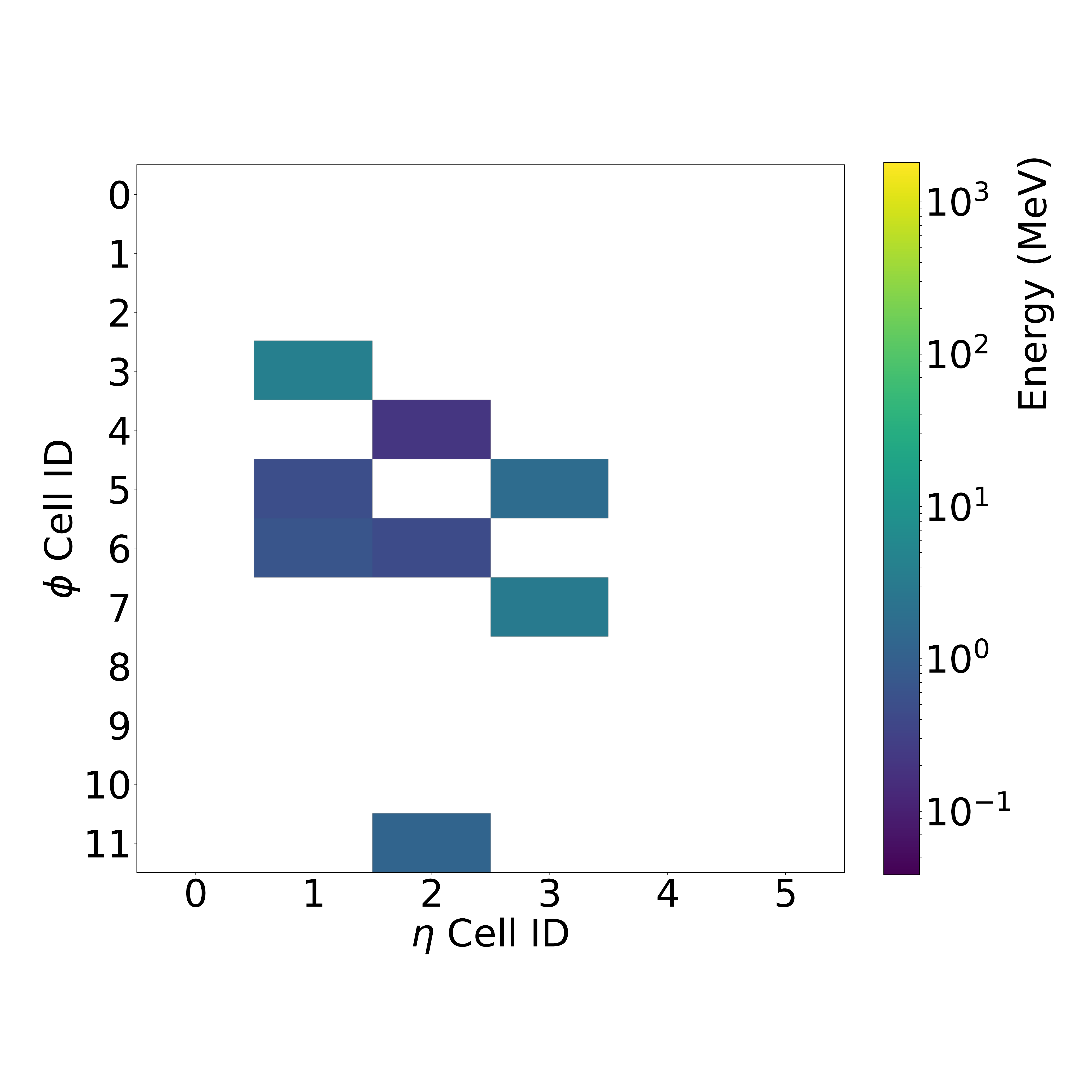}
    \caption{Two-dimensional, per-layer representation of the same shower as in Fig.~\ref{fig:3d}.}
    \label{fig:2d}
\end{figure}
}

The training data set~\cite{dataset} is prepared as follows. \textsc{Geant4} 10.2.0~\cite{Geant} is used to generate particles and simulate their interaction with our calorimeter using the \textsc{Ftfp\_Bert} physics list based on the Fritiof~\cite{ANDERSSON1987289,Andersson1996,NilssonAlmqvist:1986rx,Ganhuyag:1997gz} and Bertini intra-nuclear cascade~\cite{Guthrie:1968ue,Bertini:1971xb,Karmanov:1979if} models with the standard electromagnetic physics package~\cite{1462617}. Positrons, photons, and charged pions with various energies are incident perpendicular on the center of the calorimeter front. Energies in the training are uniform in the range between 1 GeV and 100 GeV. Fig.~\ref{fig:calo_image} shows an example 10 GeV electron event with the exact energy deposits from \textsc{Geant4} (Fig.~\ref{fig:calo_image2}) and after descretizing them according to our calorimeter geometry (Fig.~\ref{fig:calo_image1}). For visualization purposes, a 3-dimensional particle energy signature (Fig.~\ref{fig:3d}) will be displayed in the rest of this paper as a series of three 2D images in $\eta$ - $\phi$ space (Fig.~\ref{fig:2d}), where the pixel intensity represents the sum of the energies of all particles incident to that cell~\footnote{For the purposes of this study, the cell energy is the sum of the energy deposited in the lead and the argon; in practice, only the LAr energy deposits are measured. Dividing out these two components is left for future work (see Sec.~\ref{sec:conclusions}).}. The first layer can be represented as a $3\times96$ image, the middle layer as a $12\times12$ image, and the last layer as a $12\times6$ image.

\section{Generative Adversarial Networks}
\label{sec:GANreview}
Since their first formulation~\cite{goodfellow2014generative}, Generative Adversarial Networks have become a rapidly increasing area of attention in the Machine Learning literature with many applications in natural image processing.  However, there are far fewer applications in basic science and prior to this work, no applications in high energy physics and nuclear physics.

Generative Adversarial Networks (GANs) cast the task of training a deep generative model as a two-player non-cooperative minimax game, in which a generator network $G$ is trained concomitantly with an adversary, the discriminator network $D$, in order to learn a target distribution $f$. The generator $G$ learns a map from a latent space $z\sim p_{z}(z)$ (usually chosen to be $N(0, 1)$) to the space of generated samples, while $D$ learns a map from the sample space to $[0, 1]$, the probability that a shown sample is real. Note that the map that the generator learns implicitly defines a density $g$. The game-theoretical basis for this framework~\cite{goodfellow2014generative,distinguishability} ensures that if we extend the space of allowed functions that $G$ and $D$ can draw from to be the space of all continuous functions, then there exists some $G$ (and, by construction, an implicit $g$) that exactly recovers the target distribution $f$, i.e., $g \rightarrow f$, while for every sample produced by the generator, the discriminator is maximally confused and admits a posterior of being real of \sfrac{1}{2}. In order to train both $G$ and $D$, the traditional formulation of GANs~\cite{goodfellow2014generative} utilizes the loss function $ \mathcal{L}_{\mathsf{adv}}$ shown in Eqn.~\ref{eqn:gan_formulation}.

\begin{equation}
\begin{aligned}
  \mathcal{L}_{\mathsf{adv}} = 
    &\underbrace{
        \mathbb{E}_{z \sim p_z(z)}[\log(\mathbb{P}(D(G(z))=0))]
    }_{\begin{subarray}{l} 
        \text{term associated with the discriminator}\\
        \text{perceiving a generated sample as fake}
       \end{subarray}} + \\
    &\underbrace{
        \mathbb{E}_{I \sim f}[\log(\mathbb{P}(D(I)=1))]
    }_{\begin{subarray}{l}
        \text{term associated with the discriminator}\\
        \text{perceiving a real sample as real}
       \end{subarray}}
\label{eqn:gan_formulation}
\end{aligned}
\end{equation}

Though the GAN framework has shown promise, stability is still a major roadblock, and various ad-hoc and theoretical improvements have been suggested, from architectural guidelines~\cite{dcgan,odena_acgan,conditional_gan,info_gan,stackgan} to reformulations of the loss specified in Eqn.~\ref{eqn:gan_formulation} to move away from the Jensen-Shannon divergence~\cite{wgan,improved_gan,fgan,lsgan,l2gan,softmaxgan}. As suggested in~\cite{evaluation}, we are able to impose task-specific metrics which allow us to move away from loss level notions of quality and focus on task-level fidelity measures. We make the conscious decision to utilize the vanilla loss formulation as we find adequate performance with this version.

\section{The \textsc{CaloGAN}}
\label{sec:calogan}
Generative Adversarial Networks are explored as a tool to speed up full simulation of particle showers in an EM calorimeter. We identify this solution with the name \textsc{CaloGAN}. 

For it to be useful in realistic physics applications, such a system needs to be able to accept requests for the generation of showers originating from an incoming particle of type $P$ at energy $E$~\footnote{This covers a significant portion of the challenge; in practice, the fast simulation must also take the $\eta$-$\phi$ position of incidence and the incidence angles. We leave this to future work (see Sec.~\ref{sec:conclusions}).}. We introduce an auxiliary task of energy reconstruction to condition on $E$, a real valued variable. The Auxiliary Classifier GAN~\cite{odena_acgan} formalism is tested to also condition on class $P$, but ultimately abandoned in favor of training a specific generative model for each particle type, as the authors expect that versioning and particle-specific improvements will be prioritized in any practical implementation.

In practice, energy is scaled by a factor of $10^2$ and multiplied to the 1024-dimensional~\footnote{The dimensionality of the latent space is a hyper-parameter that need not be the same as the dimensionality of the target space.} latent space vector $z\in\mathbb{R}^{1024}$. The generator $G$ then maps this input to three gray-scale image outputs with different numbers of pixels, which represent the energy patterns collected by the three calorimeter layers as the requested particle propagates through them. The discriminator $D$ accepts the three images as inputs, along with $E$, the chosen value for the particle energy. The inputs are mapped to a binary output that classifies showers into real and fake, and a continuous output which calculates the total energy deposited in the three layers, then compares it with the requested energy $E$. 

\subsection{Model Architecture}

\label{ssec:calogan_arch}
Given the sparsity levels and high dynamic range in the data described in Section~\ref{sec:dataset}, we follow the LAGAN guidelines~\cite{deOliveira:2017pjk} to modify the DCGAN~\cite{dcgan} architecture for this specific regime.

\begin{figure*}
    \centering
    \includegraphics[width=0.9\textwidth]{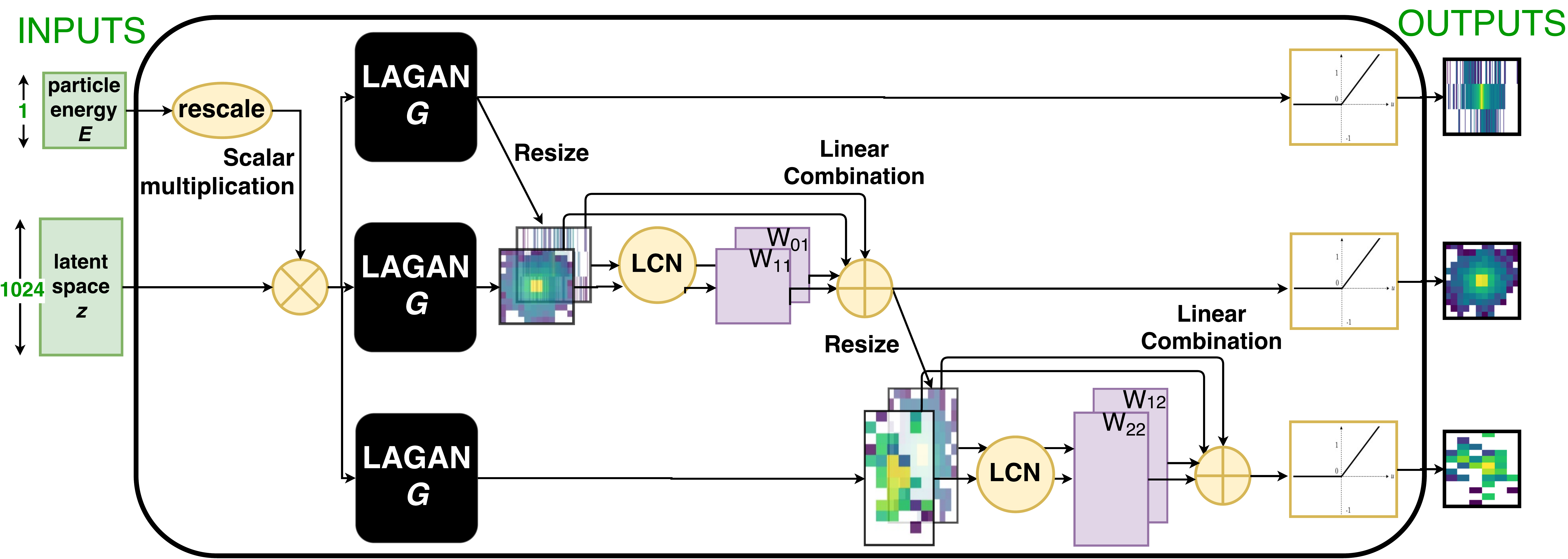}
    \caption{Composite Generator, illustrating three stream with attentional layer-to-layer dependence.}
    \label{fig:G}
\end{figure*}

In the generator (shown in Fig.~\ref{fig:G}), our design combines parallel LAGAN-like processing streams with a trainable attention mechanism that encodes the sequential connection among calorimeter layers. The LAGAN submodules are composed of a 2D convolutional unit followed by two locally-connected units with batch-normalization~\cite{batchnorm} layers in between. The dimensionality and granularity mismatch among the three longitudinal segmentations of the detector demand separate streams of operations with suitably sized kernels. Towards providing a readily adaptable tool, we provide an architecture construction that is simply a function of the desired output image size, as we seek a common denominator that can be readily applied to a variety of particles in order to obtain reasonable baselines in a quick R\&D cycle. 


Modelling the sequential nature of the relationship among the energy patterns collected by the three layer requires extra care. Drawing inspiration from~\cite{stackgan}, we choose an attention mechanism to allow dependence among layers, in which we define trainable transfer functions to optimally resize and apply knowledge of the energy pattern in previous layers to the generation of the subsequent layer readout. More specifically, in-painting takes as input a resized image from a previous layer, $\mathcal{I}^{'}$, and the hypothesized image from the current layer, $\mathcal{I}$, and learns a per-pixel attention weight $W$ via a weighting function $\omega(\mathcal{I}, \mathcal{I}^{'})$ such that the pre-ReLU version of the current layer is $W\odot \mathcal{I} + (1 - W) \odot \mathcal{I}^{'}$, where $\odot$ is the Hadamard product. This end-to-end trainable unit can utilize information about the two layers to decide what information to propagate through from the previous particle deposition.
An alternative architectural choice that includes a recurrent connection will be subject of future studies.

Leaky Rectified Linear Units~\cite{leaky} are chosen as activation functions throughout the system, with the exception of the output layers of $G$, in which we prefer Rectified Linear Units~\cite{RELU} for the creation of sparse samples~\cite{deOliveira:2017pjk}.

\begin{figure*}
    \centering
    \includegraphics[width=0.9\textwidth]{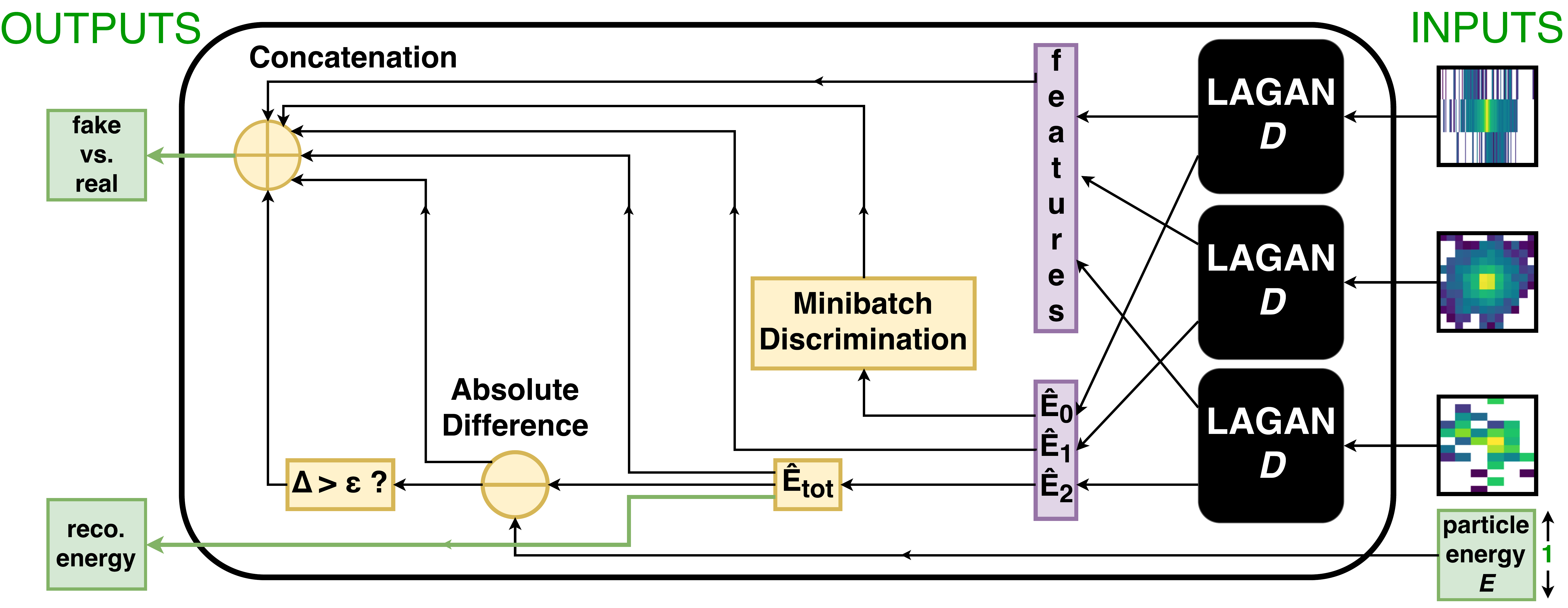}
    \caption{Composite Discriminator, depicting additional domain specific expressions included in the final feature space.}
    \label{fig:D}
\end{figure*}

In the discriminator (shown in Fig.~\ref{fig:D}), the feature space produced by each LAGAN-style output stream is augmented with a sub-differentiable version of sparsity percentage~\footnote{This is the fraction of hit pixels (occupancy).  Sub-derivatives are a generalization of derivatives to cases with kinks.}, as well as minibatch discrimination~\cite{improved_gan} on both the standard locally connected network-produced features and the output sparsity itself, to ensure a well examined space of sparsities. These are represented in Fig.~\ref{fig:D} by the `features' vector.

The discriminator is further customized with domain-specific features to ensure fidelity of samples. Given the importance of matching the requested energy $E$, $D$ directly calculates the empirical energy per layer $\hat{E}_i$, $i\in\{0, 1, 2\}$, as well as the total energy $\hat{E}_\mathrm{tot}$. Minibatch discrimination is performed on this vector of per-layer energies to ensure a proper distributional understanding. We also add $\vert E - \hat{E}_\mathrm{tot}\vert$ as a feature, as well as $\mathbb{I}_{\{\vert E - \hat{E}_\mathrm{tot}\vert > \varepsilon\}}$ with $\varepsilon=5$ GeV -- a binary, sub-differentiable feature which encodes the tolerance for GAN-produced scatterings to be incorrect in their reconstructed energy.

Further specifications of the exact hyper-parameter and architectural choices as well as software versioning constraints are available in the source code~\cite{code_new}.

Two additional architectural modifications were tested in order to build a particle-type conditioning system directly into the learning process. Neither the AC-GAN~\cite{odena_acgan}  nor the conditional GAN~\cite{conditional_gan} frameworks were able to handle the substantial differences among the three particle types. 

We suspect that both a significantly richer model and a larger latent space could alleviate some problems associated with conditioning using the investigated approaches. Although building a fully joint model is an interesting Machine Learning challenge, the practicality and flexibility of this application may suffer from having one single model for all particle showers.

\subsection{Loss Formulation}
\label{ssec:calogan_loss}
In this work, we augment the classical adversarial loss term $\mathcal{L}_\mathsf{adv}$ (Eqn.~\ref{eqn:gan_formulation}) -- which penalizes the system whenever $D$ fails to classify samples originating from generated or target distributions -- with a mean absolute error term:
\begin{equation}
  \mathcal{L}_{E} = \mathbb{E}_{z\sim p_z(z)}[\delta(E, \hat{E}(G(z)))] + \mathbb{E}_{I\sim f}[\delta(E, \hat{E}(I))]
\label{eqn:energy_task}
\end{equation}
where $\delta(e, e')= \vert e - e' \vert$, $E$ is the requested energy, and $\hat{E}$ is the reconstructed energy. This allows us to penalize instances of too little or too much deposited energy. This solution not only helps ensuring the confinement of the generated energy to a desirable range, but it also allows to encode a `soft' physical notion of conservation of energy, according to which no more energy than the initial $E$ of the incoming particle can be physically collected by the detector. 

Note, however, that this formulation discourages, but does not forbid, a deposition of more energy than was requested. We can remedy this unphysical result by sampling from a conditional distribution until energy preservation is met. This issue is further addressed in Sec.~\ref{ssec:showershapes}.

During training, the generator will maximize Eqn.~\ref{eqn:G_loss}, and the discriminator will maximize Eqn.~\ref{eqn:D_loss}.

\begin{equation}
    \mathcal{L}_{\mathsf{generator}} =  \lambda_E\mathcal{L}_{E} - \mathcal{L}_{\mathsf{adv}}
    \label{eqn:G_loss}
\end{equation}

\begin{equation}
    \mathcal{L}_{\mathsf{discriminator}} =  \lambda_E\mathcal{L}_{E} + \mathcal{L}_{\mathsf{adv}}
    \label{eqn:D_loss}
\end{equation}

\subsection{Training Strategy}

$\mathcal{L}_{E}$ is set to 0.05 to down-weight the importance of $\mathcal{L}_{E}$ compared to $\mathcal{L}_{\mathsf{adv}}$ and rescale the absolute error, which is measured in GeV, to a comparable range with respect to $\mathcal{L}_{\mathsf{adv}}$. This hyper-parameter can be tuned in a systematic way, but with minimal tuning, we were able to find a reasonable value.

The weights in the generator and discriminator are optimized in an alternating fashion over a set of 100,000 \textsc{Geant4}-simulated events for each particle type in batches of 256, using the \textsc{Adam} optimizer~\cite{adam}. The discriminator has a learning rate of $2\times10^{-5}$, and the generator has a learning rate of $2\times10^{-4}$. We note that outside of initial rough hyper-parameter tuning, we perform no dedicated optimization per particle type, and simply apply the same training parameters to all three networks. We expect significant performance improvements (especially for pions) with dedicated training.

Each system is trained for 50 epochs. Sixteen NVIDIA K80 graphics cards are used for initial hyper-parameter sweeps, with two Titan X Pascal Architecture cards used for final training. \textsc{Keras} \texttt{v2.0.3}~\cite{keras} is used to construct all models, with the \textsc{TensorFlow} \texttt{v1.1.0} backend~\cite{tensorflow2015-whitepaper}.


\section{Performance}
\label{sec:performance}
As discussed in~\cite{evaluation}, there exist several methods to qualitatively and quantitatively assess the performance of generative networks, but not all evaluation criteria are equally suitable and reliable for all applications. In this paper, we choose application-driven methods focused on sample quality. A first qualitative assessment will be accompanied by a quantitative evaluation based on physics-driven similarity metrics. The choice reflects the domain specific procedure for data-simulation comparison. These similarity metrics are based on one-dimensional statistics of the shower probability distribution.  Visualizing and verifying the performance in higher dimensions is a challenge.  One way to probe study the modeling of higher dimensions is to study ability to classify showers from different particles.  This is studied in Sec.~\ref{sec:classification}.

\subsection{Qualitative Assessment}
\label{ssec:qualitative_assessment}

\begin{figure}
    \centering
    \includegraphics[width=0.15\textwidth]{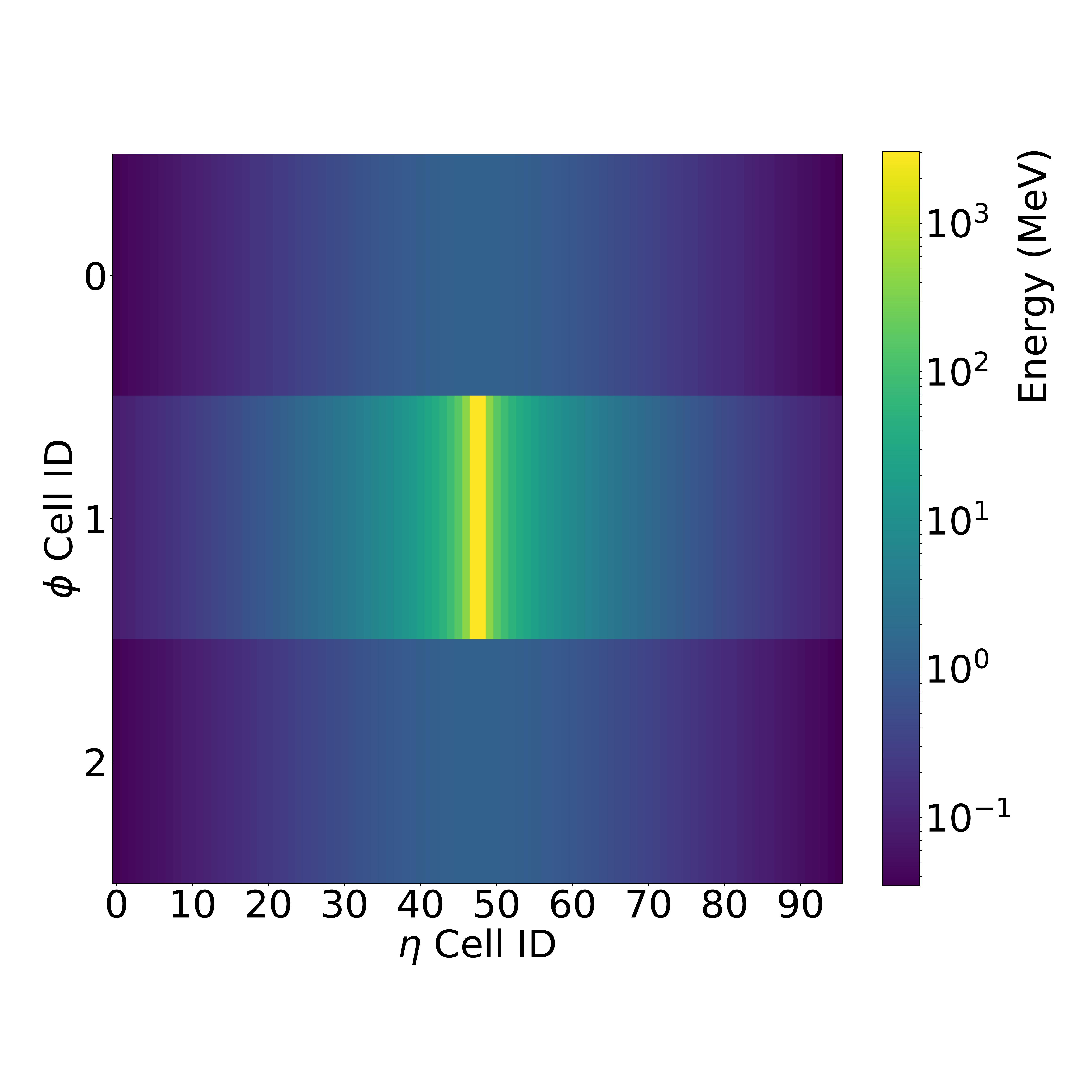}
    \includegraphics[width=0.15\textwidth]{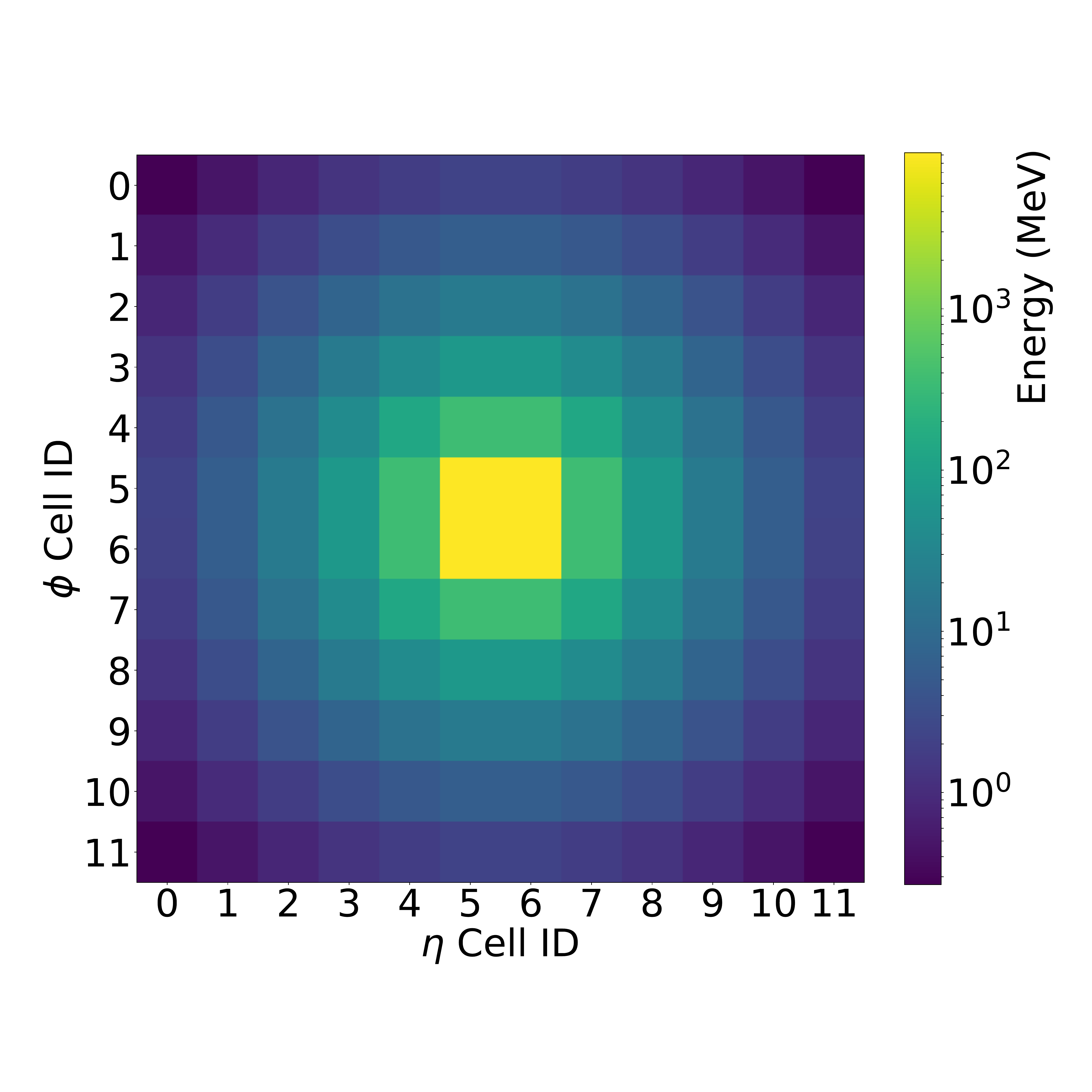}
    \includegraphics[width=0.15\textwidth]{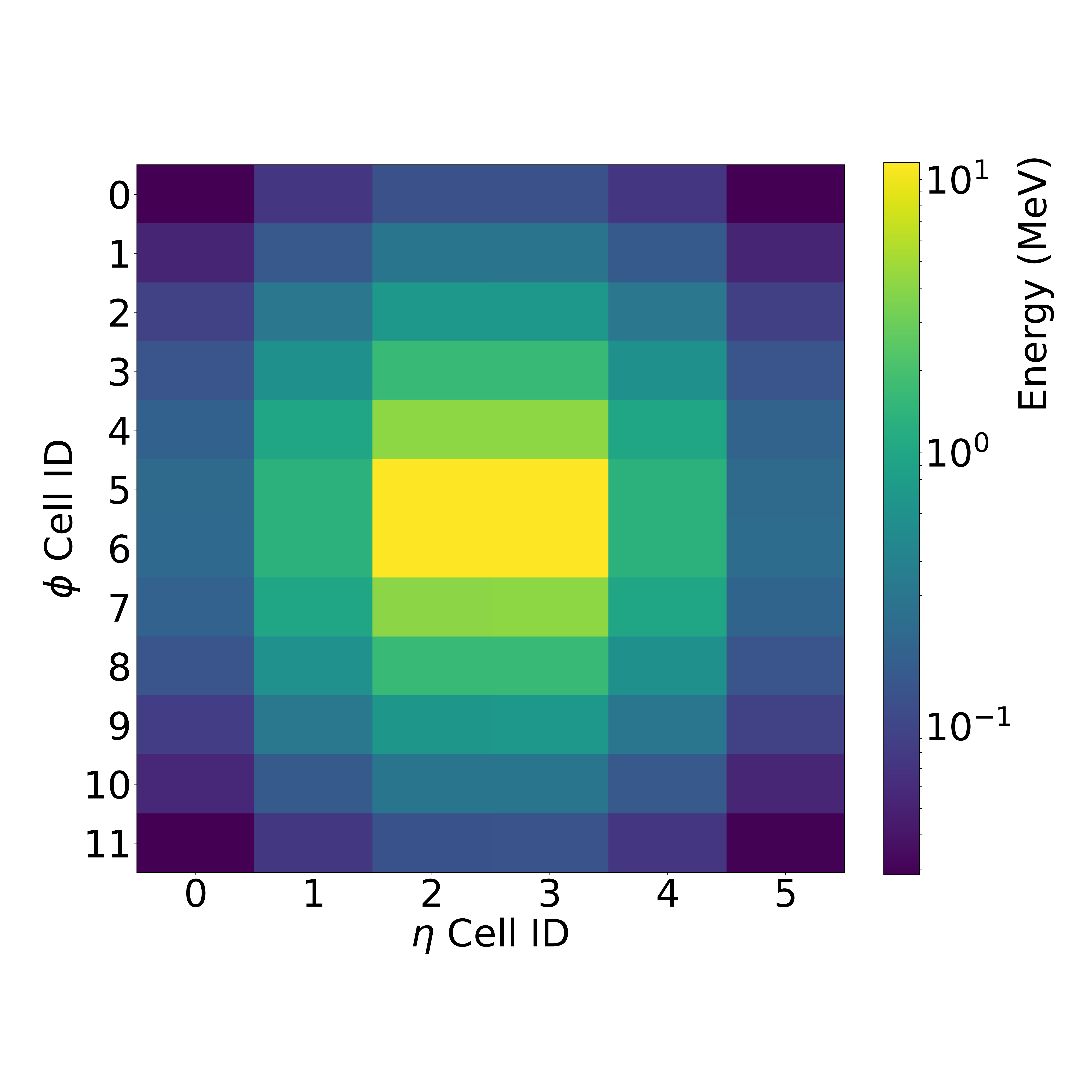}\\
    \includegraphics[width=0.15\textwidth]{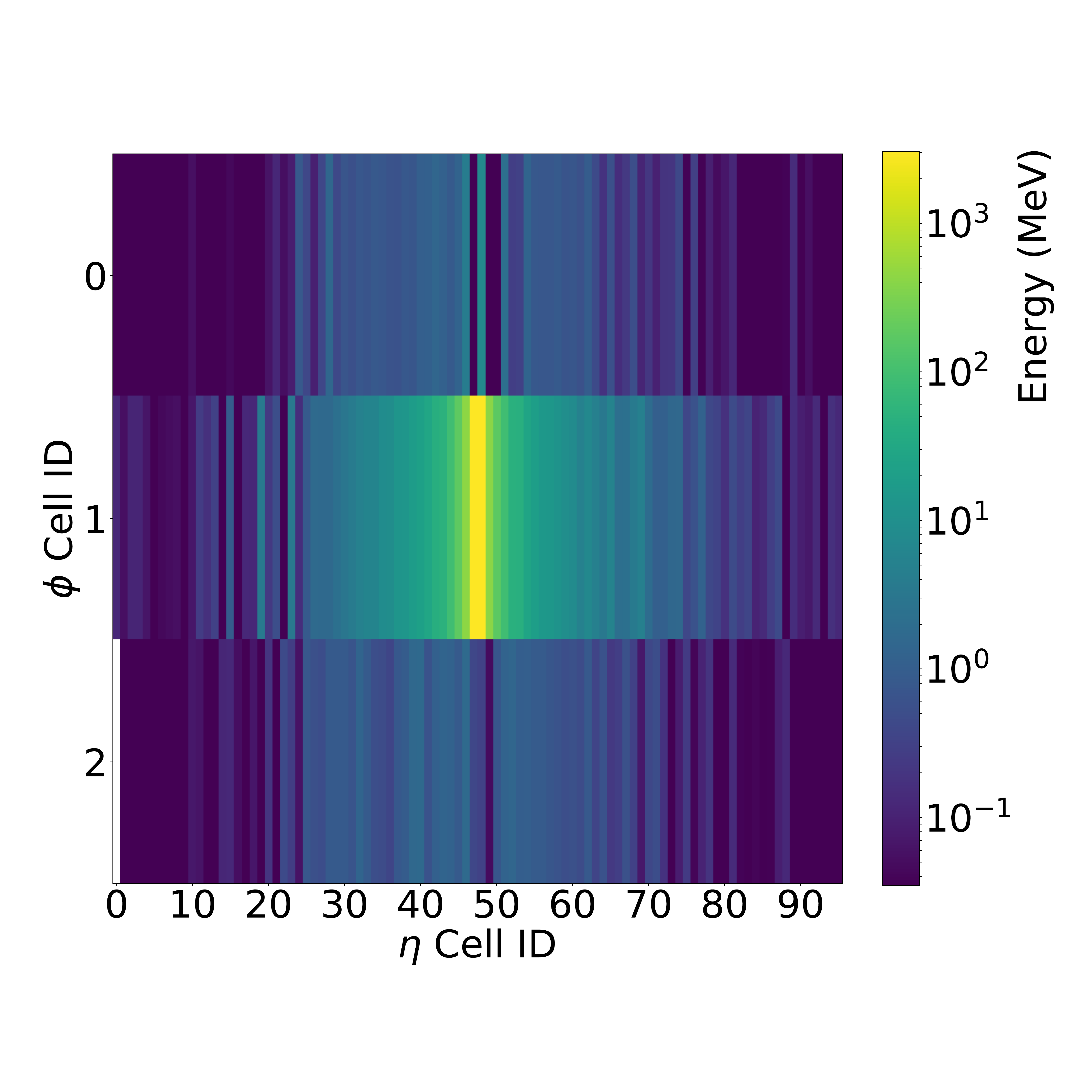}
    \includegraphics[width=0.15\textwidth]{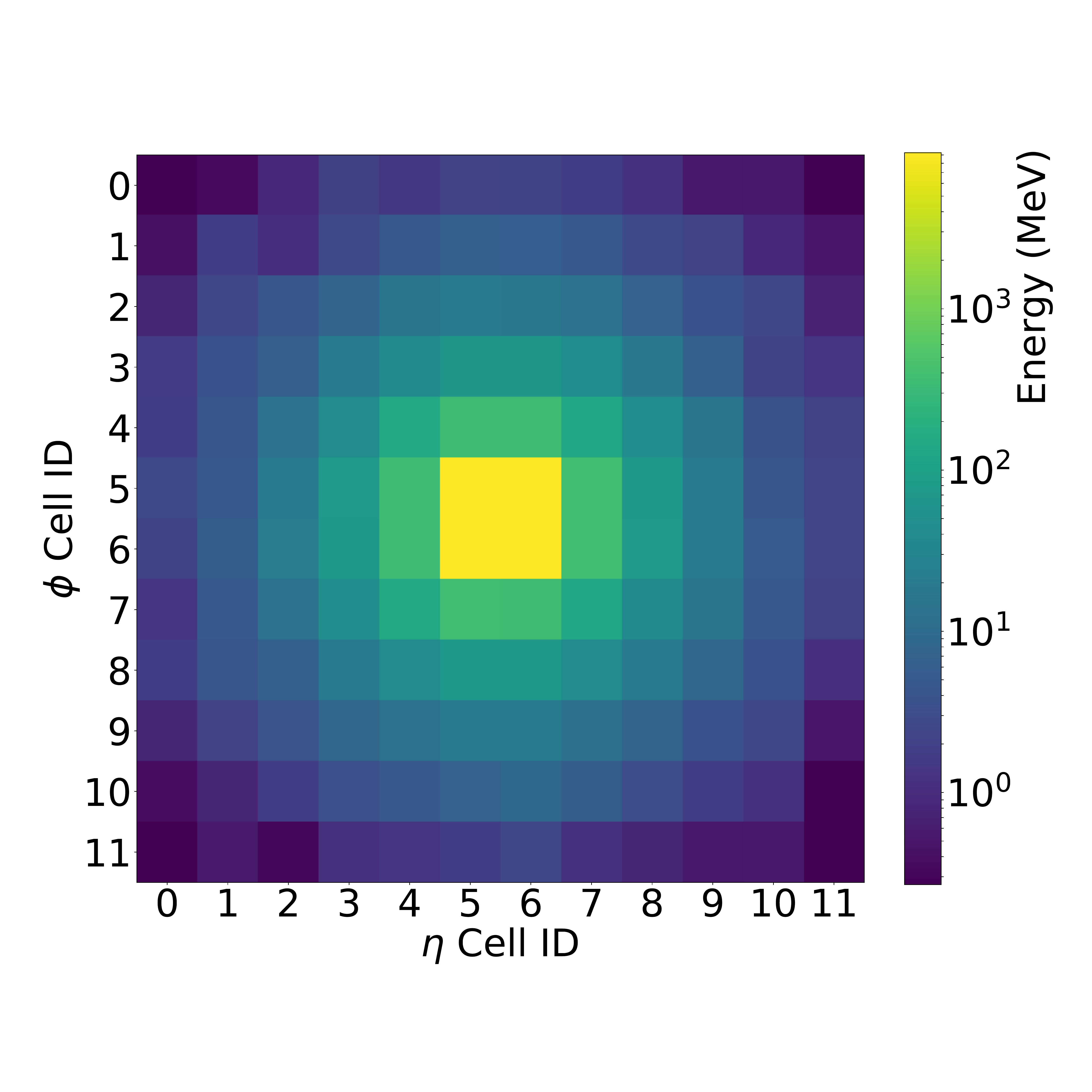}
    \includegraphics[width=0.15\textwidth]{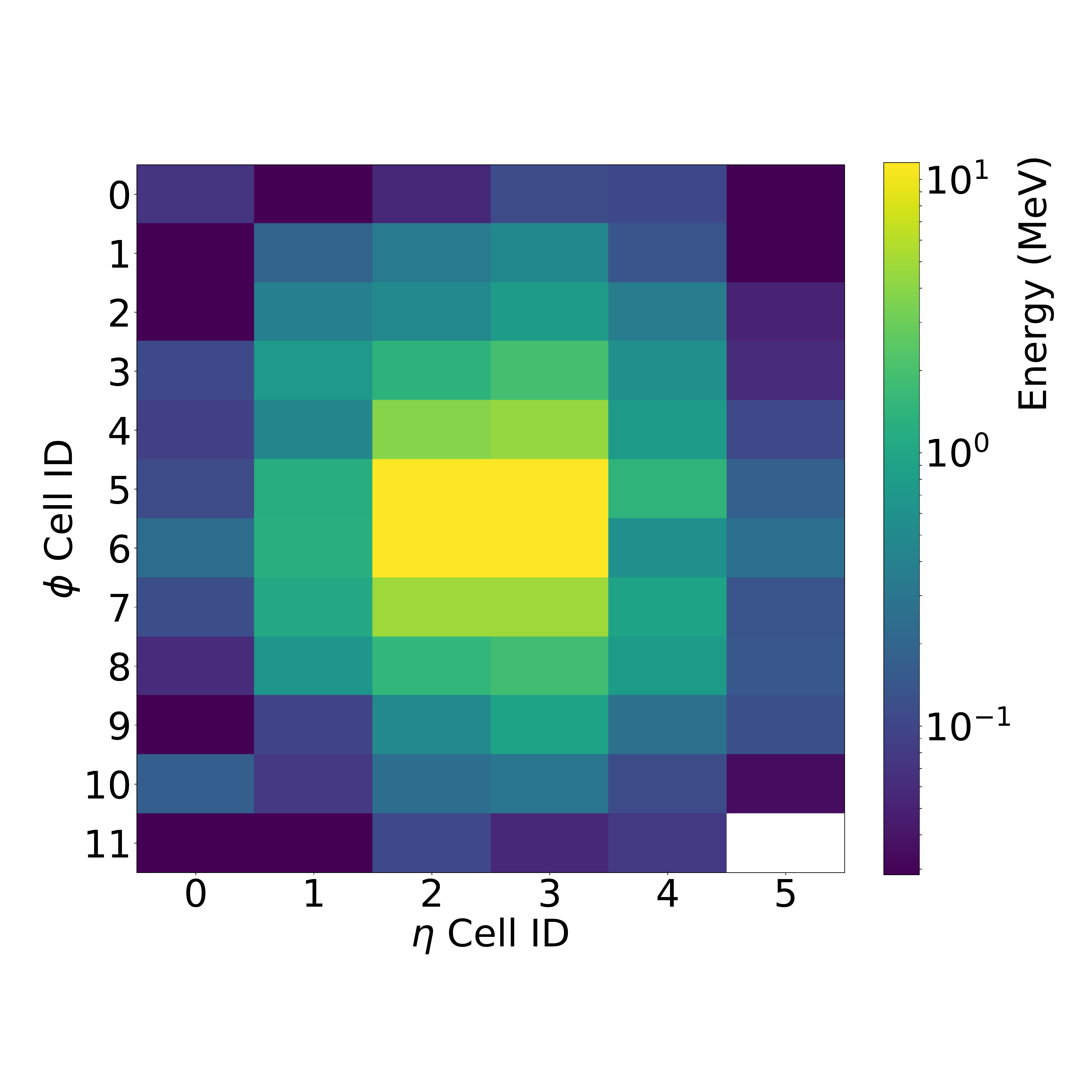}\\
    \caption{Average $e^+$ \textsc{Geant4} shower (top), and average $e^+$ \textsc{CaloGAN} shower (bottom), with progressive calorimeter depth (left to right).}
    \label{fig:eplus_avg}
\end{figure}

\begin{figure}
    \centering
    \includegraphics[width=0.15\textwidth]{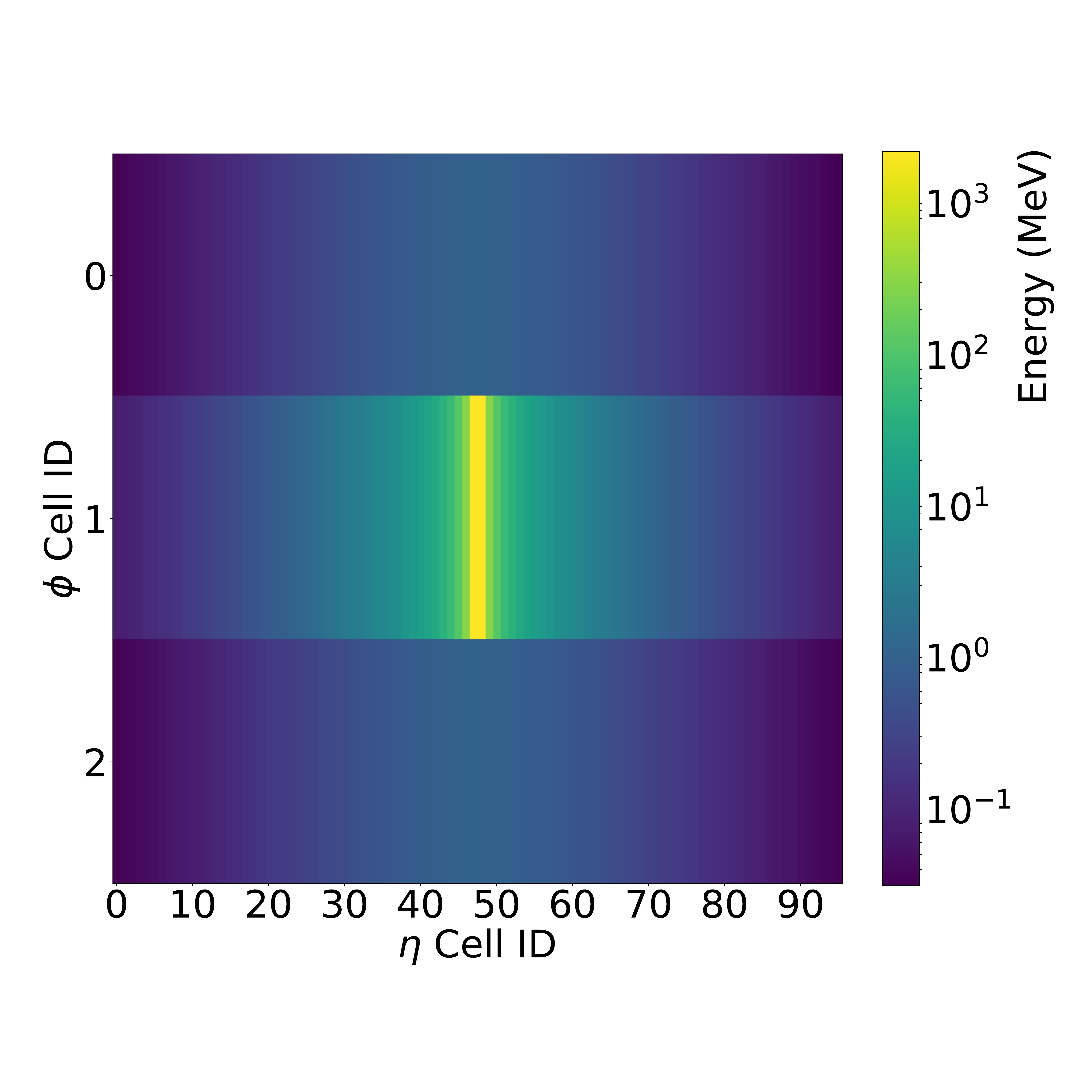}
    \includegraphics[width=0.15\textwidth]{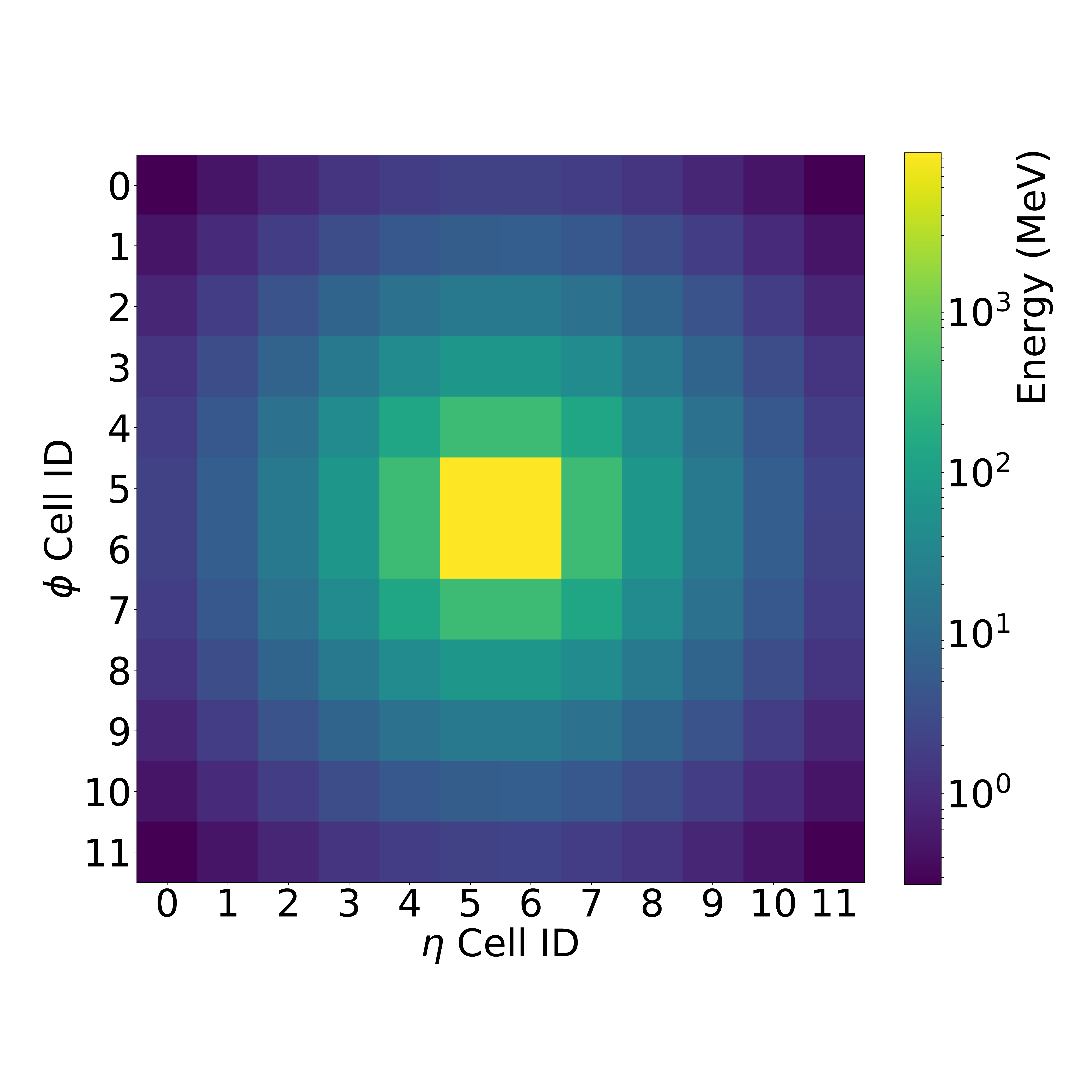}
    \includegraphics[width=0.15\textwidth]{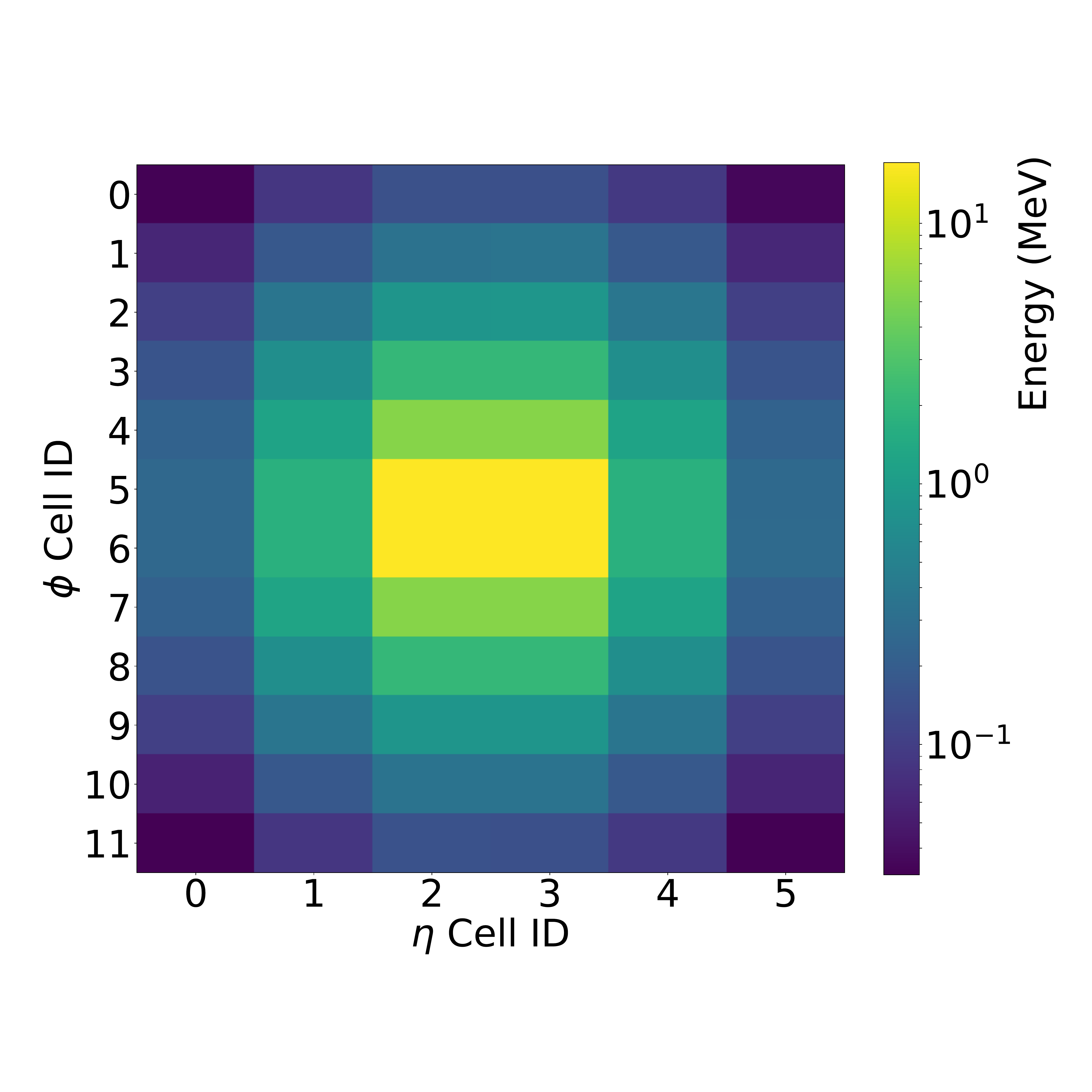}\\
    \includegraphics[width=0.15\textwidth]{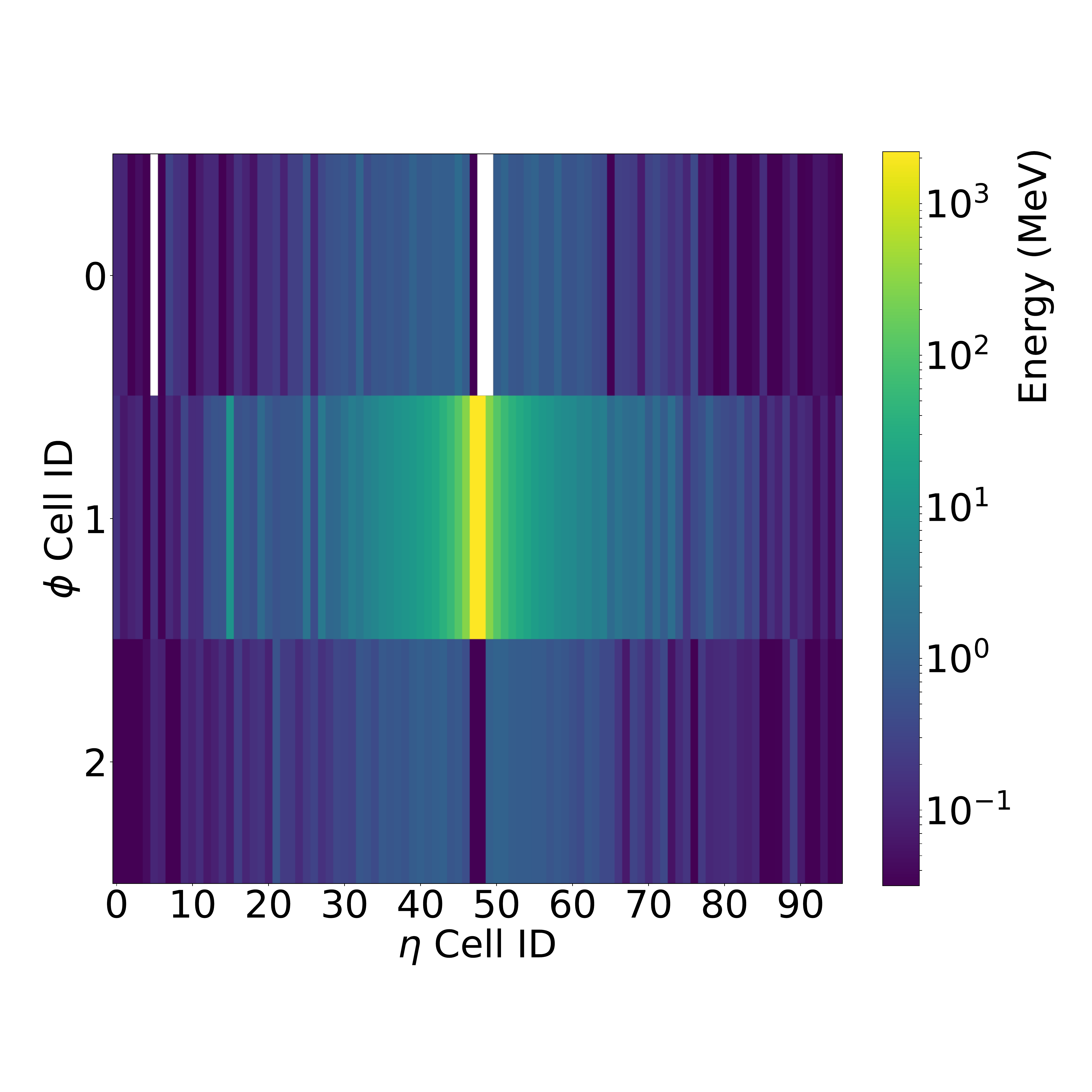}
    \includegraphics[width=0.15\textwidth]{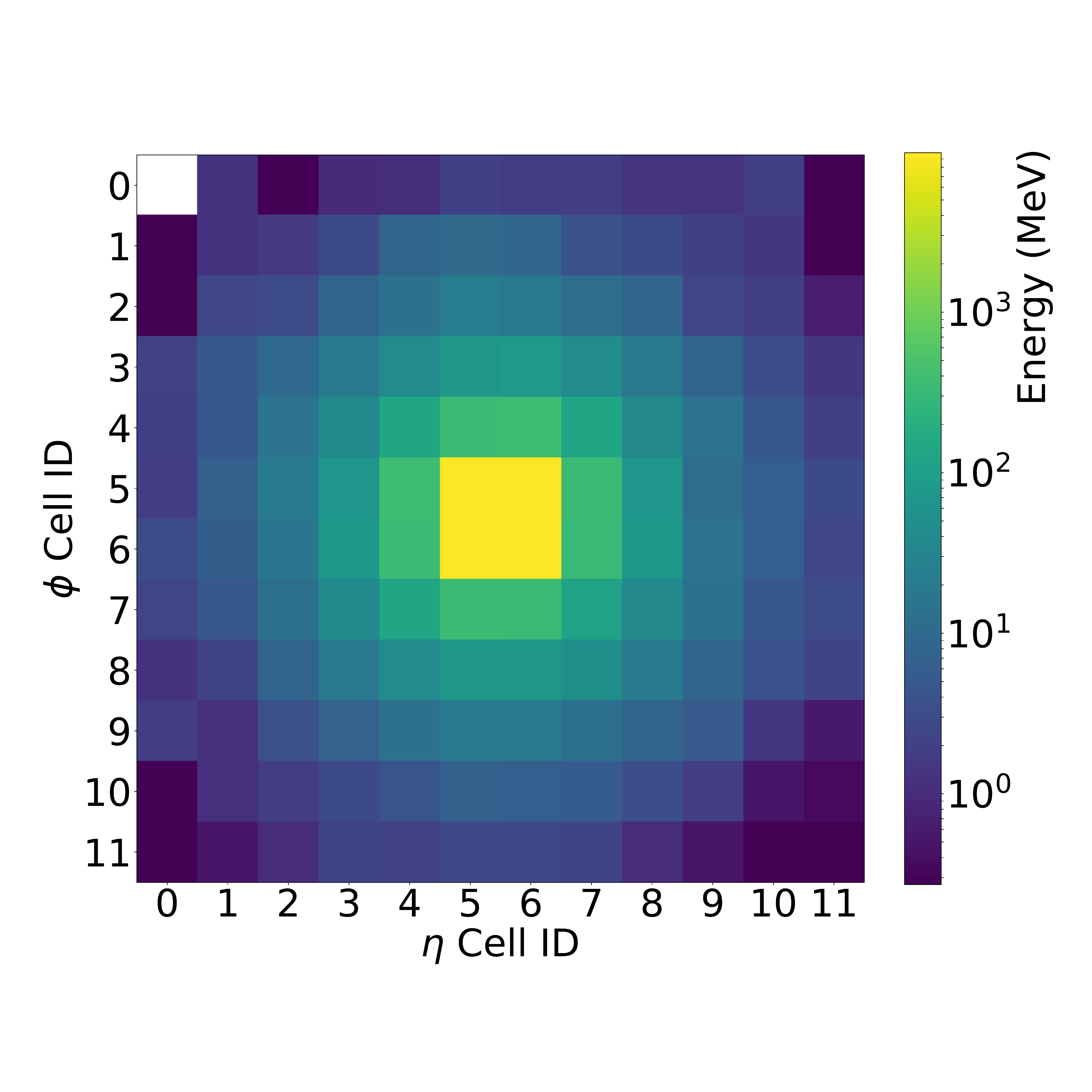}
    \includegraphics[width=0.15\textwidth]{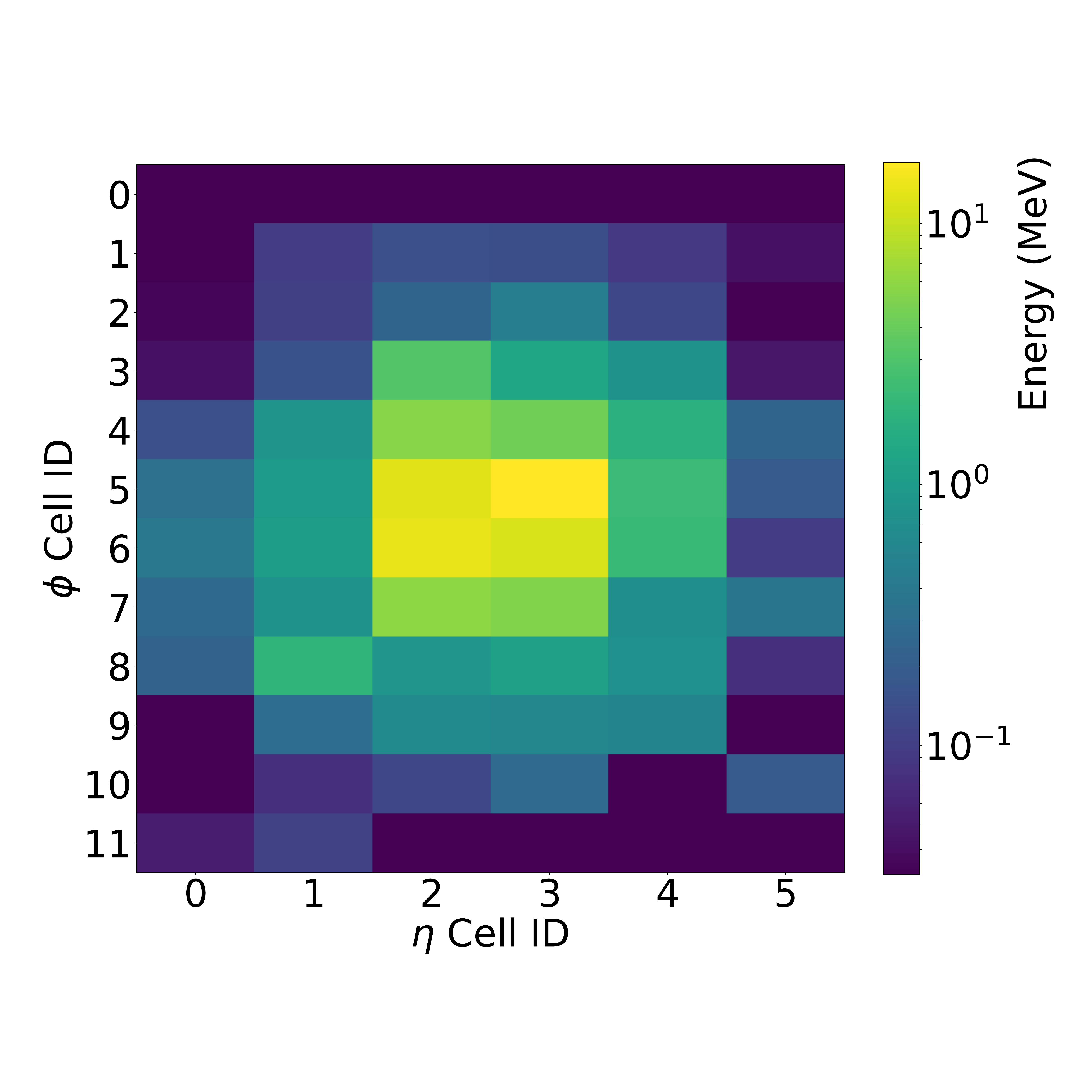}\\
    \caption{Average $\gamma$ \textsc{Geant4} shower (top), and average $\gamma$ \textsc{CaloGAN} shower (bottom), with progressive calorimeter depth (left to right).}
    \label{fig:gamma_avg}
\end{figure}

\begin{figure}
    \centering
    \includegraphics[width=0.15\textwidth]{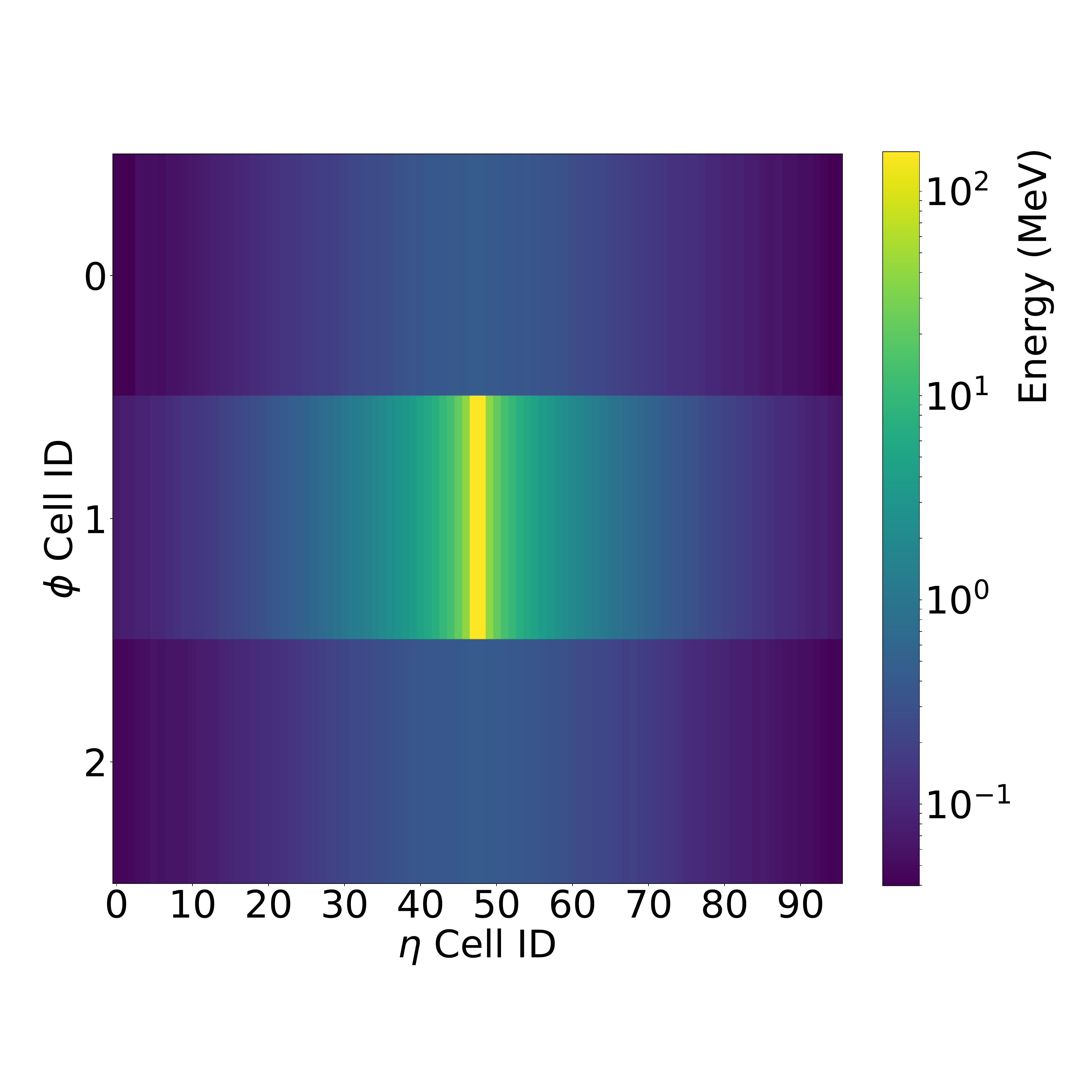}
    \includegraphics[width=0.15\textwidth]{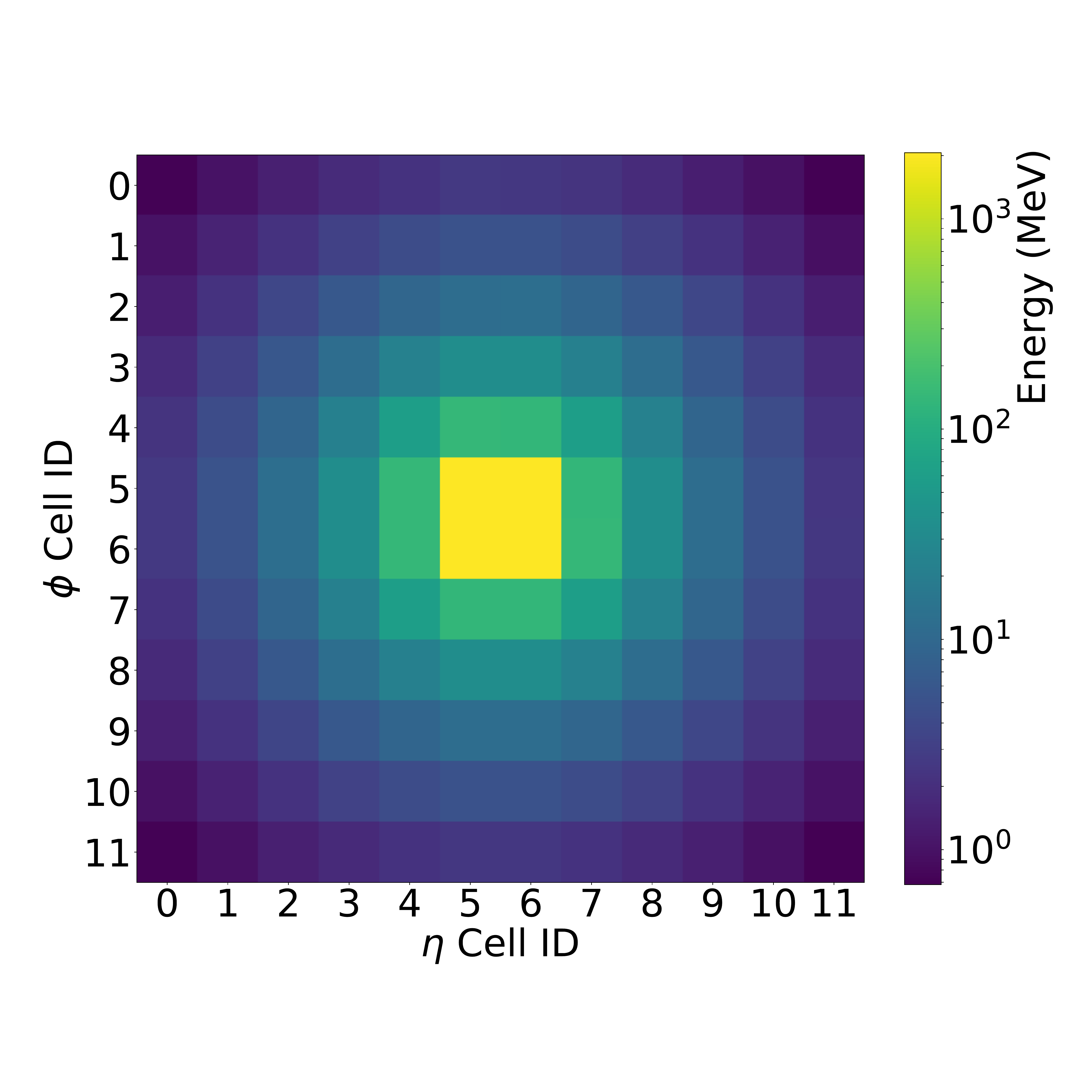}
    \includegraphics[width=0.15\textwidth]{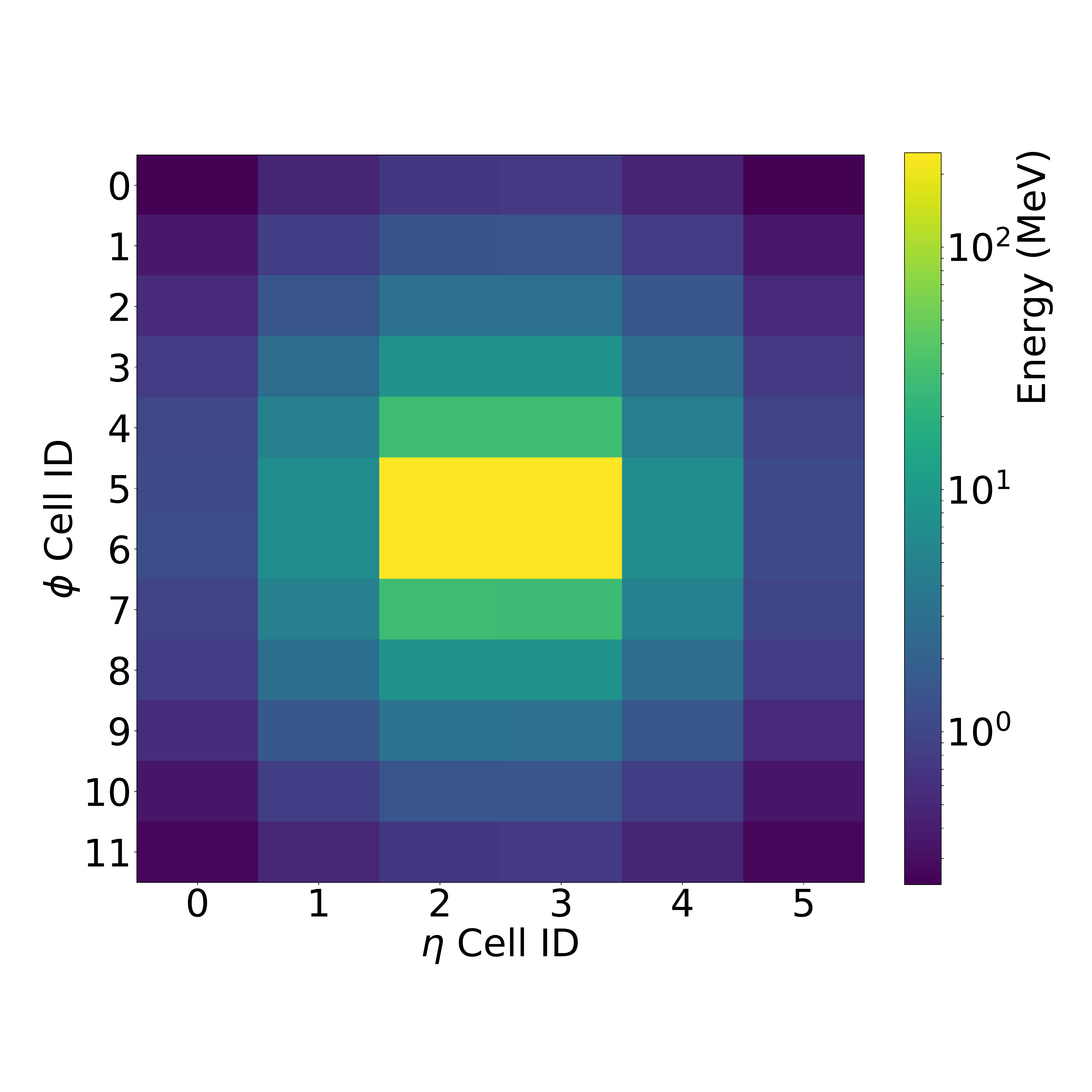}\\
    \includegraphics[width=0.15\textwidth]{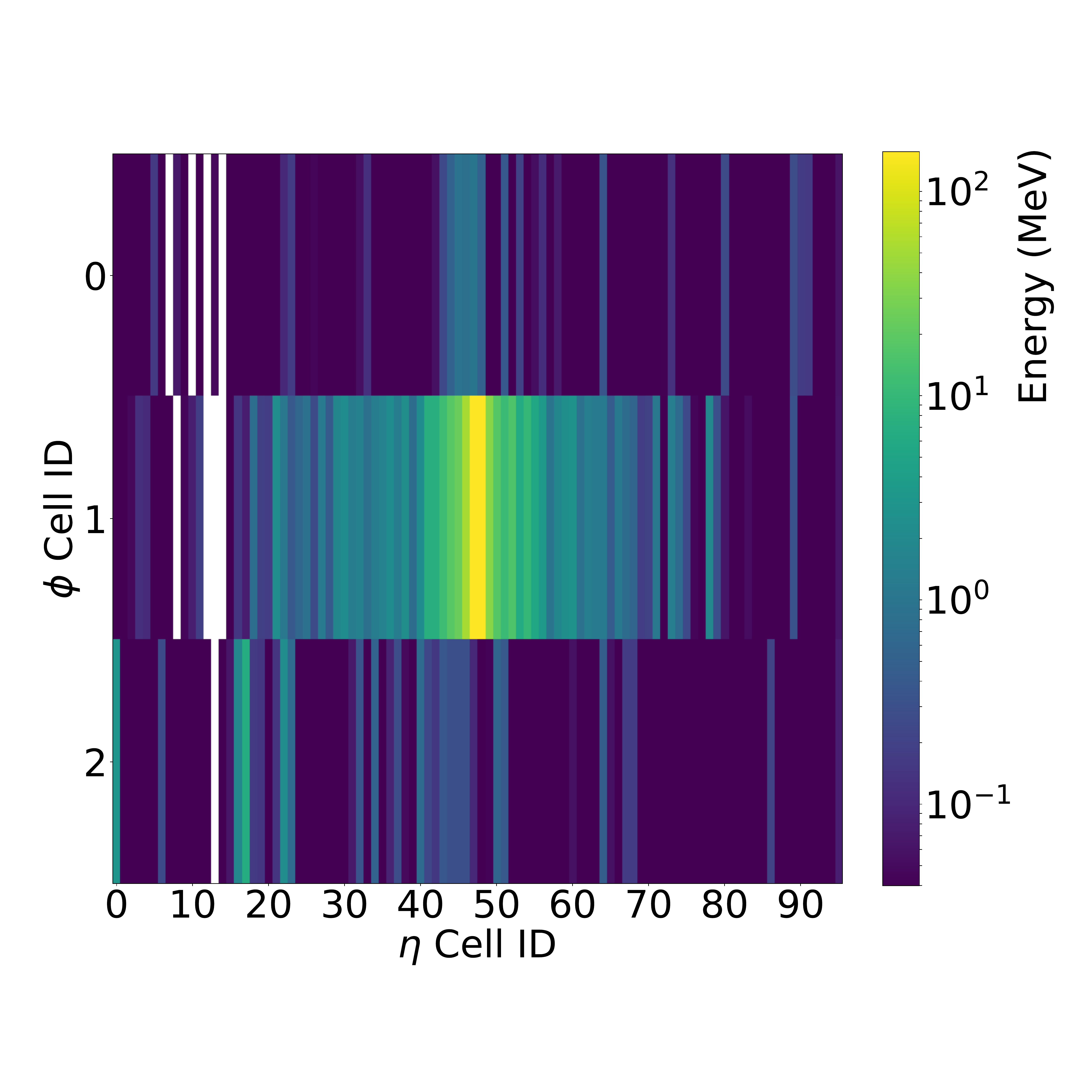}
    \includegraphics[width=0.15\textwidth]{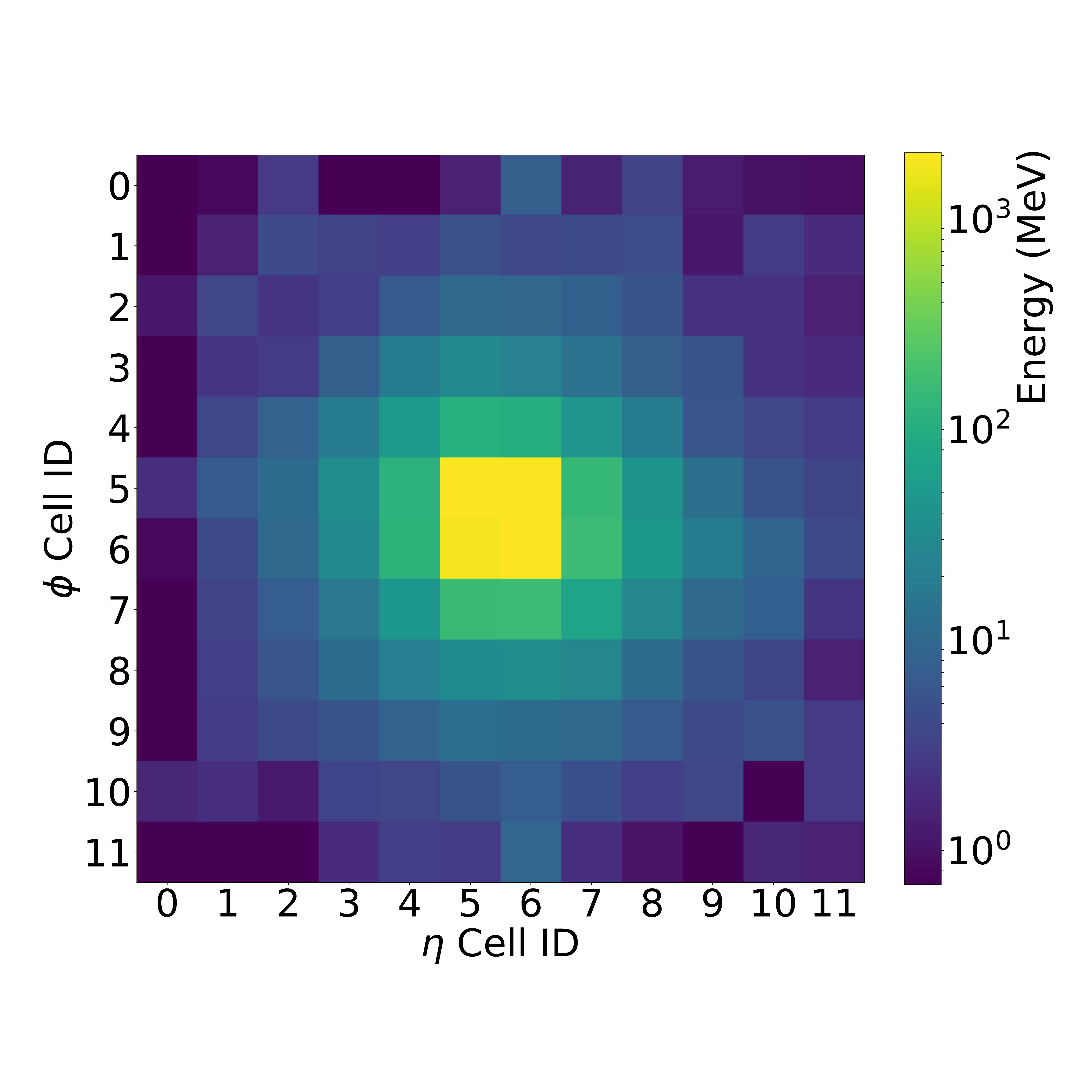}
    \includegraphics[width=0.15\textwidth]{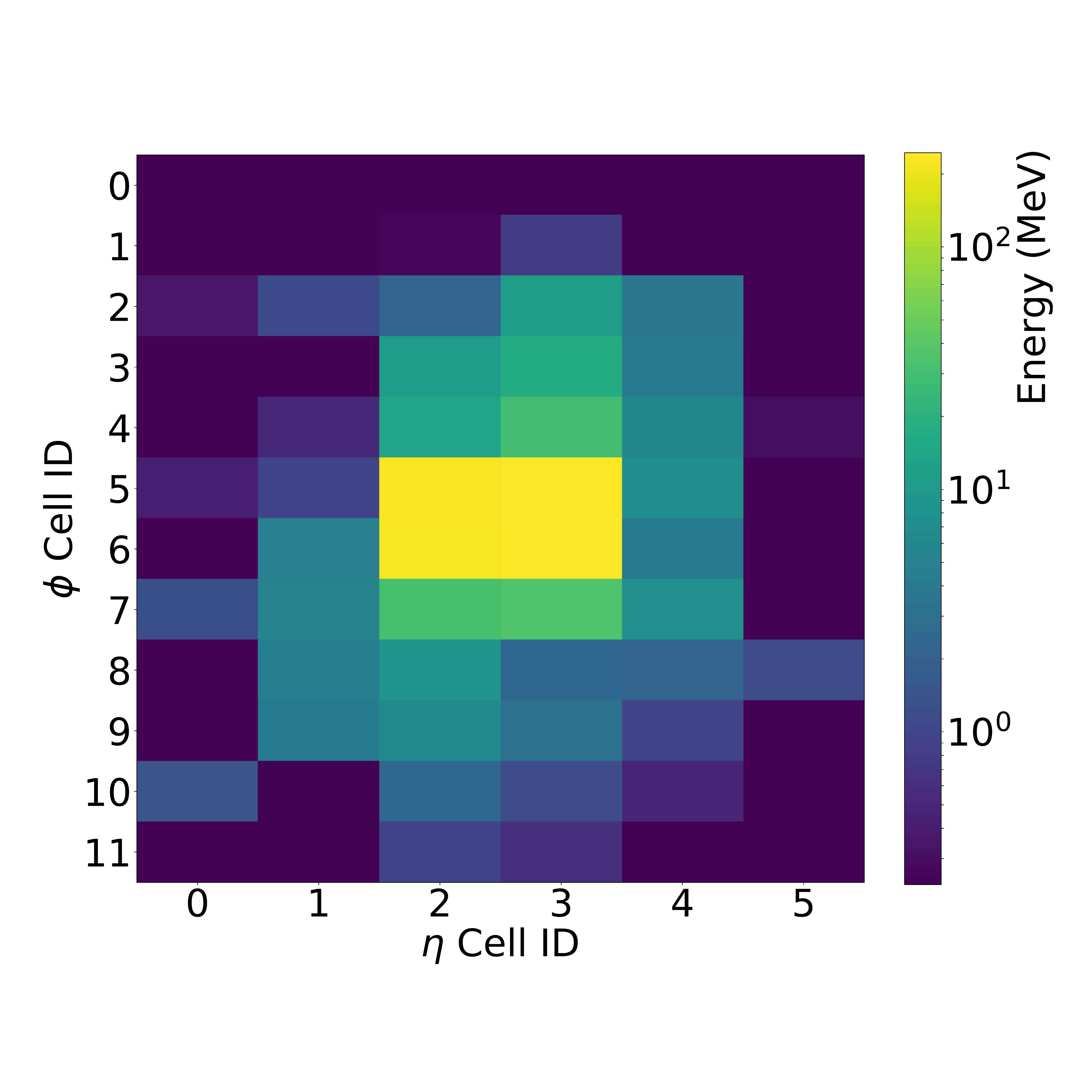}\\
    \caption{Average $\pi^+$ \textsc{Geant4} shower (top), and average $\pi^+$ \textsc{CaloGAN} shower (bottom), with progressive calorimeter depth (left to right).}
    \label{fig:piplus_avg}
\end{figure}

We first examine the average calorimeter deposition per voxel (a volumetric pixel). On average, the systems learn a complete picture of the underlying physical processes governing the cascades of $e^+$, $\gamma$, and $\pi^+$ with uniform energy between $1$ GeV and $100$ GeV (Figs.~\ref{fig:eplus_avg}, ~\ref{fig:gamma_avg}, and~\ref{fig:piplus_avg}).

\begin{figure}
    \centering
    \includegraphics[width=0.45\textwidth, trim={0cm, 0cm, 0cm, 0cm}]{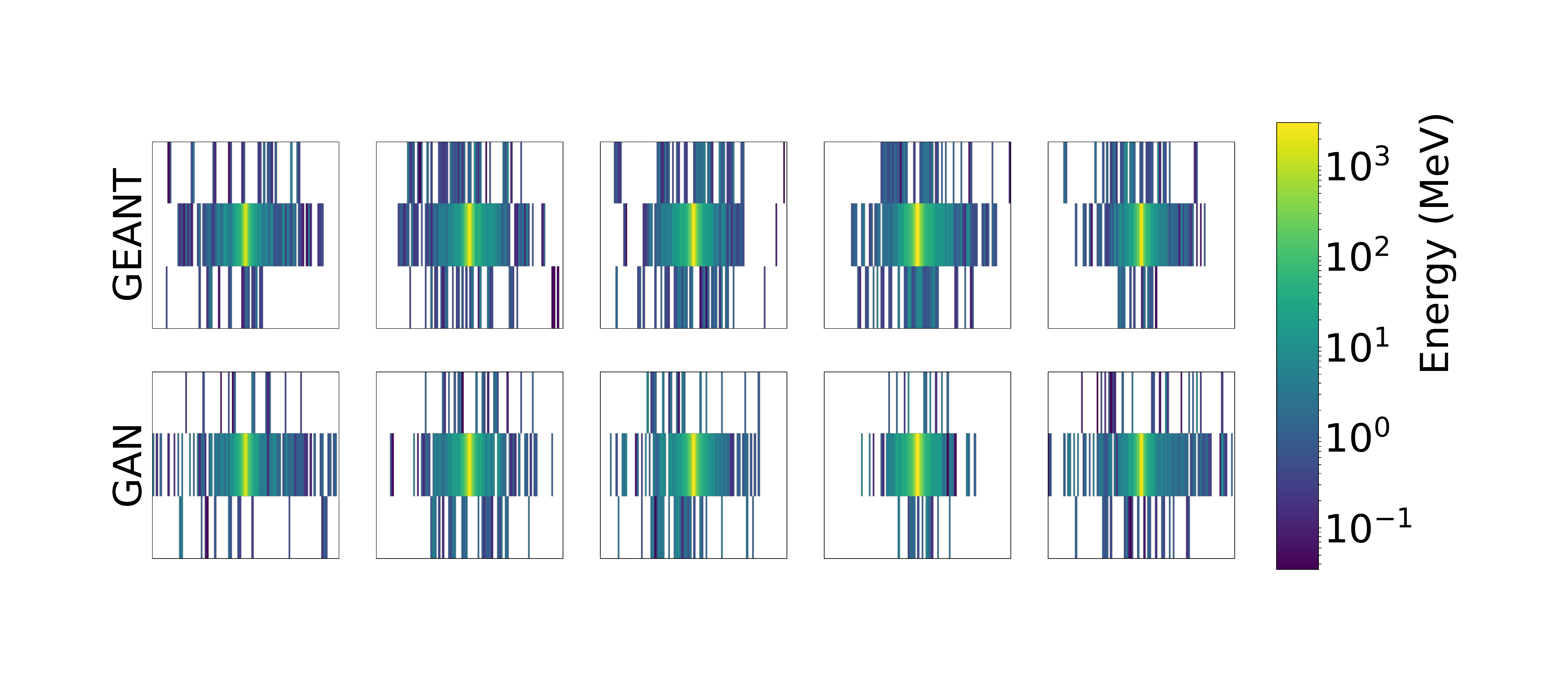}\hfill
    \includegraphics[width=0.45\textwidth, trim={0cm, 0cm, 0cm, 0cm}]{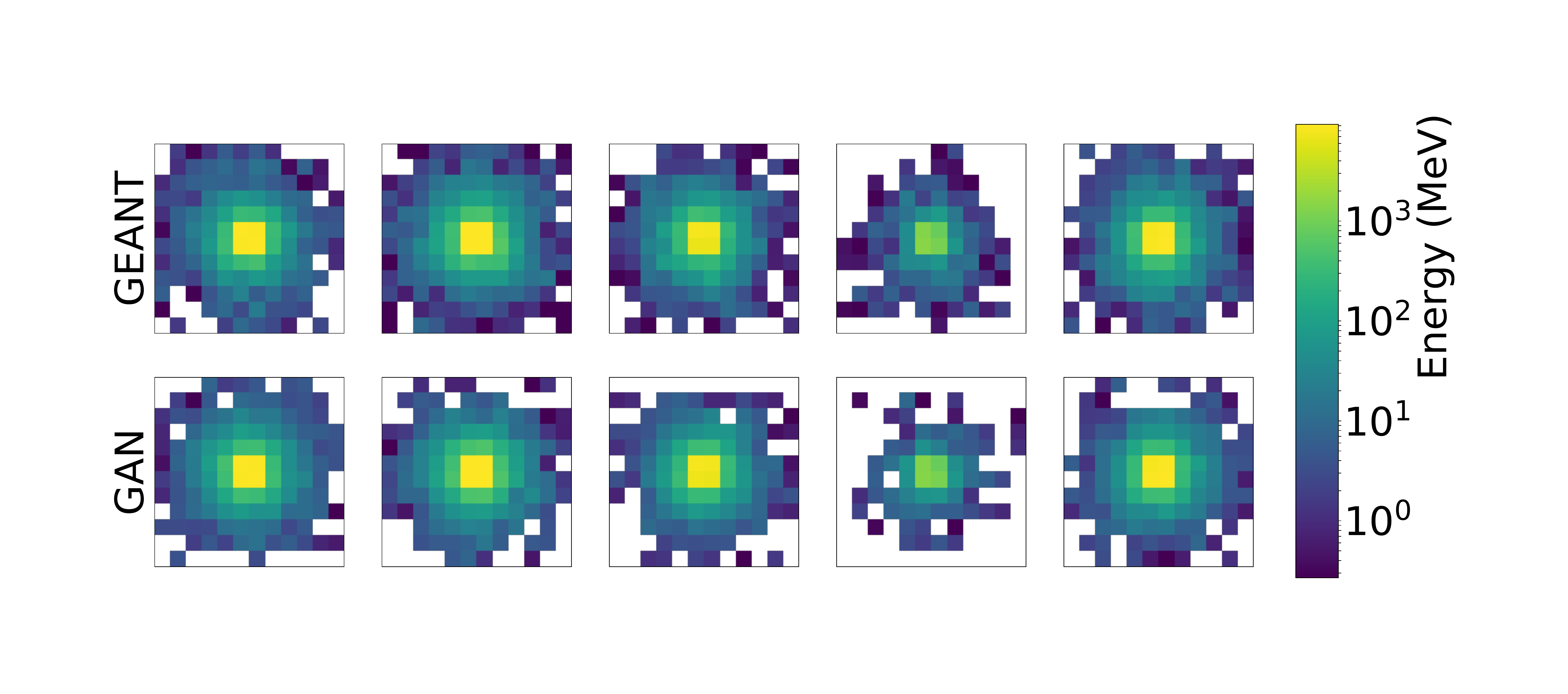}\hfill
    \includegraphics[width=0.45\textwidth, trim={0cm, 0cm, 0cm, 0cm}]{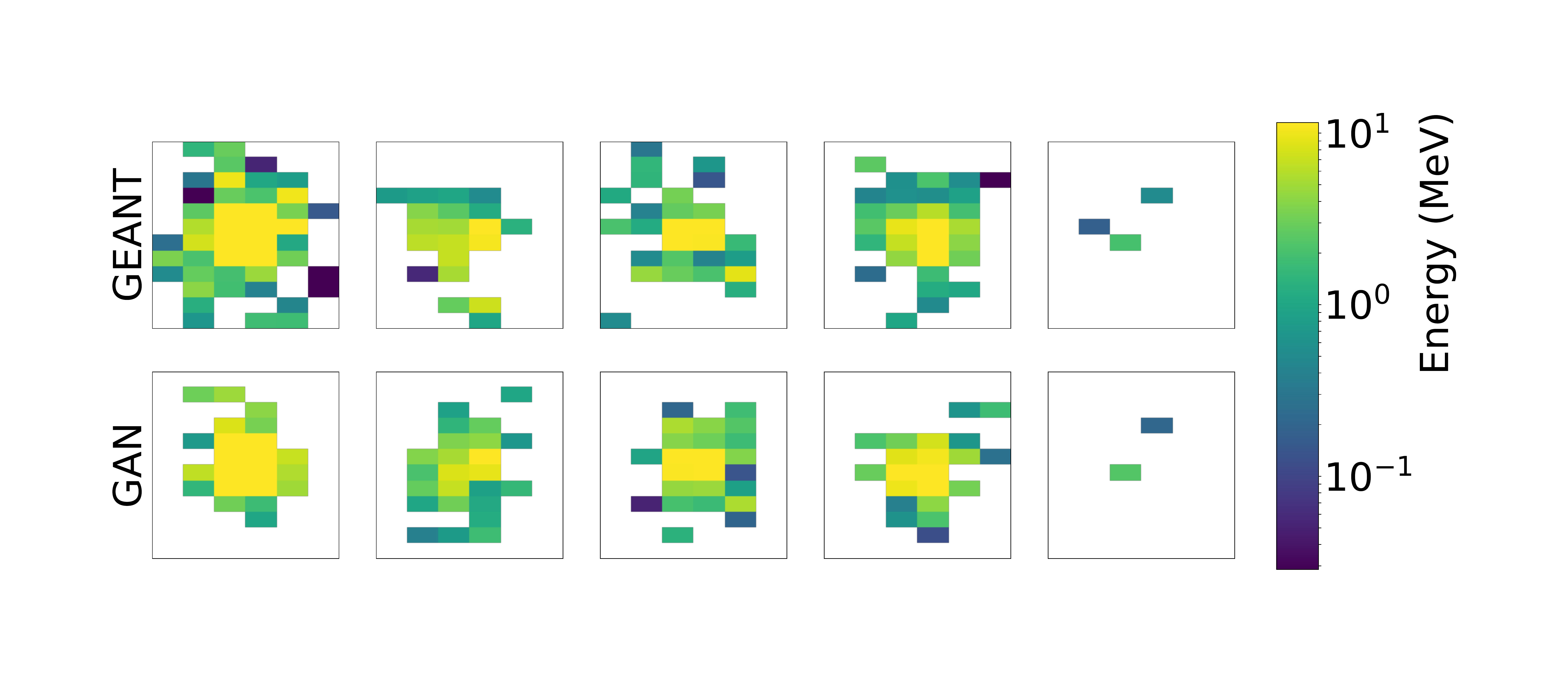}\\
    \caption{Five randomly selected $e^+$ showers per calorimeter layer from the training set (top) and the five nearest neighbors (by euclidean distance) from a set of \textsc{CaloGAN} candidates.}
    \label{fig:eplus_nn}
\end{figure}

\begin{figure}
    \centering
    \includegraphics[width=0.45\textwidth, trim={0cm, 0cm, 0cm, 0cm}]{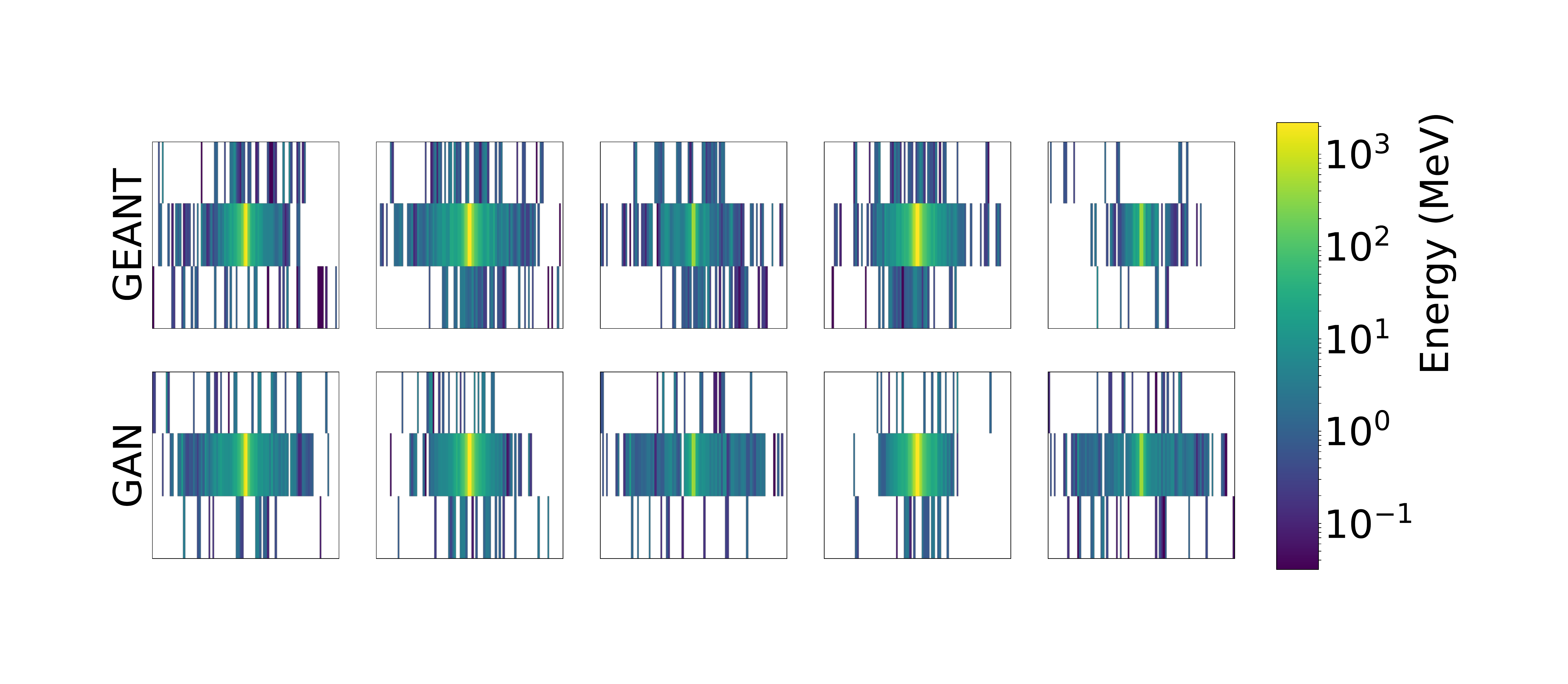}\hfill
    \includegraphics[width=0.45\textwidth, trim={0cm, 0cm, 0cm, 0cm}]{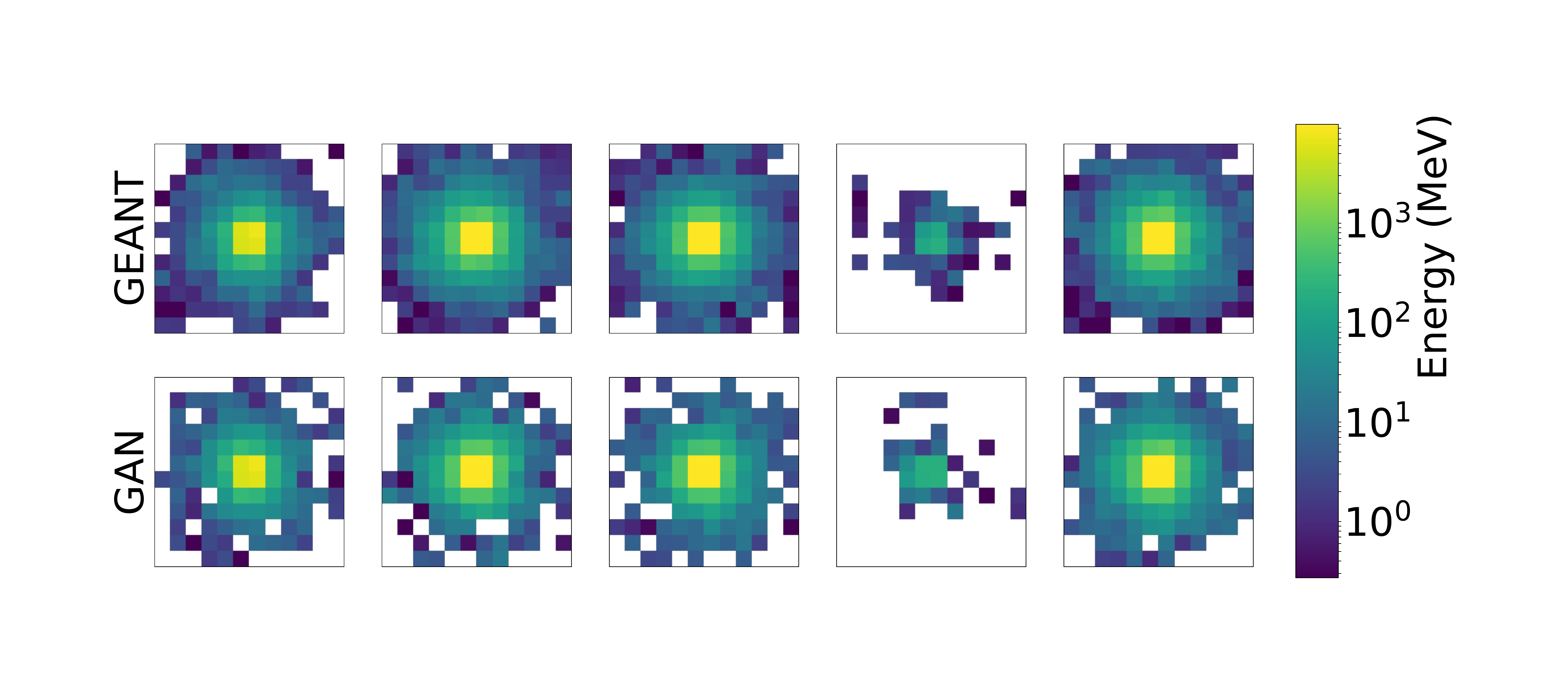}\hfill
    \includegraphics[width=0.45\textwidth, trim={0cm, 0cm, 0cm, 0cm}]{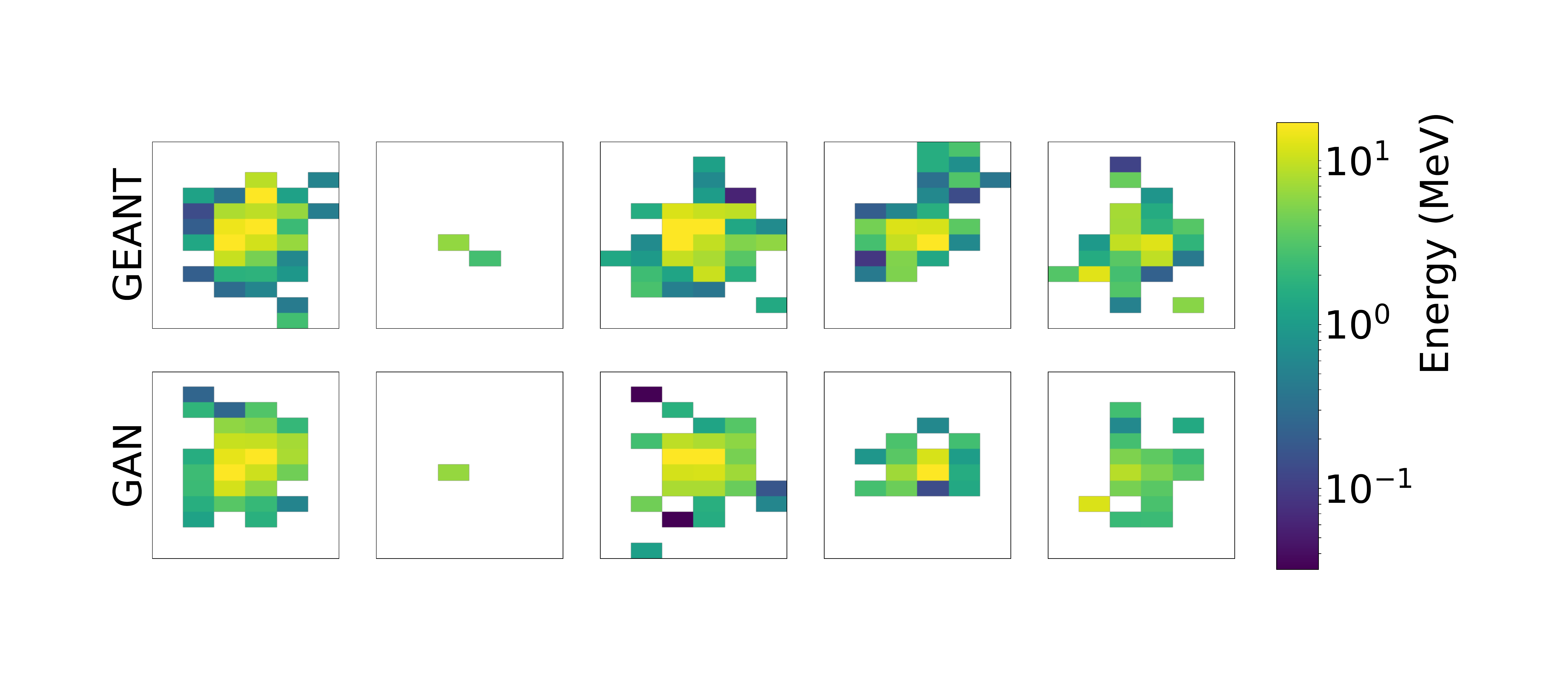}\\
    \caption{Five randomly selected $\gamma$ showers per calorimeter layer from the training set (top) and the five nearest neighbors (by euclidean distance) from a set of \textsc{CaloGAN} candidates.}
    \label{fig:gamma_nn}
\end{figure}

\begin{figure}
    \centering
    \includegraphics[width=0.45\textwidth, trim={0cm, 0cm, 0cm, 0cm}]{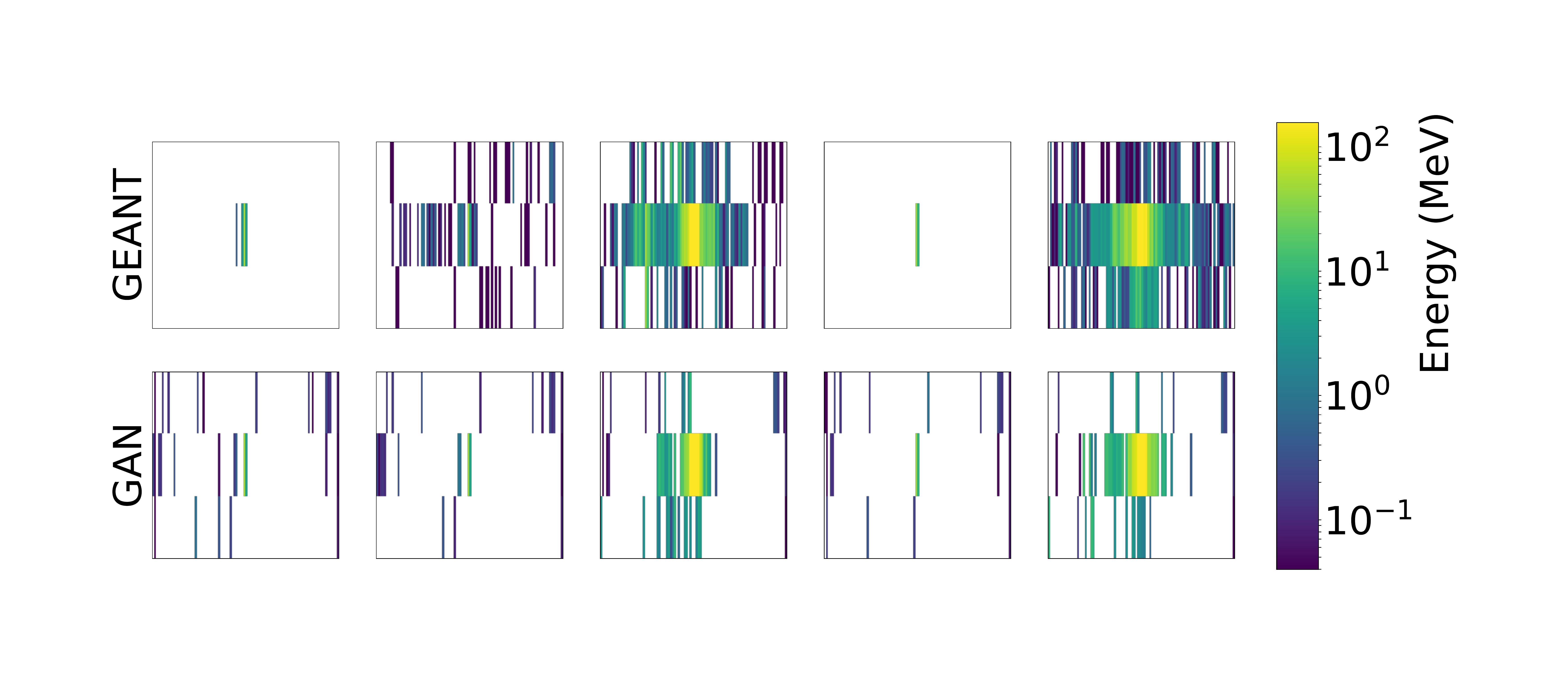}\hfill
    \includegraphics[width=0.45\textwidth, trim={0cm, 0cm, 0cm, 0cm}]{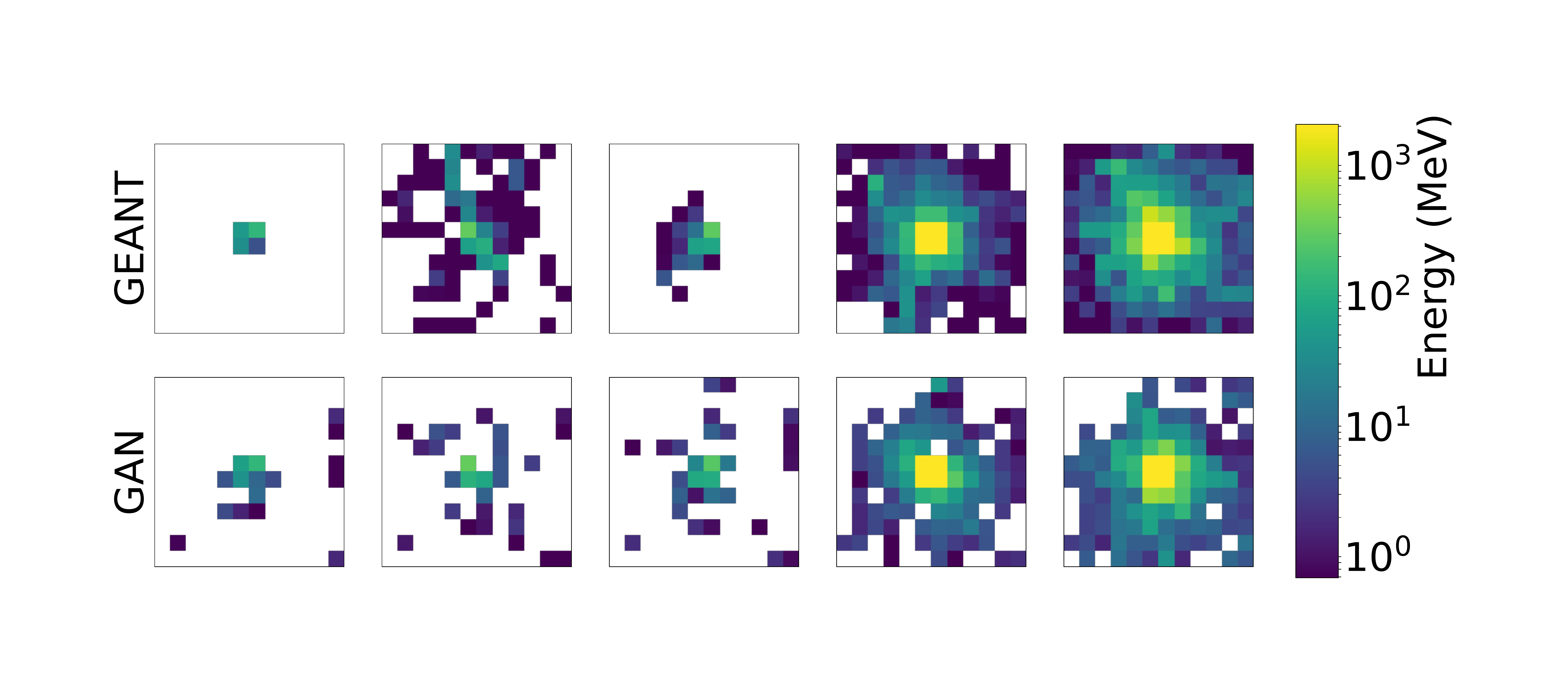}\hfill
    \includegraphics[width=0.45\textwidth, trim={0cm, 0cm, 0cm, 0cm}]{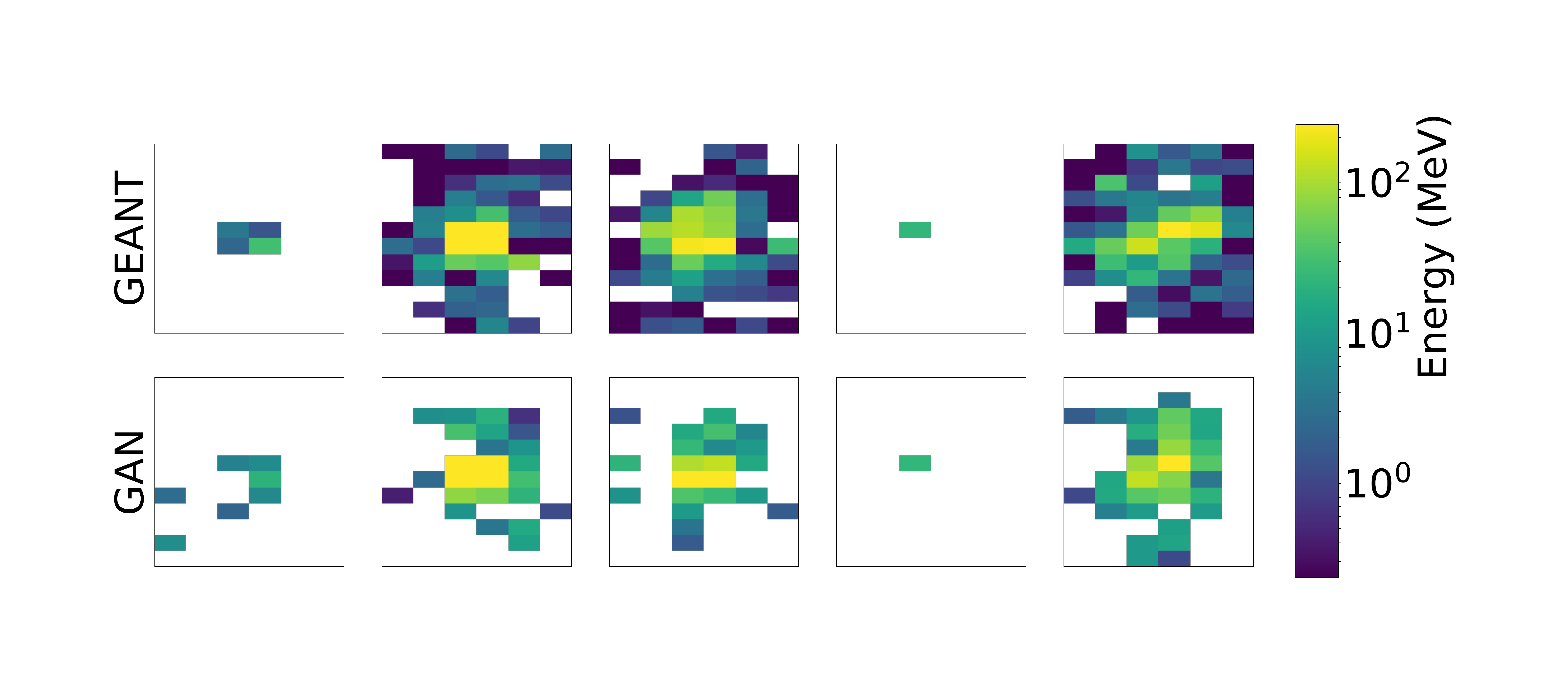}\\
    \caption{Five randomly selected $\pi^+$ showers per calorimeter layer from the training set (top) and the five nearest neighbors (by euclidean distance) from a set of \textsc{CaloGAN} candidates.}
    \label{fig:piplus_nn}
\end{figure}

Diversity and overtraining concerns can be investigated by considering the nearest neighbors among the training and generated datasets. 
Figs.~\ref{fig:eplus_nn},~\ref{fig:gamma_nn}, and~\ref{fig:piplus_nn} shows five randomly selected events and their GAN-generated nearest neighbors for all three calorimeter layers for $e^+$, $\gamma$ and $\pi^+$ showers respectively. Good qualitative agreement can be found between the two distributions across all layers, without obvious signs of \textit{mode collapse}: a failure mode in which the generator learns to produce a small subset of samples from the distribution. Compared to the other two particle types explored in this application, at the individual image level, charged pions clearly display a higher degree of complexity and diversity in their showers. Some $\pi^+$ deposit energy in all cells of a given layer, some only hitting a handful of them. This is because charged pions undergo nuclear interactions in addition to electromagnetic interactions.



\subsection{Shower Shapes}
\label{ssec:showershapes}

Electron and photon classification and energy calibration use properties of the calorimeter shower~\cite{Aad:2011mk,Aad:2014fxa,Aaboud:2016yuq,Aaboud:2016vfy}. These same features can be used to quantitatively assess the quality of the GAN samples.  The list of features used for evaluation is provided in Table~\ref{table:qualityvariables} in Appendix~\ref{shower_shape_appendix}. The key physical quantity that governs the shapes of these distributions is the number of radiation lengths $X_0$ that are traversed by the particle. By definition, $X_0$ is the distance an electron will travel before its energy is reduced to $1/e$ on average. The equivalent distance for photons is slightly further (by 9/7~\cite{Agashe:2014kda}) and is set by the mean free path for pair production. The transverse shower size is also proportional to $X_0$. For a brief review, see e.g.~\cite{Agashe:2014kda}.

\begin{figure*}[]
    \centering

    \includegraphics[width=0.2\textwidth]{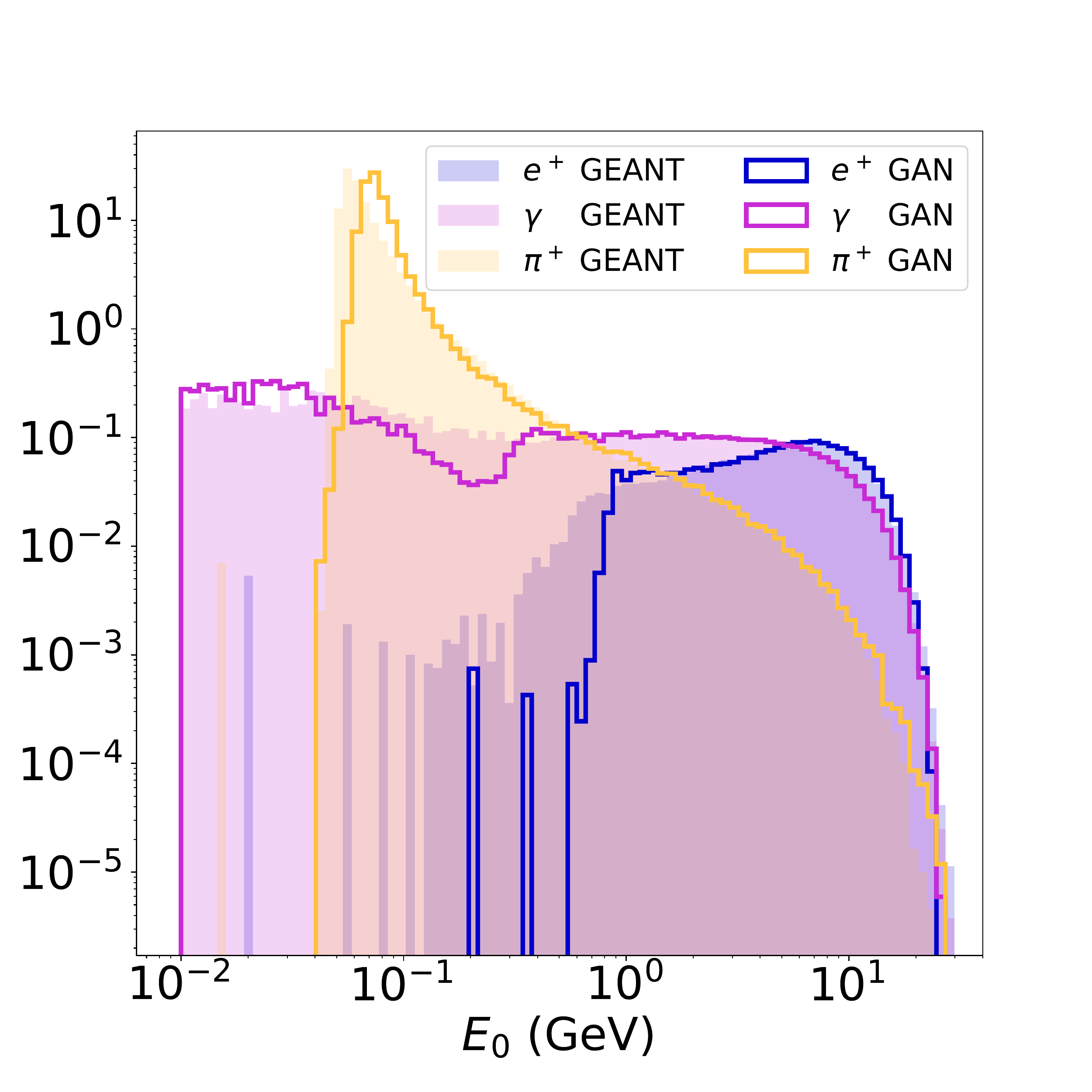}
    \includegraphics[width=0.2\textwidth]{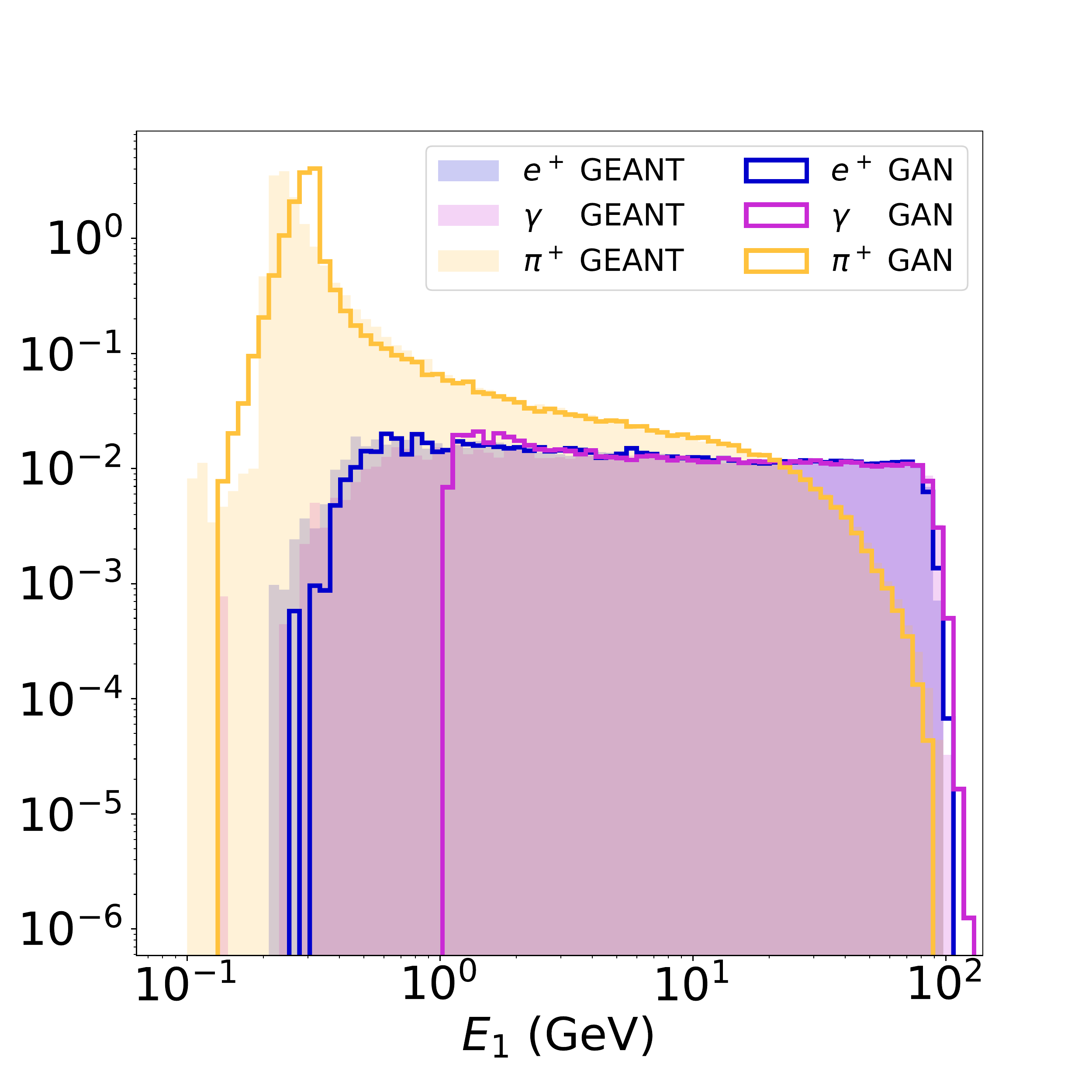}
    \includegraphics[width=0.2\textwidth]{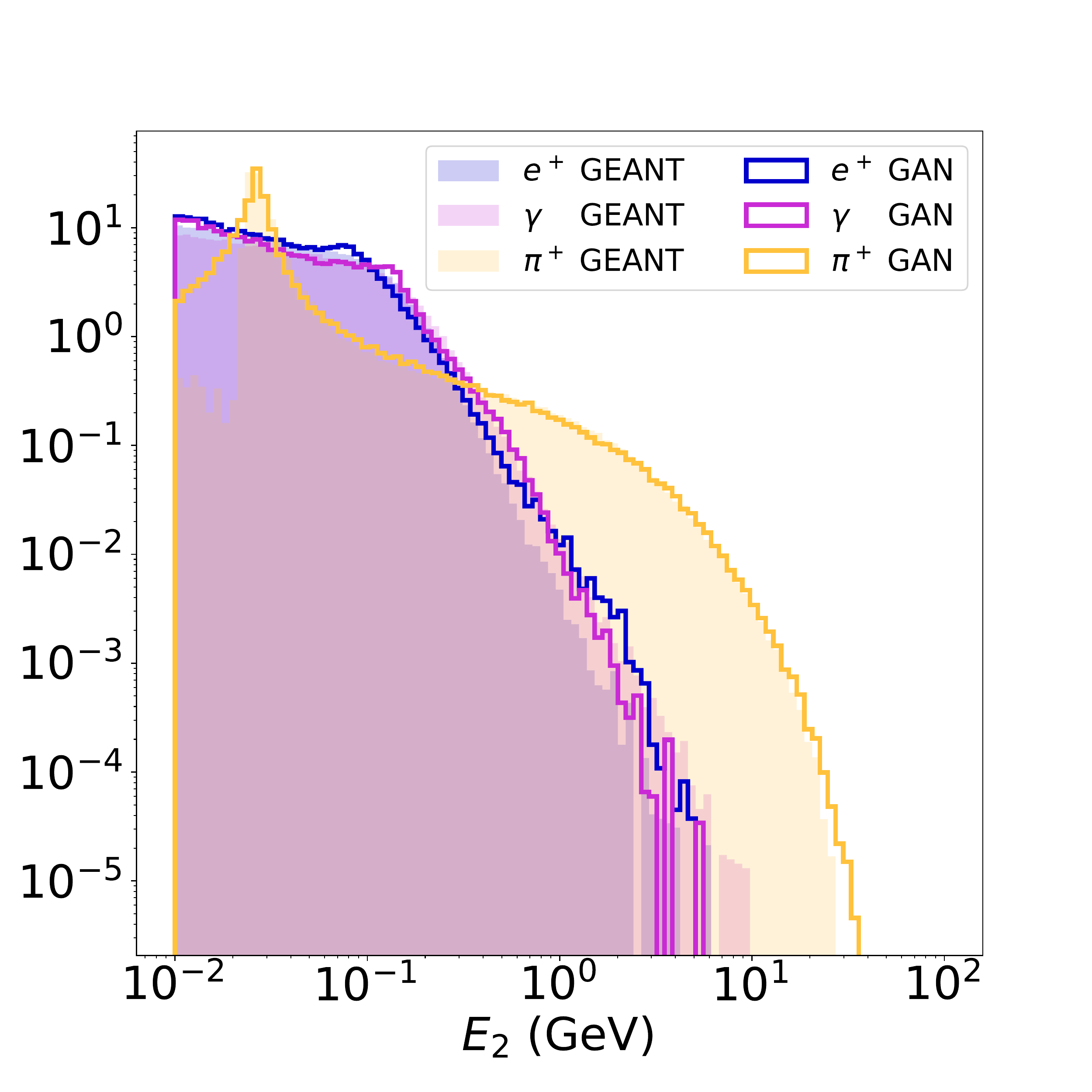}
    \includegraphics[width=0.2\textwidth]{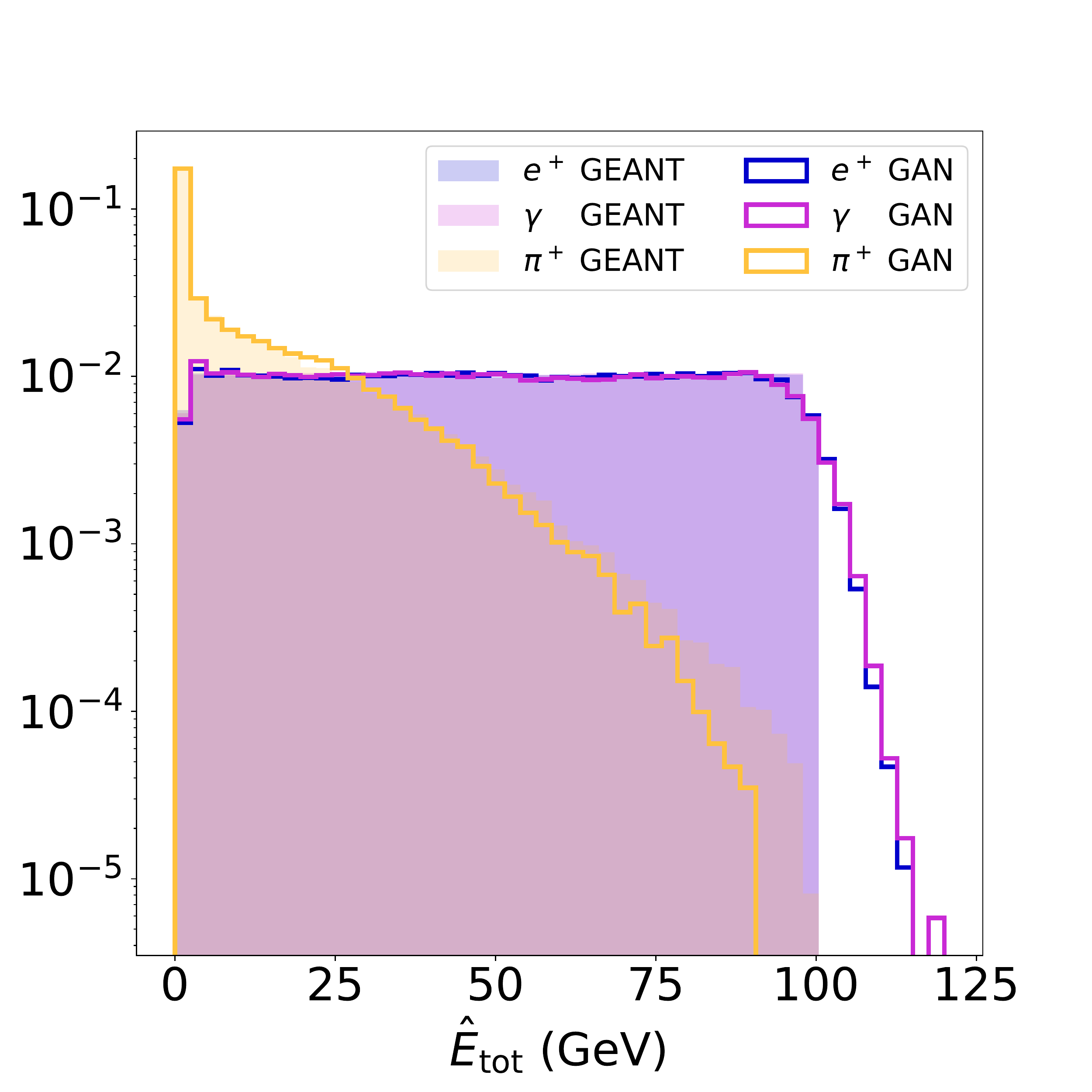}
    
    \includegraphics[width=0.2\textwidth]{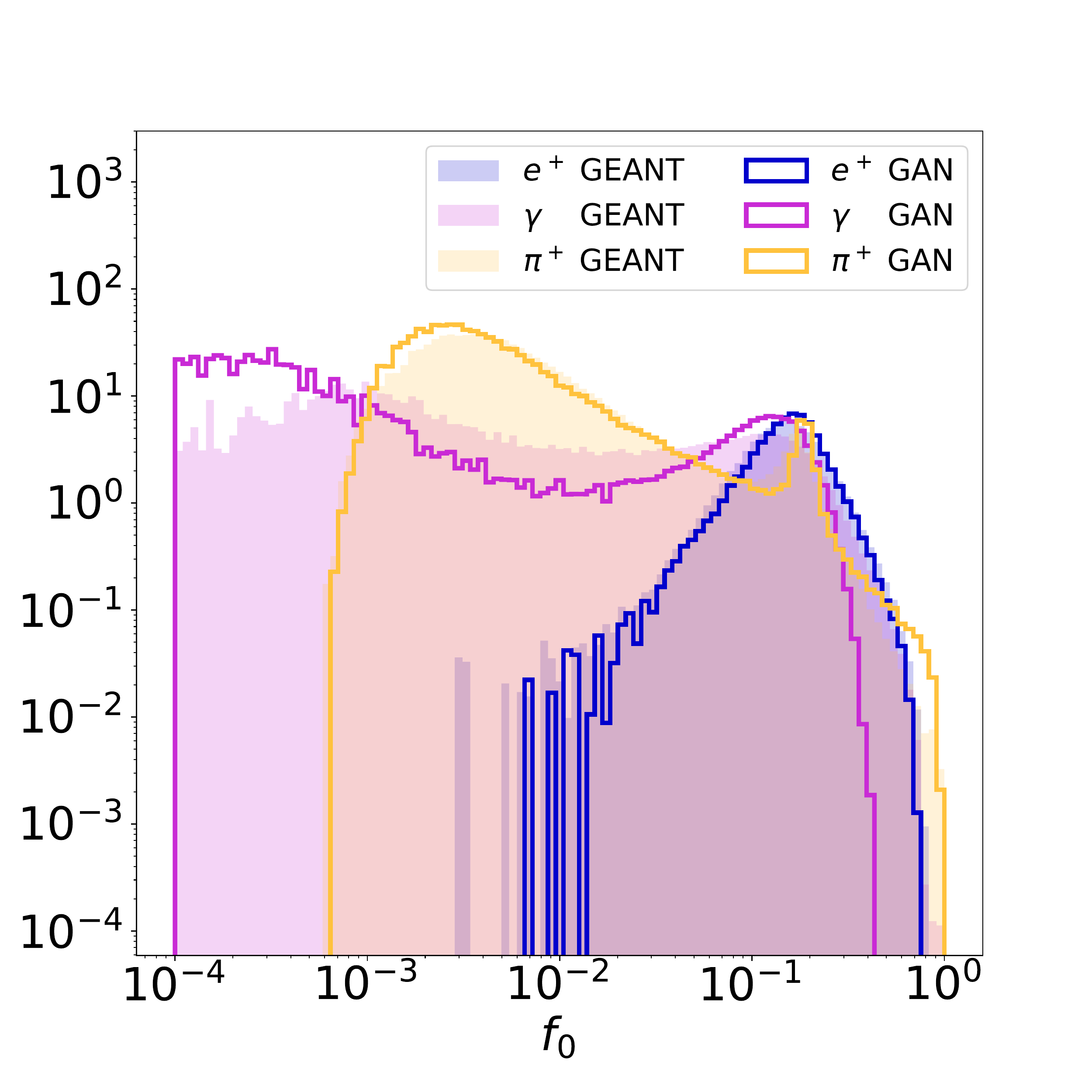}
    \includegraphics[width=0.2\textwidth]{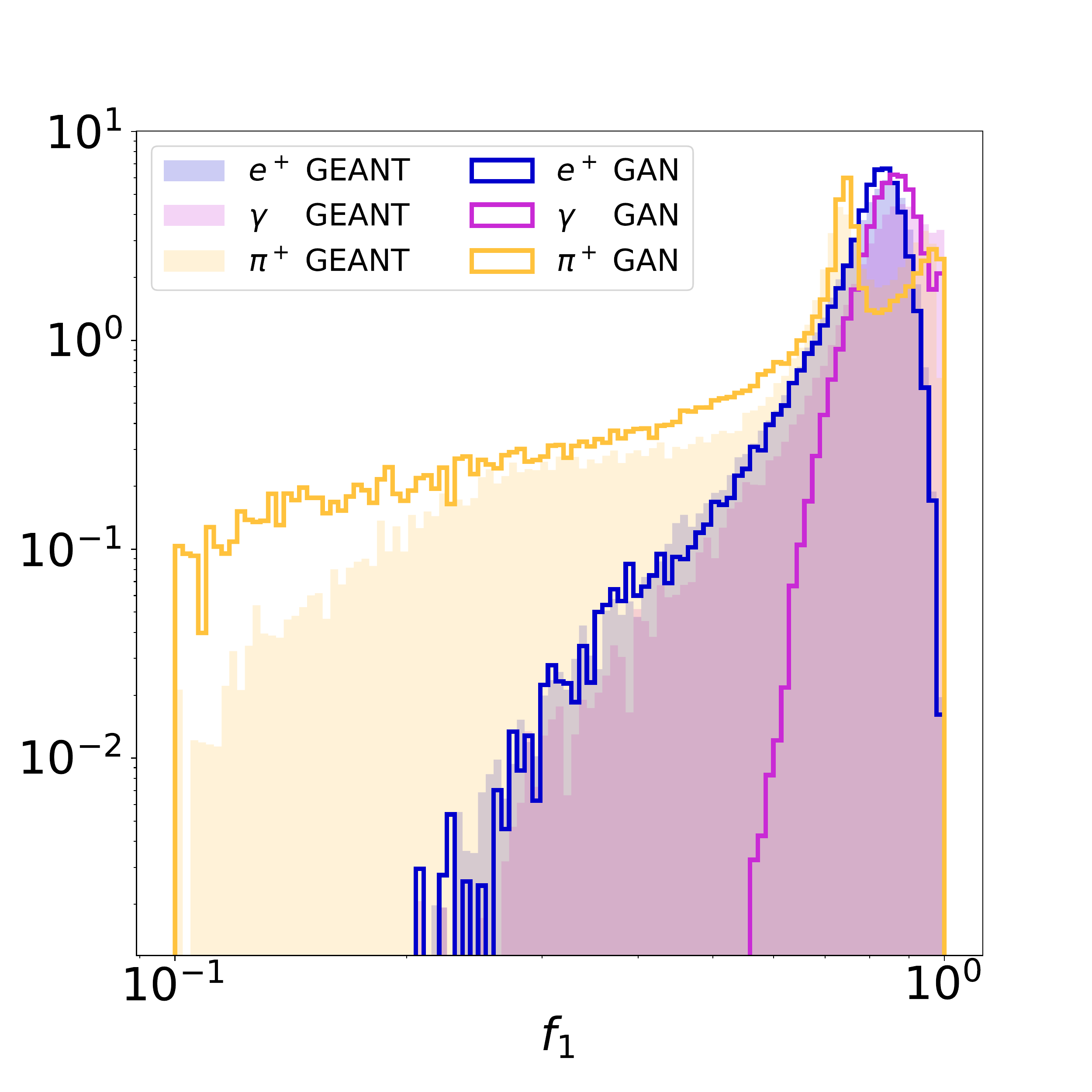}
    \includegraphics[width=0.2\textwidth]{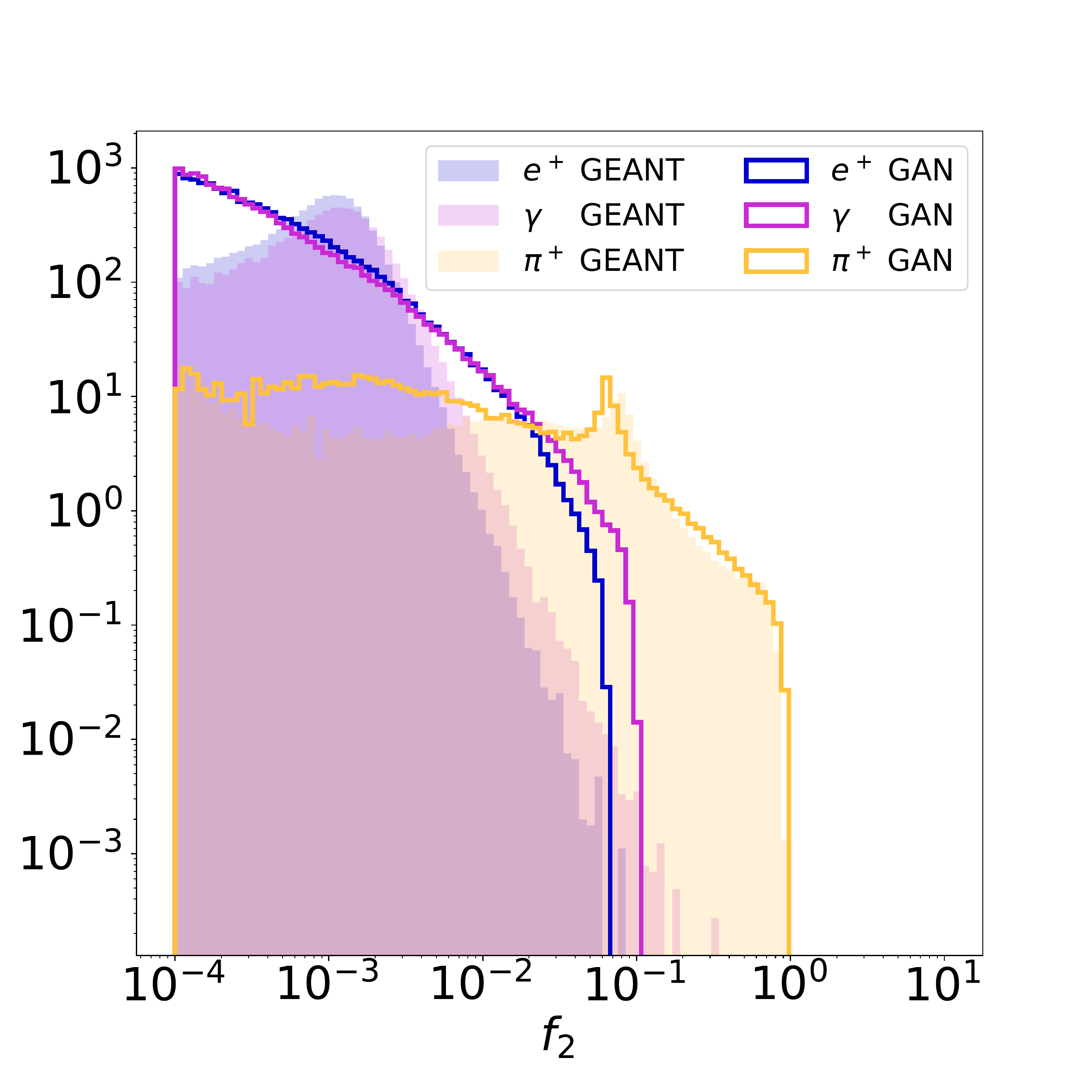}
    \includegraphics[width=0.2\textwidth]{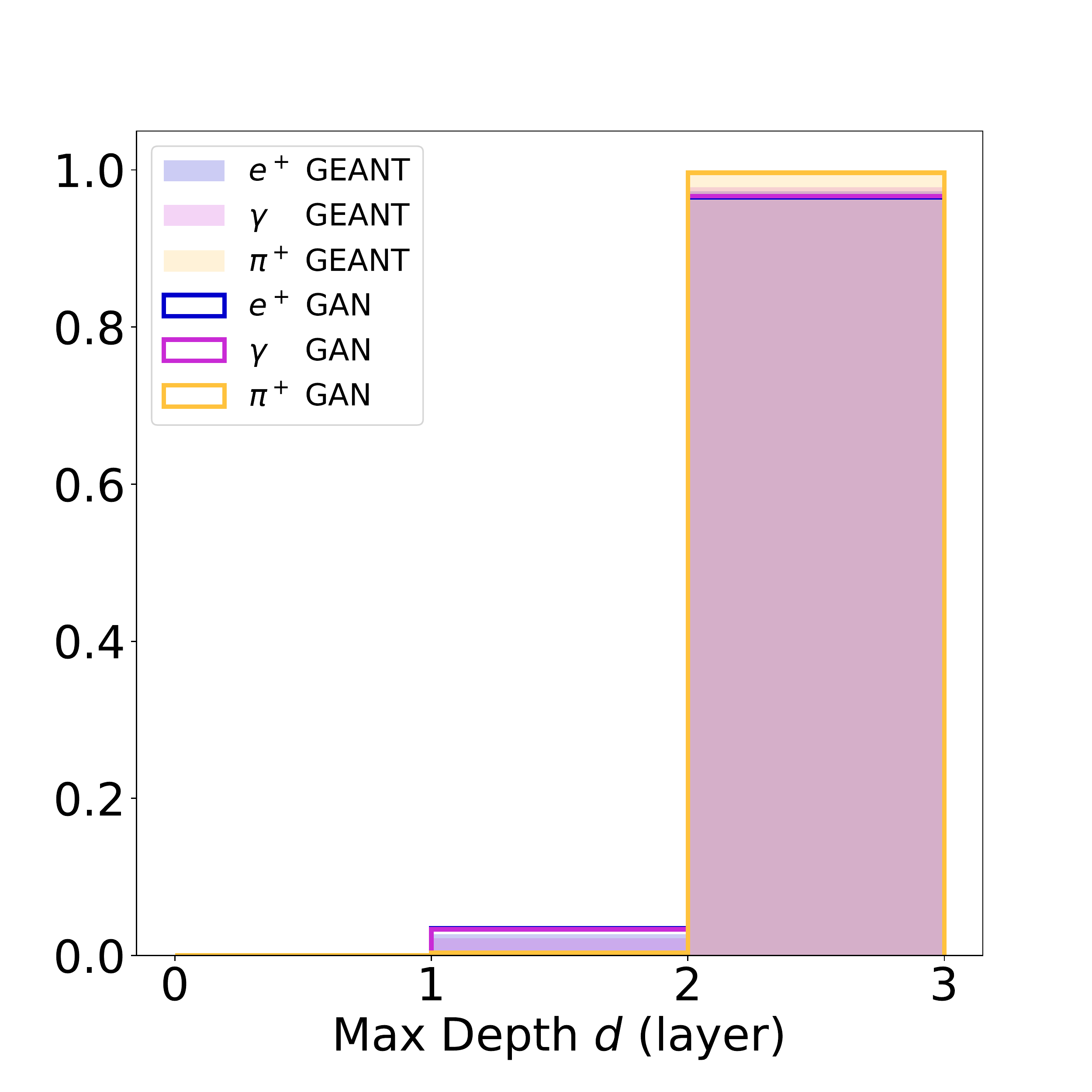}
    
    \includegraphics[width=0.2\textwidth]{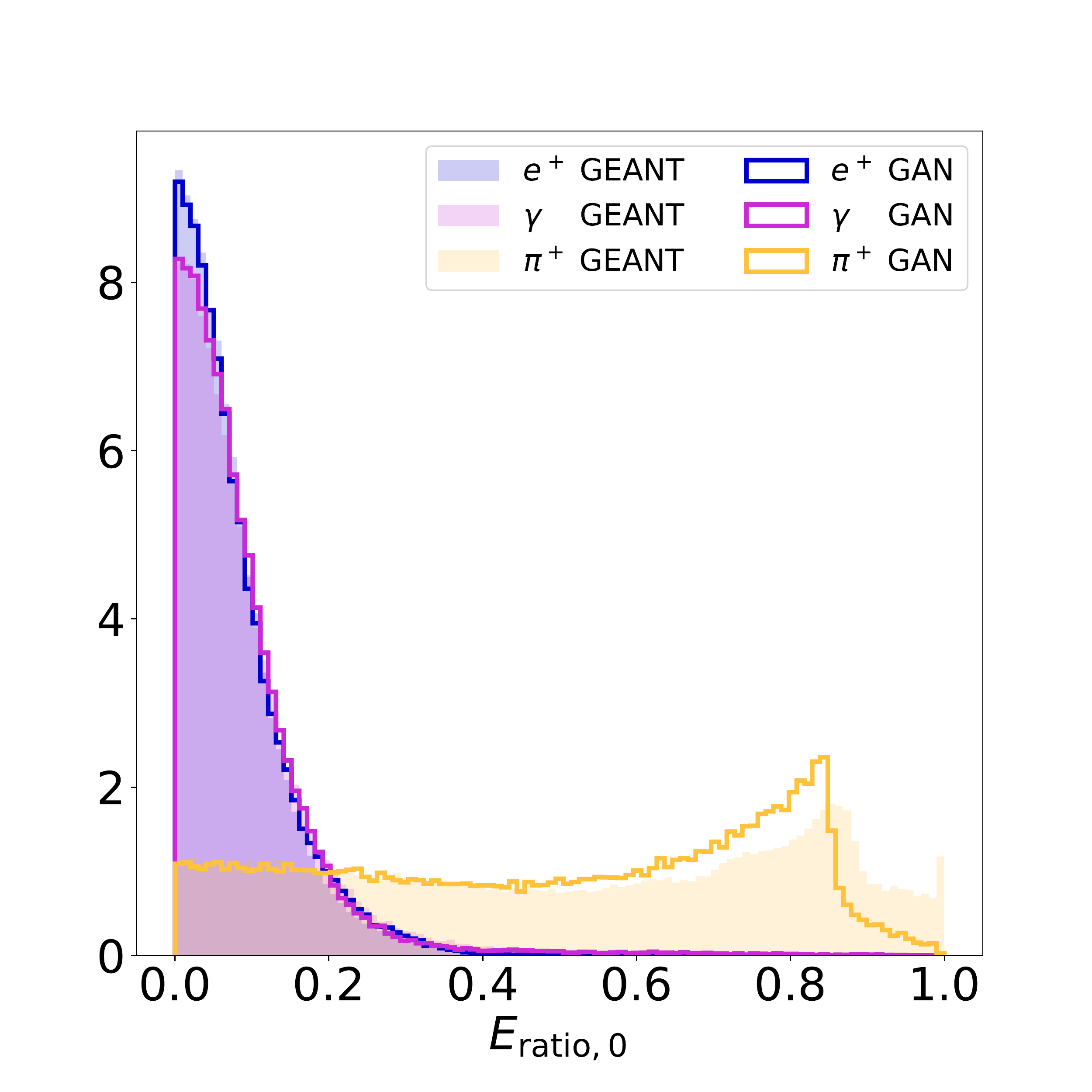}
    \includegraphics[width=0.2\textwidth]{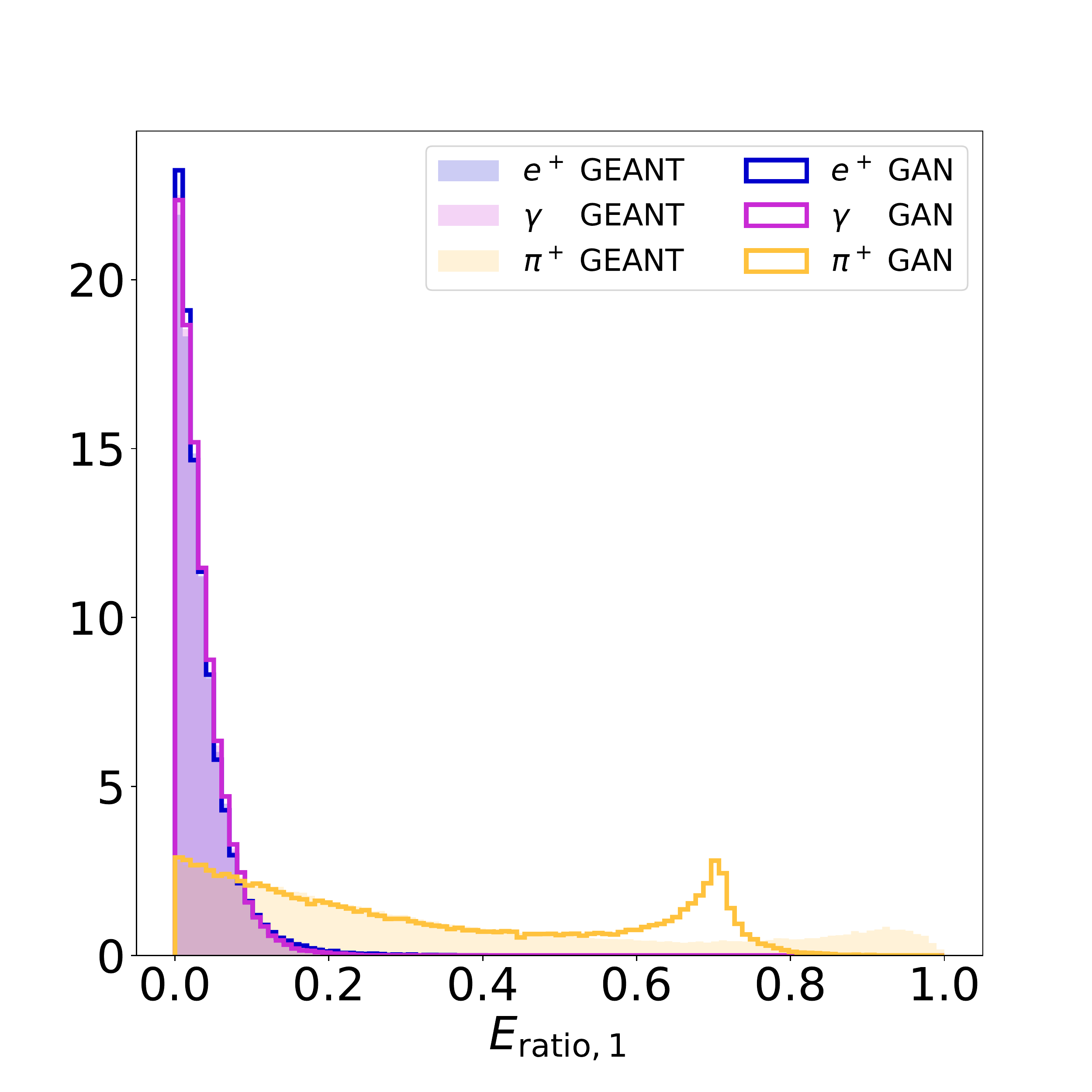}
    \includegraphics[width=0.2\textwidth]{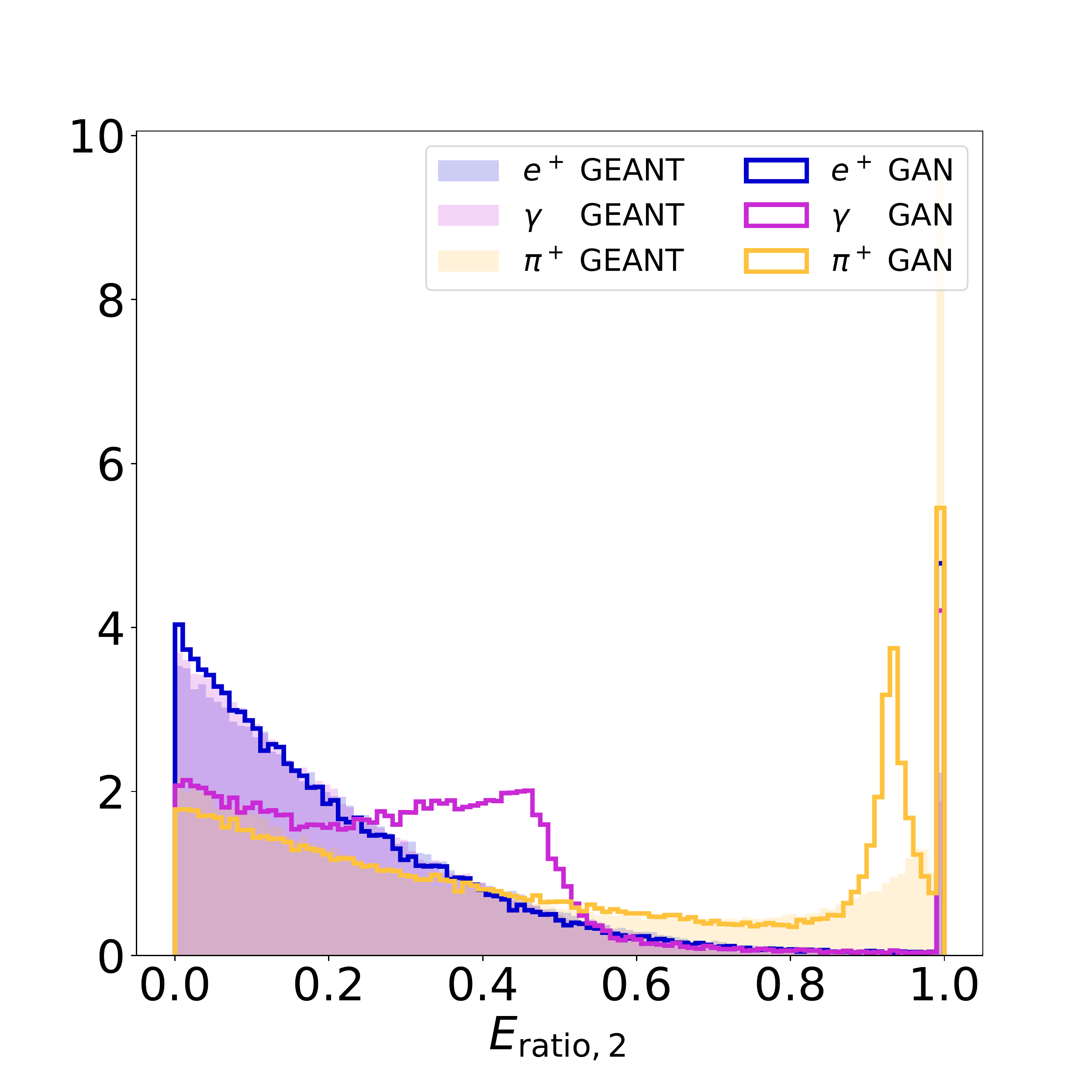}
    \includegraphics[width=0.2\textwidth]{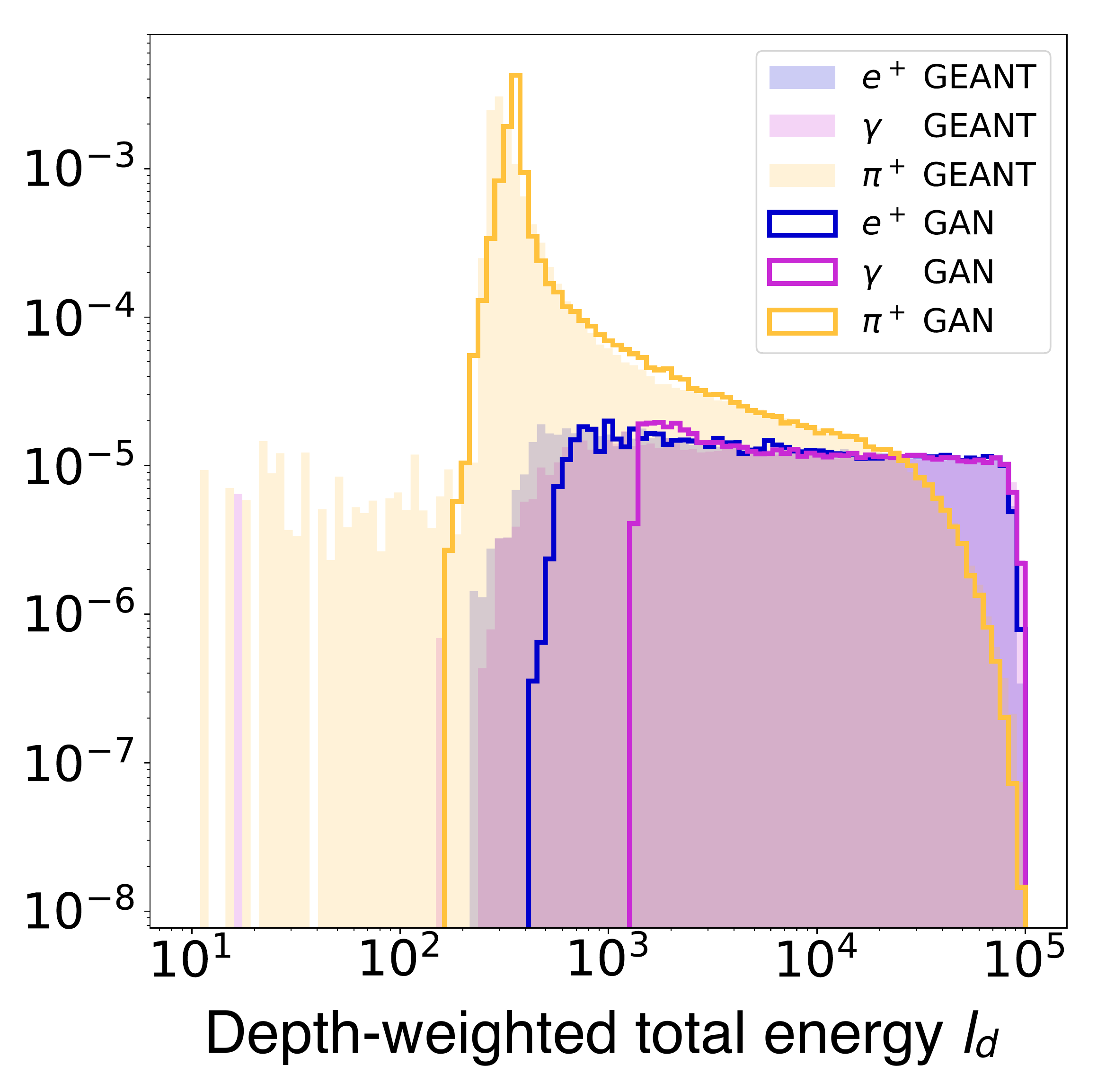}
    
    \includegraphics[width=0.2\textwidth]{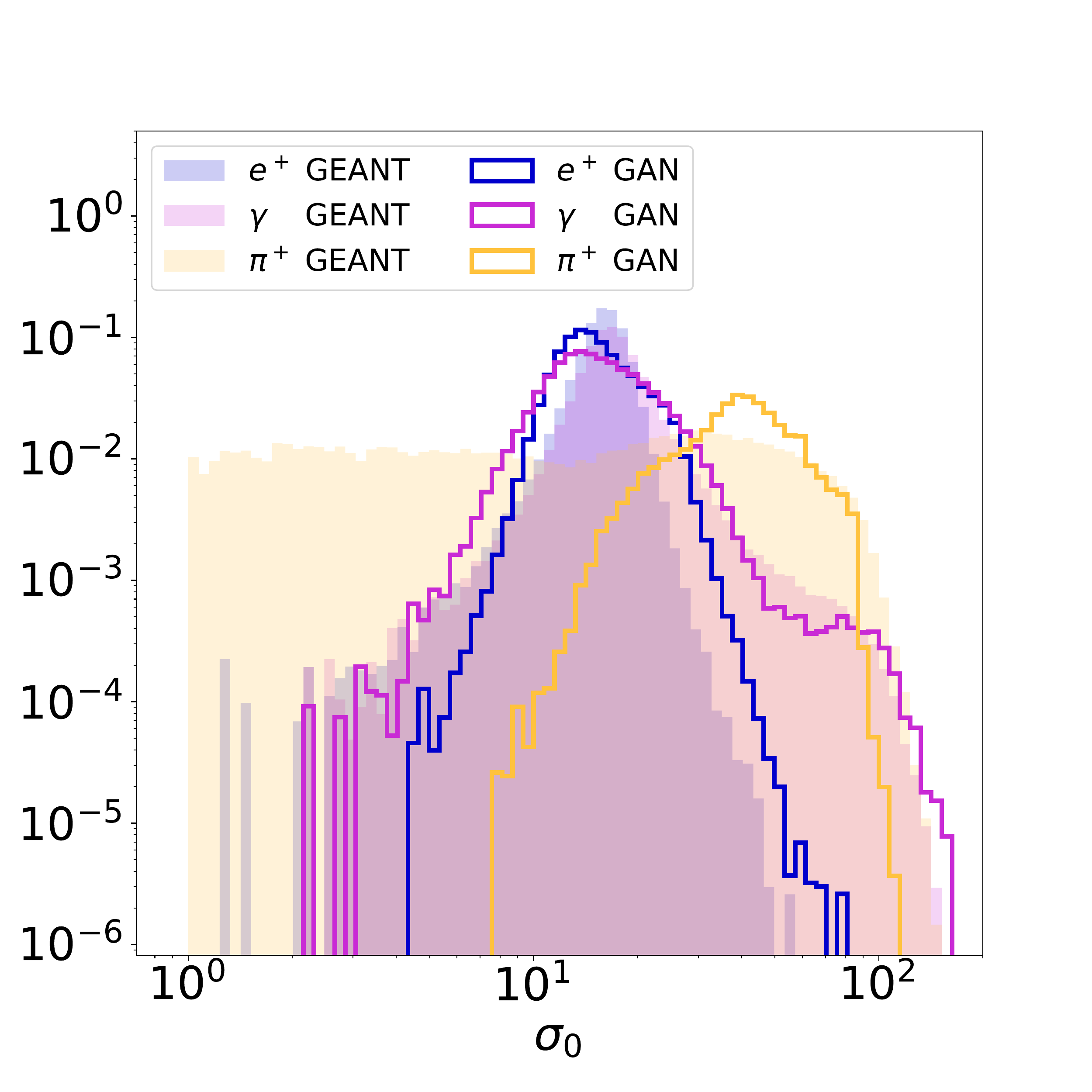}
    \includegraphics[width=0.2\textwidth]{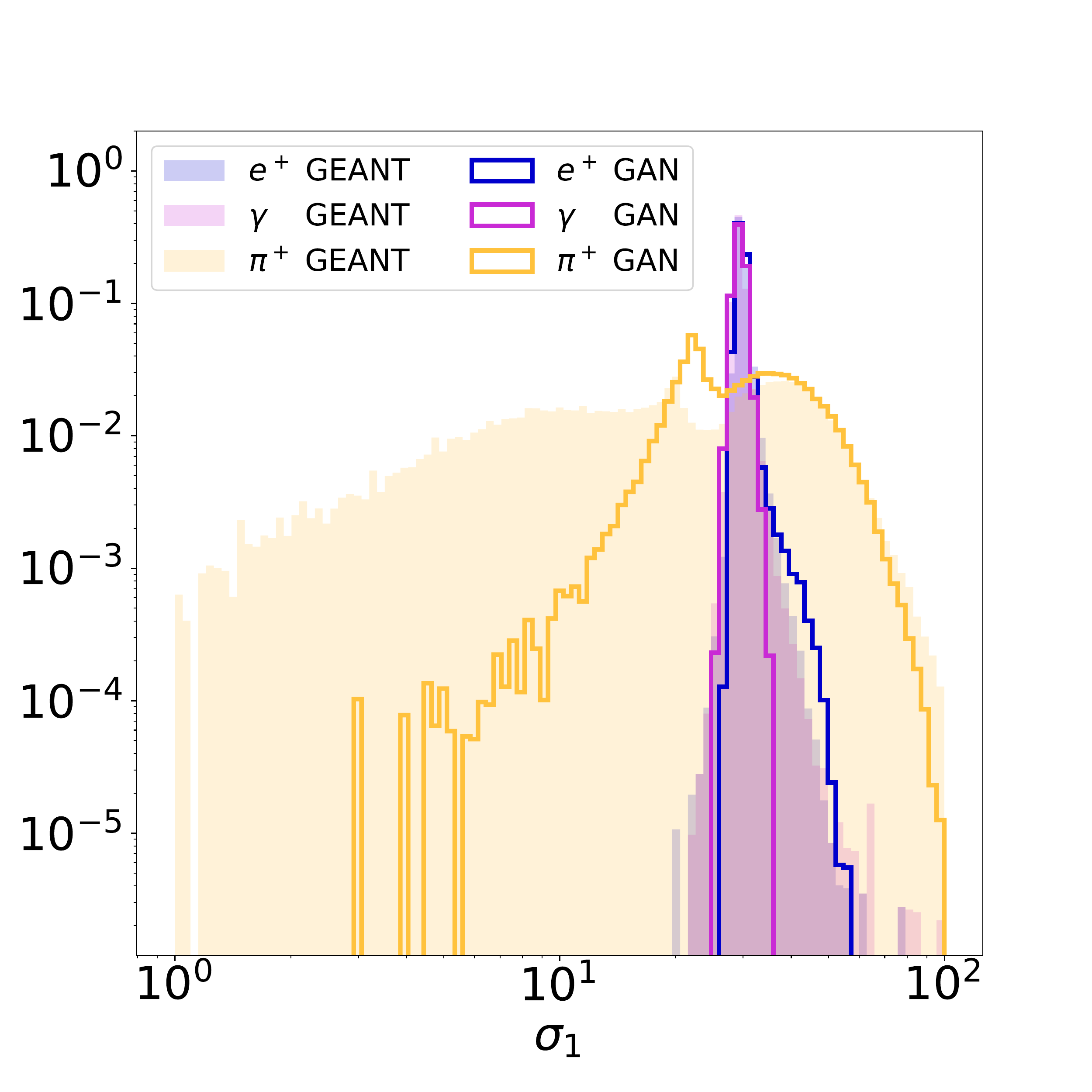}
    \includegraphics[width=0.2\textwidth]{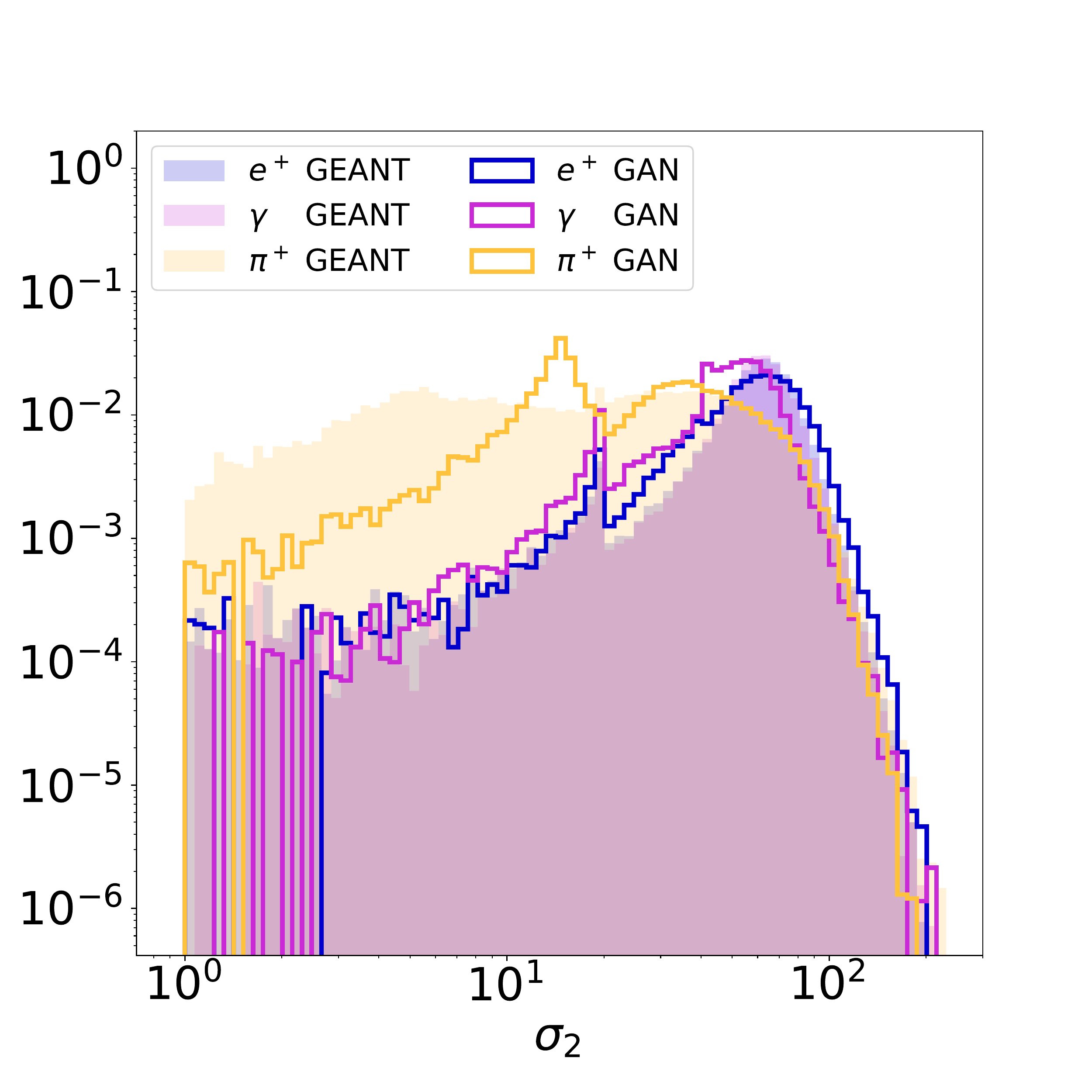}
    \includegraphics[width=0.2\textwidth]{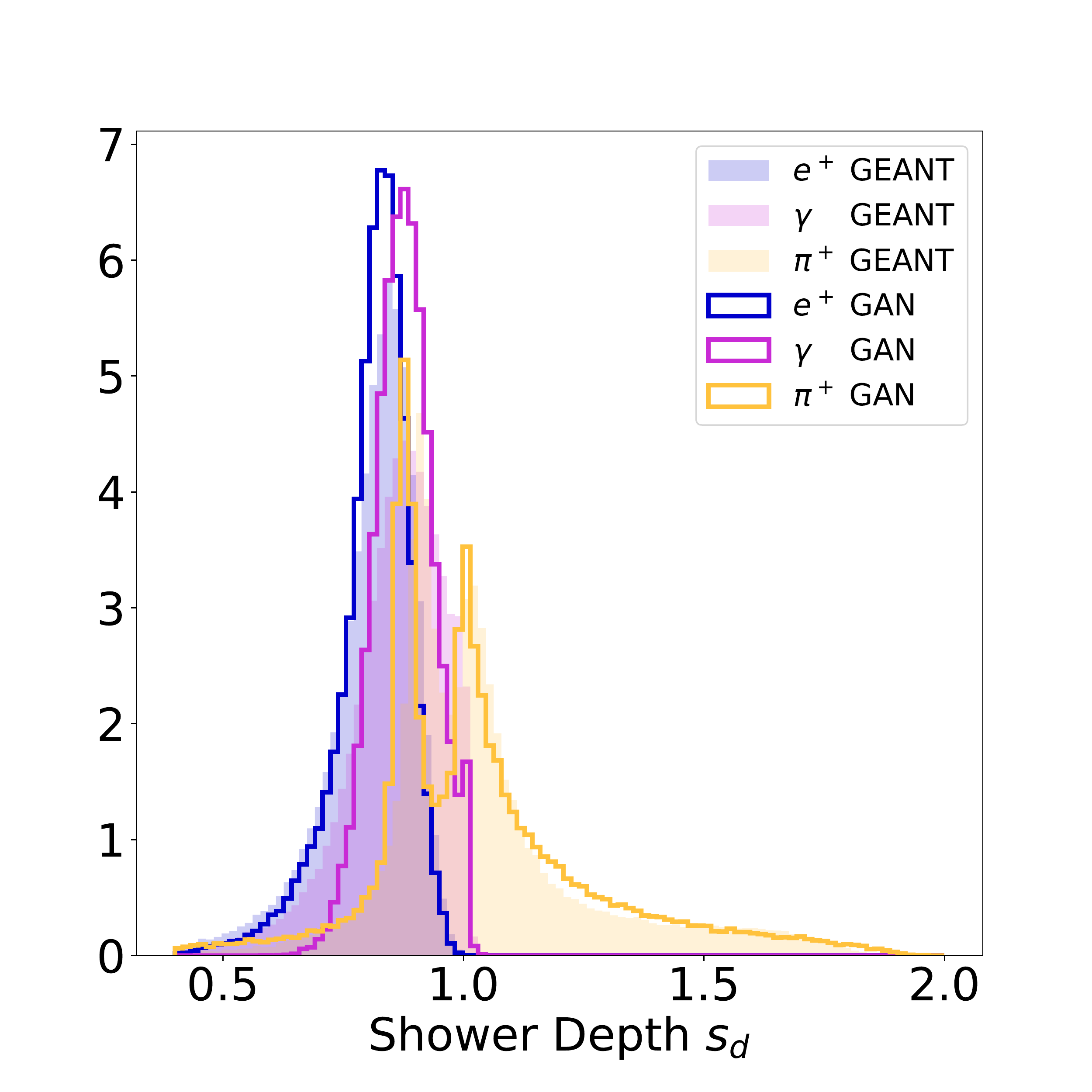}
    
    \includegraphics[width=0.2\textwidth]{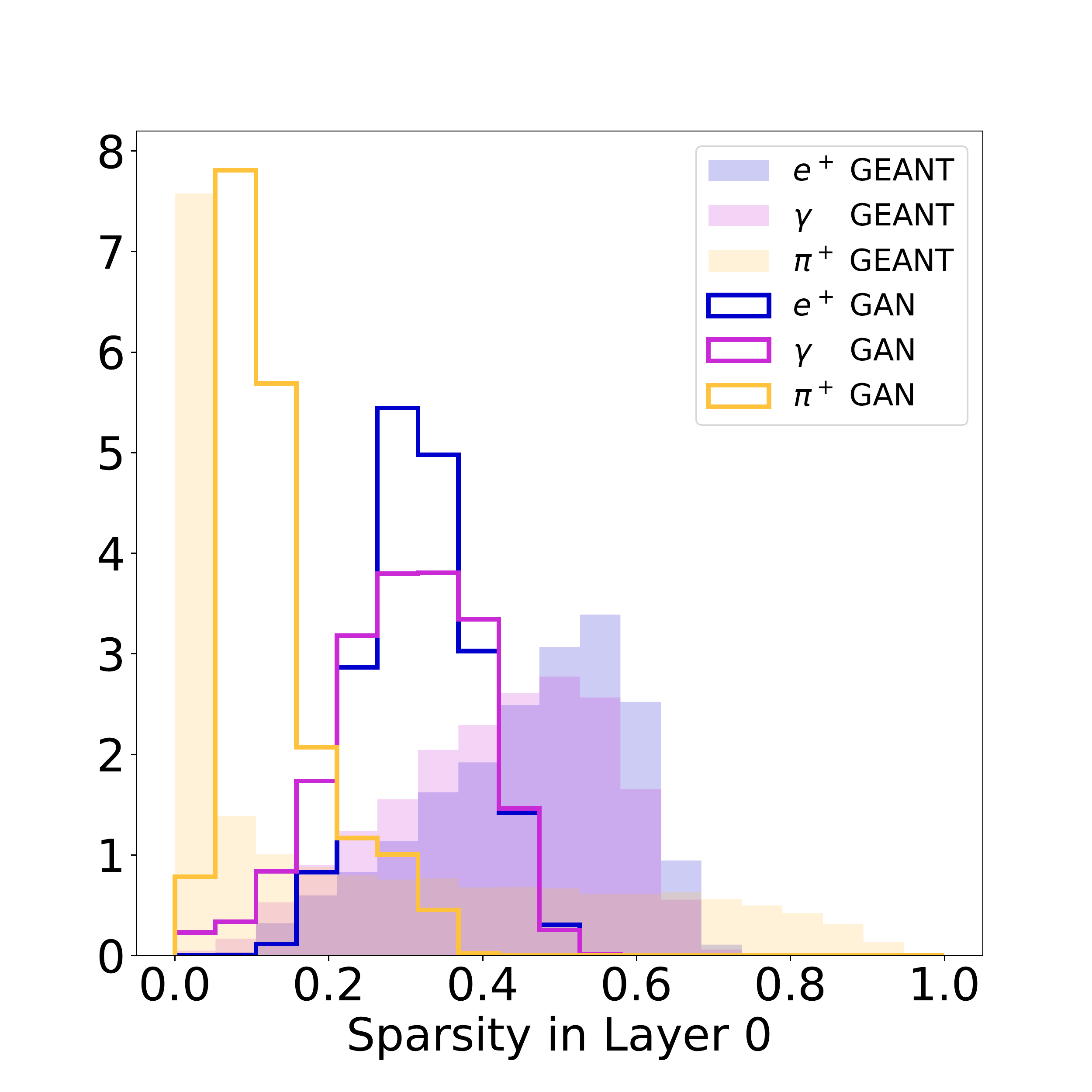}
    \includegraphics[width=0.2\textwidth]{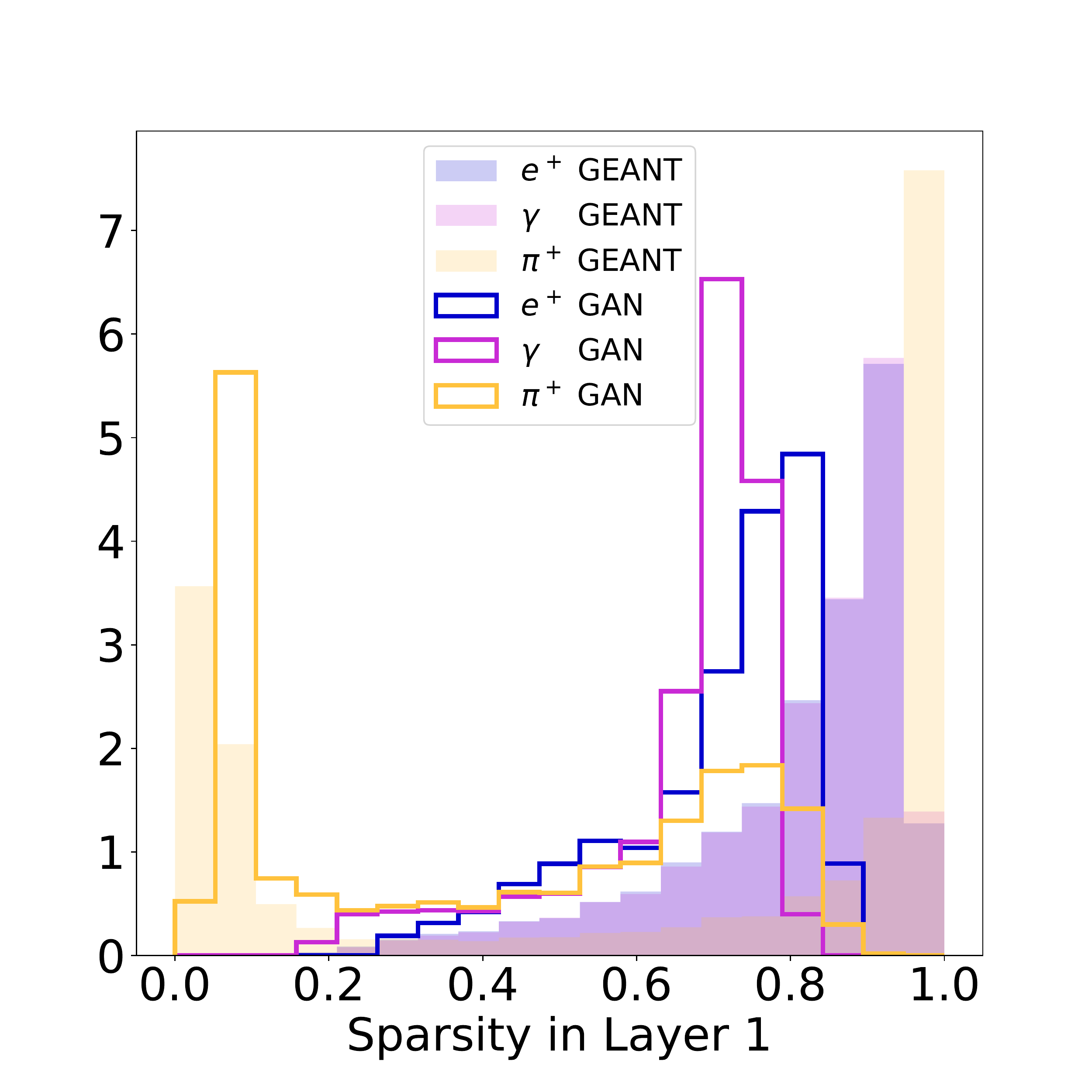}
    \includegraphics[width=0.2\textwidth]{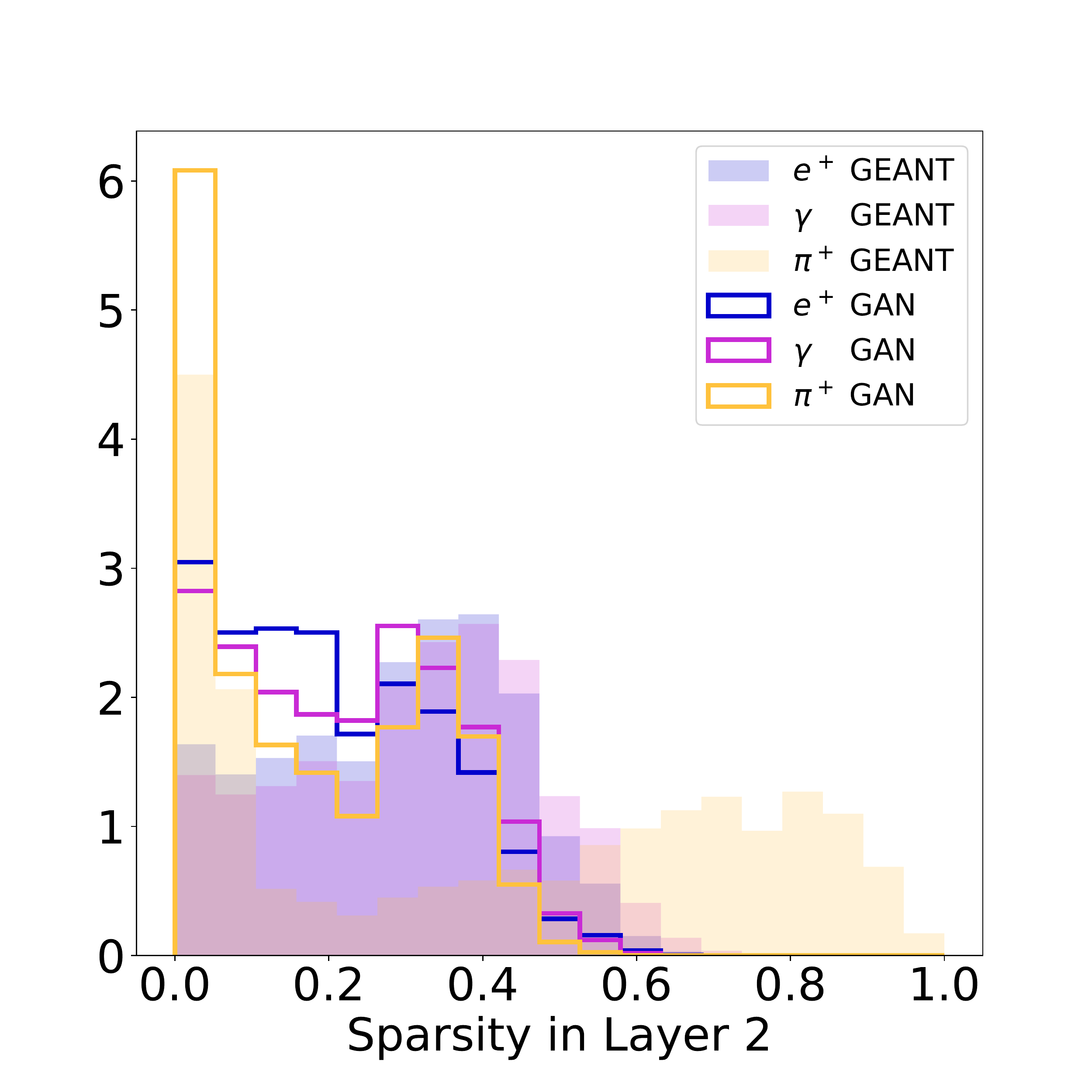}
    \includegraphics[width=0.2\textwidth]{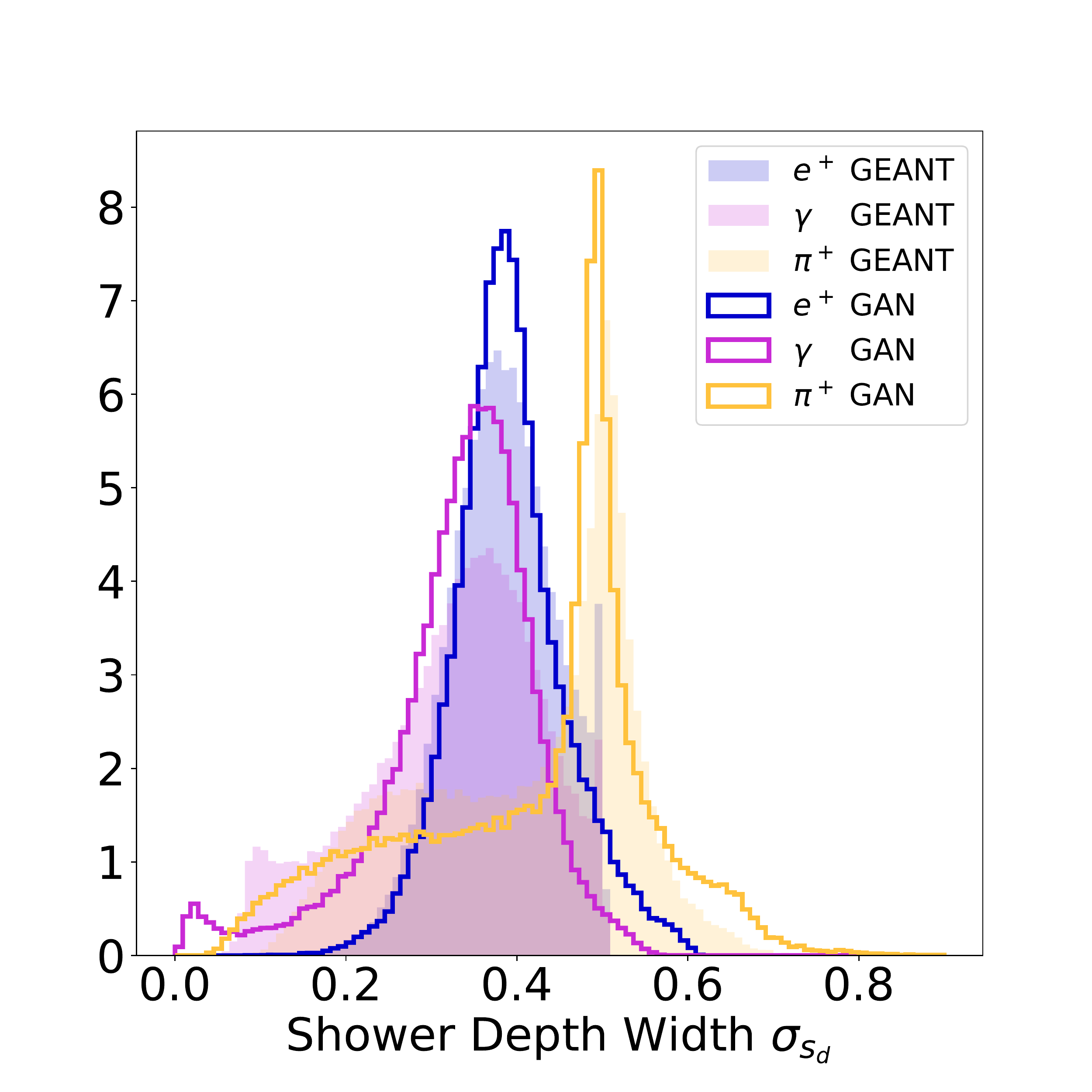}
    \caption{Comparison of shower shape variables, introduced in Table~\ref{table:qualityvariables}, and other variables of interest, such as the sparsity level per layer, for the \textsc{Geant4} and \textsc{CaloGAN} datasets for $e^+$, $\gamma$ and $\pi^+$.}
    \label{fig:shower_shapes}
\end{figure*}

The 1-dimensional distributions for \textsc{Geant4}- and GAN-generated samples are available in Fig.~\ref{fig:shower_shapes}.
Although the sparsity levels per layer are only roughly matched, note that, for the majority of the remaining variables, the GAN picks up on complex features in the distributions across several orders of magnitude and all particles types. The unique features that pions exhibit, compared to the other particles, make it unfavorable to train a single model for multiple particle types.

Note that shower shape variables were not explicitly part of the training, which is based only on the distribution of pixel intensities and energy. In the future, one can integrate the shower shape distributions into the loss-function itself. For now, we have left them out for a comprehensive validation assessment.

In addition to comparing shower shapes to reference distribution, we want to measure the quality of conditioning on energy. As outlined in Sec.~\ref{ssec:calogan_loss}, we cannot explicitly impose conservation of energy, but one can devise a simple sampling system to only keep simulated showers that obey this constraint. 

\begin{figure}
    \centering
    \includegraphics[width=0.48\textwidth]{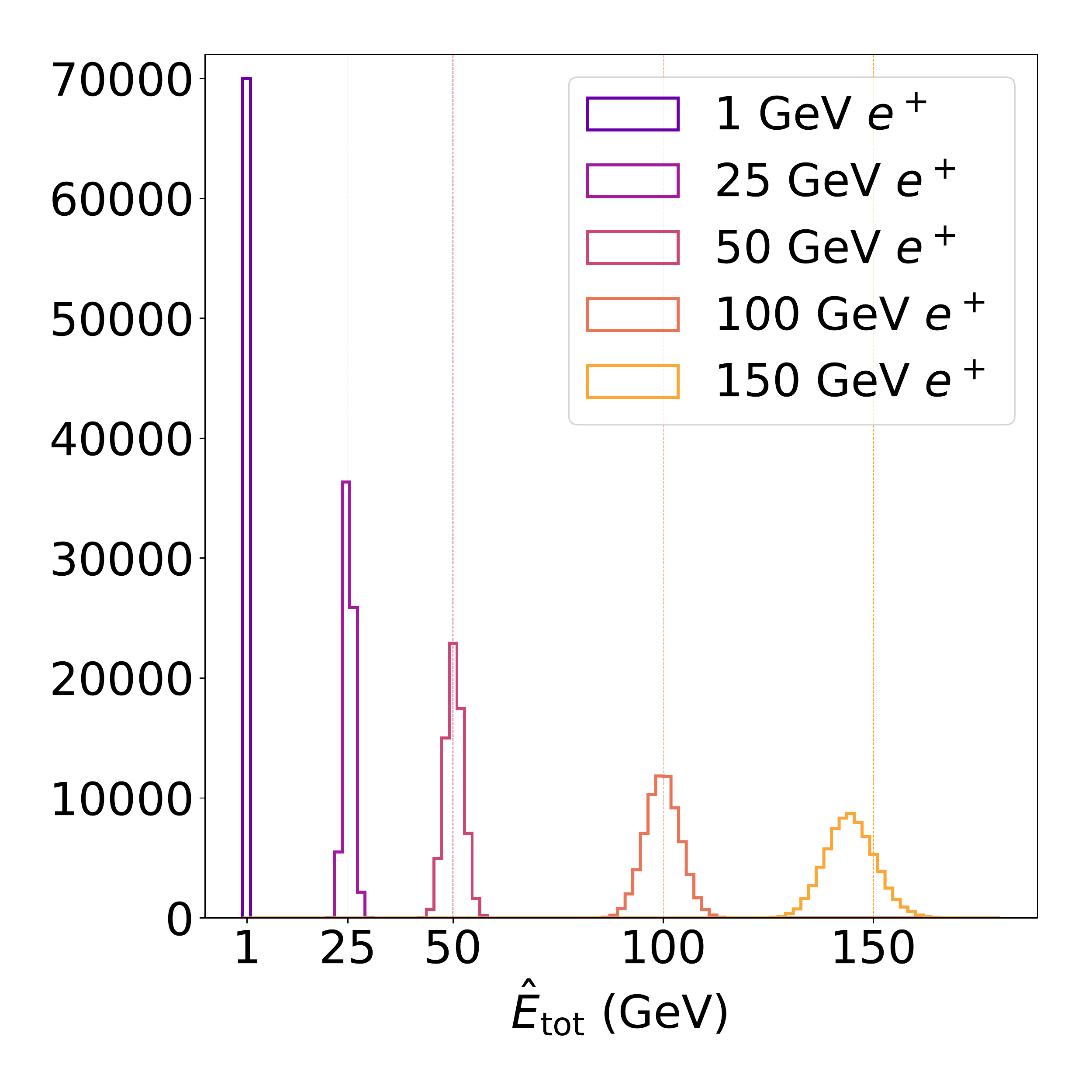}
    \caption{Post energy-conditioned empirical energy response for $e^+$ incident at 1, 25, 50, 100, and 150 GeV. Though our model is only trained on the unform range between 1 and 100 GeV, it still admits a compelling peak at 150 GeV.}
    \label{fig:single_request}
\end{figure}

As can be noted in the $e^+$ example in Fig.~\ref{fig:single_request}, our loss formulation coupled with the uniform training distribution admits an approximately symmetric conditional output energy distribution. In Fig.~\ref{fig:single_request}, note that the vertical lines that approximately coincide with the mode of each distribution represent the requested energy, and could easily be used as a threshold on selecting physical events. A noteworthy feature of this system is that one can request energies that lie outside the trained region (capped at 100 GeV in this application), to which a trained \textsc{CaloGAN} will return samples around the requested energy level -- though with broader width, and mode shifted towards the training domain. Whether or not these extrapolated samples obey shower shape distributions and other metrics is left as future work.

\subsection{Classification as a Performance Proxy}
\label{sec:classification}
Transferability of classification performance from GAN-generated samples to \textsc{Geant4}-generated samples can be used as a proxy both for \textsc{CaloGAN} image quality and potential utility in a practical fast simulation setting. 

We perform ten identical trainings of simple six-layer fully-connected $e-\gamma$ and $e-\pi$ classifiers, and we report the accuracies for in-domain and out-of-domain testing (Table~\ref{classification}) along with the following observations:
\begin{itemize}
\item when training on \textsc{Geant4}, testing on the generated \textsc{CaloGAN} data set yields similar results to testing on a separate \textsc{Geant4} data set, leading us to believe that the GAN has learned most of the discriminating physics between the classes of particles.  Note that percent-level differences in accuracy may however be relevant for particular applications;
\item the significantly higher performance obtained on the \textsc{CaloGAN}-generated test set when training on a separate dataset of \textsc{CaloGAN}-generated images  highlights a greater inter-class differentiation in the GAN synthetic dataset than originally present in the target \textsc{Geant4} distribution.
\end{itemize}

This could either be due to new unphysical, class-dependent features produced by the GAN, or to the inability of the GAN to cover the entire feature space for at least one of the particle classes. It is likely that both of these contribute.  To some extent, unphysical features are mitigated by the discriminator network of the GAN training itself, but both physical and unphysical features that are not very useful for distinguishing real from fake could turn into very useful features for the two-particle classification case. Such information would therefore appear discriminative in GAN images but not in \textsc{Geant4}. While classification is a useful metric for probing the high-dimensional feature space and shows promising results, there are still challenges for interpreting and improving upon the outcome.


\begin{table*}
\caption{Mean and standard deviation over 10 particle classification trials using a six-layer fully-connected network with dropout. The networks are trained using a dataset from the domain specified in the first column, and tested on an independent dataset from the domain specified in the header.}
\label{classification}
\renewcommand{\arraystretch}{1.5}
\begin{minipage}{0.45\linewidth}
\centering
\caption*{$e^+$ vs. $\pi^+$}
\begin{tabular}{@{}p{.16\textwidth}p{.21\textwidth}p{.27\textwidth}p{.27\textwidth}@{}}
\toprule
&                                                    & \multicolumn{2}{c}{\textbf{Test on}}                                                                             \\
                           & \multicolumn{1}{c}{}                              & \multicolumn{1}{c}{ \textsc{Geant4}} & \textsc{CaloGAN} \\ \cline{3-4} 
                           & \multicolumn{1}{c}{\textsc{Geant4}} & \multicolumn{1}{c}{99.6\% $\pm$ 0.1\%}                                                 & 96.5\% $\pm$ 1.1\%                           \\ 
\multirow{-2}{*}{\textbf{Train on}} & \multicolumn{1}{c}{\textsc{CaloGAN}}   & \multicolumn{1}{c}{98.2\% $\pm$ 0.9\%}                                                 & 99.9\% $\pm$ 0.2\% \\  
\bottomrule
\end{tabular}
\end{minipage}%
\begin{minipage}{.45\linewidth}
\centering
\caption*{$e^+$ vs. $\gamma$}
\renewcommand{\arraystretch}{1.5}
\begin{tabular}{@{}p{.16\textwidth}p{.21\textwidth}p{.27\textwidth}p{.27\textwidth}@{}}
\toprule
&                                                    & \multicolumn{2}{c}{\textbf{Test on}}                                                                             \\
                           & \multicolumn{1}{c}{}                              & \multicolumn{1}{c}{ \textsc{Geant4}} & \textsc{CaloGAN} \\ \cline{3-4} 
                           & \multicolumn{1}{c}{\textsc{Geant4}} & \multicolumn{1}{c}{66.1\% $\pm$ 1.2\%}                                                 & 70.6\% $\pm$ 2.6\%                           \\ 
\multirow{-2}{*}{\textbf{Train on}} & \multicolumn{1}{c}{\textsc{CaloGAN}}   & \multicolumn{1}{c}{54.3\% $\pm$ 0.8\%}                                                 & 100.0\% $\pm$ 0.0\% \\  
\bottomrule
\end{tabular}
\end{minipage}
\end{table*}

\subsection{Computational Performance}
\label{ssec:comp_perf}

In addition to the promise of being a high-fidelity fast simulation paradigm and respecting many shower shape variables, the \textsc{CaloGAN} affords many orders of magnitude in computational speedups~\footnote{Note that non-distributed training can take day(s), depending on the total number of training epochs, but it always executes in constant time regardless of the number of showers requested at generation time, so it is a fixed cost that is not relevant for the majority of the total computing budget.}. We benchmark generation time on $e^+$ with incident energy drawn uniformly between 1 GeV and 100 GeV. \textsc{Geant4} and \textsc{CaloGAN} on CPU are benchmarked on nearly identical compute-nodes on the PDSF distributed cluster at the National Energy Research Scientific Computing Center (NERSC), and numerical results are obtained over an average of 100 runs. \textsc{CaloGAN} on GPU hardware is benchmarked on an Amazon Web Service (AWS) \texttt{p2.8xlarge} instance, where a single NVIDIA\textsuperscript{\textregistered} K80 is used for the purposes of benchmarking. 

In Table~\ref{perf-table}, we show the time-to-generate a single particle shower in milliseconds. We provide different batch sizes for \textsc{CaloGAN}, as we expect different use-cases will have different demands around batching computation. We note that a batch can accept any number of different requested energies. With the largest batch sizes on GPU, our method admits a speedup of 5 orders of magnitude compared to the single-threaded \textsc{Geant4} benchmark. In addition, generation time with \textsc{Geant4} scales with incident energy, whereas computational time is flat as a function of incident energy for the \textsc{CaloGAN}. 

\begin{table}[]
\centering
\caption{Total expected time (in milliseconds) required to generate a single shower under various algorithm-hardware combinations}
\renewcommand{\arraystretch}{1.3}
\begin{tabular}{ccll@{}}
\toprule
\textbf{Simulator } & \textbf{Hardware } & \multicolumn{1}{c}{\textbf{Batch Size \ }} & \multicolumn{1}{c}{\textbf{ms/shower }} \\ \hline
\textsc{Geant4} & CPU & N/A & \cellcolor[HTML]{084081}{\color[HTML]{EFEFEF} 1772} \\ \hline
 &  & 1 & \cellcolor[HTML]{3294C2}{\color[HTML]{EFEFEF} 13.1} \\ \cline{3-4} 
 &  & 10 & \cellcolor[HTML]{8DD3BE}5.11 \\ \cline{3-4} 
 &  & 128 & \cellcolor[HTML]{BFE6BF}2.19 \\ \cline{3-4} 
 & \multirow{-4}{*}{CPU} & 1024 & \cellcolor[HTML]{C3E7C1}2.03 \\ \cline{2-4} 
 &  & 1 & \cellcolor[HTML]{2889BC}{\color[HTML]{EFEFEF} 14.5} \\ \cline{3-4} 
 &  & 4 & \cellcolor[HTML]{A5DCB6}3.68 \\ \cline{3-4} 
 &  & 128 & \cellcolor[HTML]{F2FAEB}0.021 \\ \cline{3-4} 
 &  & 512 & \cellcolor[HTML]{F3FAEC}0.014 \\ \cline{3-4} 
\multirow{-9}{*}{\textsc{CaloGAN}} & \multirow{-5}{*}{GPU} & 1024 & \cellcolor[HTML]{F3FBED}0.012 \\ \bottomrule
\end{tabular}
\label{perf-table}
\end{table}

\subsubsection{Implementation Notes}

As noted previously in Sec.~\ref{ssec:calogan_arch}, separating per-particle-type \textsc{CaloGAN} architectures and implementations affords many benefits. It is easy to imagine a situation where the life cycles surrounding models for different particle types are very different. In addition, this allows for total independence of versioning, framework, or language. 

 When possible, any GAN should maximally employ batching -- we imagine most applications can request all showers from one event simultaneously, maximally taking advantage of CPU/GPU while minimizing data transfer overhead.

\section{Conclusions and Future Outlook}
\label{sec:conclusions}
Using modern generative deep neural network techniques, we have generated three-dimensional electromagnetic showers in a multi-layer sampling LAr calorimeter with uneven spatial segmentation, while attempting to preserve spatio-temporal relation among layers.  Our approach infused Physics domain knowledge and reproduced many aspects of key shower shape properties comparable to the ones in the \textsc{Geant4} full simulation. We showed the possibility of up to five orders of magnitude decrease in computing time.

Future work will focus on improving performance by drawing from the recent Machine Learning developments in GAN training procedures, as well as testing the direct inclusion of important shower shape variables as constraints at training time.
Further developments will build on this result and continue expanding the complexity of the training dataset to include incoming particles at different locations and angles within the detector, as well as the hadronic calorimeter.
Concurrent plans include contributing to testing the computational performance on high performance computing (HPC) clusters, and porting these solutions into the simulation packages used by the nuclear and particle physics communities, in order for the various experiments to be able to maximally benefit from this new technology.

\begin{acknowledgments}
This work was supported in part by the Office of High Energy Physics of the U.S. Department of Energy under contracts DE-AC02-05CH11231 and DE-FG02-92ER40704. The authors would like to thank Wahid Bhimji, Zach Marshall, Mustafa Mustafa, and Prabhat, for helpful conversations.
\end{acknowledgments}

\bibliography{myrefs}

\appendix
\section{Shower Shape Variables}
\label{shower_shape_appendix}
Table~\ref{table:qualityvariables} contains the description and mathematical definition of the shower shape variables used to compare the generated and target distributions. These are defined as functions of $\mathcal{I}_i$, the vector of pixel intensities for an image in layer $i$, where $i \in \{0, 1, 2\}$.
\begin{table*}
\caption{One-dimensional observables used to assess the quality of the GAN samples}
\renewcommand{\arraystretch}{1.2}
\begin{tabular}{ m{4.2cm}  m{5.1cm}  m{4.3cm}} 
    \toprule
     {Shower Shape Variable} & {Formula} & \textit{Notes}
     \\
     \midrule
     $E_i$ & $E_i = \sum_\mathrm{pixels} \mathcal{I}_i $ & Energy deposited in the $i^{th}$ layer of calorimeter\\
     \midrule
     $E_\mathrm{tot}$ & $E_\mathrm{tot}=\sum\limits_{i=0}^2 E_i$ & Total energy deposited in the electromagnetic calorimeter\\
     \midrule
     $f_i$ & $f_i = E_i / E_\mathrm{tot}$ & Fraction of measured energy deposited in the $i^{th}$ layer of calorimeter\\
    \midrule
     $E_{\mathrm{ratio}, i}$ & $
\dfrac{\mathcal{I}_{i,(1)} - \mathcal{I}_{i,(2)}}{\mathcal{I}_{i,(1)} + \mathcal{I}_{i,(2)}}
$ & Difference in energy between the highest and second highest energy deposit in the cells of the $i^{th}$ layer, divided by the sum\\
\midrule
$d$ & $d = \max\{i: \max(\mathcal{I}_i) > 0\}$ & Deepest calorimeter layer that registers non-zero energy\\
\midrule
Depth-weighted total \newline energy, $l_d$ & $l_d=\sum\limits_{i=0}^2 i\cdot  E_i$ & The sum of the energy per layer, weighted by layer number \\
\midrule
Shower Depth, $s_d$ & $s_d = l_d / E_\mathrm{tot}$ & The energy-weighted depth in units of layer number \\
\midrule
Shower Depth Width, $\sigma_{s_d}$ & $$\sigma_{s_d} = \sqrt{\frac{\sum\limits_{i=0}^2 i^2\cdot  \mathcal{I}_i}{E_\mathrm{tot}} - \left(\frac{\sum\limits_{i=0}^2 i\cdot \mathcal{I}_i}{E_\mathrm{tot}}\right)^2}$$ &The standard deviation of $s_d$ in units of layer number \\
\midrule
$i^{\mathrm{th}}$ Layer Lateral Width, $\sigma_i$& $\sigma_i =\sqrt{\frac{\mathcal{I}_i \odot H^2}{E_i} - \left(\frac{\mathcal{I}_i \odot H}{E_i}\right)^2}$& The standard deviation of the transverse energy profile per layer, in units of cell numbers \\
\bottomrule
\end{tabular}
\label{table:qualityvariables}
\end{table*}

\end{document}